\documentclass[10pt,a4paper]{article}   

\usepackage{graphicx}   
\usepackage{amsmath,amssymb,bm}  
\usepackage{siunitx}    
\usepackage[colorlinks=true,allcolors=blue]{hyperref} 
\usepackage{caption}
\usepackage{subcaption}
\usepackage{booktabs} 
\usepackage{float}
\usepackage{geometry}   
\usepackage{setspace}   
\usepackage[colorlinks=true,allcolors=blue]{hyperref}
\newcommand{\shortauthors}{S. Das, L. Lachhvani and Others}
\newcommand{\shorttitle}{Nonlinear energy transfer between coherent modes}

\usepackage{fancyhdr}
\usepackage{titlesec}

\usepackage{titlesec}

\titleformat{\section}
  {\normalfont\normalsize\bfseries\centering}
  {\thesection}{0.5em}{\MakeUppercase}

\titlespacing*{\section}
  {0pt}      
  {1.6ex}    
  {0.85ex}    

\titleformat{\subsection}
  {\normalfont\normalsize\bfseries\centering}
  {\thesubsection}{0.5em}{\MakeUppercase}

\titlespacing*{\subsection}
  {0pt}{1.6ex}{0.85ex}

\titleformat{\subsubsection}
  {\normalfont\normalsize\bfseries\centering}
  {\thesubsubsection}{0.5em}{}

\titlespacing*{\subsubsection}
  {0pt}{1.3ex}{0.75ex}

\begin{document}
\fancypagestyle{firstpage}{
  \fancyhf{}
  \lhead{\footnotesize\itshape Under consideration for publication in J. Plasma Phys.}
  \cfoot{\thepage}
}

\thispagestyle{firstpage}

\begin{center}
{\LARGE\bfseries\linespread{0.95}\selectfont
\textcolor{black}{Nonlinear Energy Transfer Analysis between Coherent Modes in Developing Plasma Turbulence}\par}\vspace{1em}

\textbf{Sandip Das}\textsuperscript{1,$\dagger$},
\textbf{Lavkesh Lachhvani}\textsuperscript{1,2,$\ddagger
$},
\textbf{Kunal Singha}\textsuperscript{1,2},
\textbf{Rosh Roy}\textsuperscript{1,2},
\textbf{Tanmay Karmakar}\textsuperscript{1,2},
\textbf{Daniel Raju}\textsuperscript{1,2},
\textbf{Prabal Chattopadhyay}\textsuperscript{1,2}
\\[0.5em]

\textsuperscript{1}\textit{Institute for Plasma Research, Bhat, Gandhinagar 382428, India} \\[0.2em]
\textsuperscript{2}\textit{Homi Bhabha National Institute, Anushakti Nagar, Mumbai, 400094, India} \\[0.5em]

\today
\end{center}


\renewcommand{\thefootnote}{\fnsymbol{footnote}}

\footnotetext[2]{Email address for correspondence: 
\href{mailto:sandip.das@ipr.res.in}{\textit{sandip.das@ipr.res.in}}}

\footnotetext[3]{Email address for correspondence: 
\href{mailto:lavkesh@ipr.res.in}{\textit{lavkesh@ipr.res.in}}}

\renewcommand{\thefootnote}{\arabic{footnote}}

\begin{abstract}

Energy transfer among various spectral components of fluctuating physical parameters in plasma occurs due to the nonlinear interactions, but these effects are typically not captured by  the traditional linear spectral methods. Plasma density fluctuations measured in the Inverse Mirror Plasma Experimental Device (IMPED) have signatures of nonlinear mode interactions among various instability modes, i.e. Rayleigh–Taylor (RT) and Drift-Wave (DW) modes. In this paper, the energy transfer among these modes as a result of nonlinear wave interactions (through the quadratic coupling processes) have been investigated in detail. The existing computational methods for single field model such as Ritz method and Kim method have been explored to understand the dynamics of nonlinear interaction between \textcolor{black}{coherent modes}. Both methods are applied and validated in simulation as well as experimental data from IMPED for \textcolor{black}{coherent modes in plasma}. We find that the validity and applicability of the methods depend on the statistical nature of the data, particularly higher-order moments such as kurtosis, and on spatial stationarity. Energy transfer analysis using these methods reveals \textcolor{black}{primarily the transfer of energy from a RT mode and a mix mode of DW and RT to a comparatively low-frequency DW mode}, demonstrating the capability of the method to quantify spectral energy transport between the \textcolor{black}{coherent modes}.

\end{abstract}


\section{Introduction}
Turbulence generally evolves through distinct stages, beginning with the growth of small-amplitude fluctuations driven by linear instabilities and progressing toward a statistically stationary, broadband state. In the early, or \textit{developing}, phase, energy is injected into a limited set of unstable modes, producing coherent spectral peaks and a time-dependent redistribution of power as nonlinear interactions gradually become important. As the fluctuation amplitudes increase, nonlinear mode coupling transfers energy away from these unstable components, broadening the spectrum and forming a cascade toward damped scales. Eventually, a balance is established between linear growth, nonlinear transfer, and dissipation, leading to a \textit{fully developed} turbulent regime in which the statistical properties of the fluctuations become approximately stationary \cite{horton1999,ritz1989}.

In fluid and plasma turbulence, the time evolution of fluctuating physical quantities, such as plasma density and electrostatic potential, are often approximated as a superposition of statistically independent wave components. Under this assumption, the fluctuations can be partially characterized through their power spectra\cite{smith1974fft}. During the 1970s, linear spectral analysis techniques based on second-order statistics\cite{smith1974nonlinear, smith1973power}, including auto-power and cross-power spectra, were introduced as diagnostic tools for plasma waves and instabilities. These methods have since been widely and successfully used to investigate wave dynamics and spectral properties in laboratory and space plasmas. However, linear spectral methods inherently assume statistical independence between modes and therefore cannot capture the phase-coherent interactions generated by nonlinear dynamics. As a result, they provide limited insight into how energy is exchanged among spectral components. To overcome this limitation, higher-order spectral techniques have been developed to reveal nonlinear coupling processes. In particular, bispectral methods, based on third-order statistics, allow the identification of phase-correlated wave triplets and the assessment of nonlinear wave--wave interactions that remain hidden in conventional power spectra. Using the third-order cumulant, it has been demonstrated in both fluids and plasmas that three-wave coupling processes play an essential role in turbulent dynamics \cite{elgar1985,kim1979,miksad1983}. \textcolor{black}{Similar three-wave coupling interactions between coherent modes and broadband turbulence have also been reported in magnetized plasmas, such as the coupling between geodesic acoustic modes (GAMs) and broadband turbulence in the T-10 tokamak \cite{Melnikov2017}.
}

While the bispectrum is effective in identifying phase-coherent triadic interactions among waves, they primarily quantify the presence of nonlinear coupling and do not, by themselves, specify the \textit{direction} of energy flow. Subsequent studies have shown that if there are only few coherent modes, then direction of energy transfer can be computed as discussed by Kim \textit{et. al.}\cite{kim1980bispectrum}. However, this approach is not readily applicable to more complex situations such as developing or fully developed plasma turbulence, where a large number of interacting modes are present. To address this limitation, Ritz \textit{et al.} \cite{ritz1986,ritz1989,Ritz1988} introduced the concept of a \textit{power transfer function} followed by \textit{energy transfer function}, which quantifies the net nonlinear energy exchange among spectral modes. This formulation provides both qualitative and quantitative insight into how fluctuation energy is redistributed across the spectrum. However, this method does not provide very accurate results for general fluctuation data because it approximates the fourth-order moments using the square of second-order moments \cite{ritz1989}. Moreover, it produces large, non-physical damping coefficients when applied to fully developed turbulent systems, both in simulations \cite{Shen2020ImprovedBispectrum} and in measured fluctuation data \cite{ritz1986}. Consequently, Ritz’s method performs reasonably well to estimate quadratic nonlinearity only for developing turbulence models, whereas it fails for fully developed turbulent systems \cite{Shen2020ImprovedBispectrum}. To address limitations of the Ritz method, Kim \textit{et al.} \cite{kim1996} introduced a modified formulation that retains fourth-order cumulants and separates the data into ideal and nonideal components. In this framework, only the ideal part is assumed to participate in linear growth (or damping) and three-wave coupling processes. Further it is observed that, in models of fully developed turbulence, unphysical behavior in the estimated damping coefficients can arise if nonideal contributions—such as higher-order nonlinear couplings, systematic errors, or measurement noise—are neglected\cite{ritz1989, Shen2020ImprovedBispectrum}. To mitigate these effects, they imposed a stationarity condition by enforcing equality between the input and output ideal powers, thereby reducing the influence of nonideal components on the inferred energy transfer\cite{kim1996}. Ritz’s method is generally not well suited for fully developed turbulent systems, whereas Kim’s method can be applied to both developing and fully developed turbulence, as discussed in Ref.~\cite{Shen2020ImprovedBispectrum}. There are, however, situations in which neither Ritz’s nor Kim’s method performs satisfactorily. For example, in modeling nonlinear drift oscillations of moored vessels under non-Gaussian random sea-wave excitation, Ritz’s method becomes ineffective due to system and measurement errors, while, Kim’s method fails in this case due to a strong power imbalance between the input and output spectra, which exhibit different spectral trends, leading to the loss of important spectral information and inaccurate transfer estimates \cite{kim1987ocean}.
 In such cases, bispectral analysis can still be applied effectively by employing the \textit{Complete Iterative Method}, as discussed by Shen \textit{et al.}~\cite{Shen2020ImprovedBispectrum}.

The physical origin of the three-wave coupling processes mainly comes from the quadratic $\mathbf{E}\times\mathbf{B}$ convective nonlinearity inherent in drift-wave dynamics. Drift waves (DW) arise due to the background density gradient $\nabla n_0$, where electrostatic potential fluctuations generate an $\mathbf{E}\times\mathbf{B}$ velocity that advects the equilibrium density through the term $\mathbf{v}_E \cdot \nabla n_0$\cite{anderson2020elucidating, hasegawa1983plasma}. In contrast, the Rayleigh--Taylor/interchange (RT) mode develops when an effective gravitational (fluids) or magnetic curvature (toroidal plasmas) or centrifugal (linear magnetized plasma) force acts on the same unfavorable density gradient, producing flute-like perturbations that extract energy from the equilibrium \cite{krall1973principles}. In both cases, the nonlinear interaction appears through the quadratic term $\mathbf{v}_E \cdot \nabla F$, where, $F$ represents advected plasma fields such as density or vorticity, which reduces in two-dimensional geometry to the Poisson bracket $\{\phi, F\}$ and is explicitly quadratic in fluctuation amplitudes \cite{TerryDiamond1985}. In Fourier space, this nonlinear term generates triadic interactions satisfying the resonance conditions $\mathbf{k} = \mathbf{k}_1 + \mathbf{k}_2$ and $f = f_1 + f_2$. Finite collisionality further enhances this process by breaking adiabatic electron shielding, increasing the density–potential phase shift, and broadening the unstable spectral bandwidth; this leads to larger fluctuation amplitudes and strengthens the nonlinear $\mathbf{E}\times\mathbf{B}$ coupling \cite{hasegawa1983plasma}.

In this work, we present the nonlinear three-wave interaction between DW and RT modes, identified using the digital bispectrum, which directly captures phase coupling arising from the quadratic Poisson-bracket nonlinearity. The efficiency of this nonlinear coupling is enhanced when the DW and RT eigen functions spatially coincide, allowing stronger nonlinear coupling and sustained phase locking, as evidenced by finite bispectrum\cite{kim1979}. The existing computational methods for single field turbulence model such as Ritz method and Kim method have been explored to understand the
turbulence dynamics. We observe that the Ritz method performs well for developing turbulent systems, its accuracy depends strongly on the statistical properties of the data. In particular, it is reliable for fluctuations that are close to Gaussian nature, characterized by low skewness and kurtosis. When the kurtosis is large, the method deteriorates because Ritz’s formulation approximates fourth-order moments using second-order statistics, whereas kurtosis directly reflects fourth-order contributions. Consequently, strong departures from Gaussianity introduce errors in the estimated transfer functions. This issue has been systematically investigated in this work. We examine the applicability of the Ritz method to both simulation data and experimental fluctuation measurements, with emphasis on the role of statistical properties in developing plasma turbulence. We show that Ritz’s method provides reliable transfer functions only when the data are close to Gaussian, and that beyond a threshold level of kurtosis it fails to recover physically meaningful transfer functions, the result verified using both simulated and measured fluctuations from the IMPED. In contrast, Kim’s method performs robustly regardless of the statistical character of the data. Even for signals with large kurtosis, Kim’s formulation yields consistent and physically correct transfer functions, as demonstrated in both simulations and experiments. Although Kim’s method was originally developed for fully developed, stationary turbulence where input and output powers balance, we apply it here to developing turbulent regimes in which the spectra are nonstationary and not fully \textcolor{black}{coincidence} but showing similar trends. Despite this, Kim’s method continues to provide accurate transfer functions for both simulated and experimental fluctuation data from IMPED.

The flow of the paper is organized as follows. In section~(\ref{section: sepectral framework}), we introduce the spectral analysis tools required to compute linear, quadratic, and nonlinear energy transfer functions. Section~(\ref{section: methods}) describes both Ritz’s and Kim’s methods, including simulation tests used to validate Ritz’s approach and to examine its applicability with respect to the statistical properties of the data. This section also presents Kim’s method, along with simulation studies demonstrating its performance and its ability to address the limitations of Ritz’s formulation. In Section~(\ref{section: application to expt data}), we apply both methods to experimental fluctuation data, including nonlinear transfer-function and growth-rate analyses at two radial locations, thereby assessing the validity of each approach in experimental data. \textcolor{black}{Finally, Sections~(\ref{section: discussion}) and (\ref{section: conclusion}) present the discussion and conclusions, respectively, while Section~(\ref{section: outlook}) presents the outlook and future directions of the present work.}

\section{Spectral Analysis Framework for Nonlinear Energy Transfer Function Estimation}
\label{section: sepectral framework}

This section establishes the spectral analysis framework required for estimating
nonlinear energy transfer functions. Linear interactions are characterized using
second-order spectral quantities such as the power spectral density, which
describe the distribution of energy across frequencies. Nonlinear interactions,
arising from quadratic phase coupling and triadic interactions, are captured
using higher-order spectral measures. We consider time-series signals divided
into $M$ ensembles, each containing $N$ data points, and estimate Fourier-based
spectral quantities through ensemble averaging. In addition to auto-spectral
measures, cross-spectral quantities are introduced to characterize inter-signal
coupling and associated energy transfer mechanisms. The definitions of these
spectral quantities are provided below.

\noindent\textit{(a) Auto Power Spectral Density (PSD): }It describes how the signal power is distributed
across frequencies. It is defined as\cite{SmithPowersCaldwell1974}
\begin{equation}
    P_{a}(f) = \frac{1}{M} \sum_{l=1}^M X_f^{*l} X_f^l ,
\label{eq:aPSD}
\end{equation}
where $X_f^l$ denotes the Fourier transform of the $l$-th ensemble and $*$ represents
complex conjugation.\\

\noindent\textit{(b) Cross Power Spectral Density(CSD): }For two time-series signals $x(t)$ and $y(t)$ with Fourier transforms
$X_f$ and $Y_f$, respectively, the cross power spectral density (CSD) is defined as\cite{SmithPowersCaldwell1974}
\begin{equation}
    P_{c}(f) = \frac{1}{M} \sum_{l=1}^M X_f^{*l} Y_f^l .
\end{equation}
The CSD quantifies the correlation and phase relationship between the two signals
at each frequency and is generally complex-valued.\\

\noindent\textit{(c) Auto Bispectrum: } The auto bispectrum is a third-order spectral quantity that captures nonlinear
interactions and phase coupling between different frequency components. It is
defined as\cite{kim1979}
\begin{equation}
    B_a(f_1,f_2) = \frac{1}{M}\sum_{l=1}^M
    X_{f_1}^l \, X_{f_2}^l \, X_{f_1+f_2}^{*l}.
\end{equation}
The auto bispectrum quantifies phase-coupled interactions among frequency triplets
$(f_1, f_2, f_1+f_2)$ and is sensitive to quadratic nonlinearities in the signal.

\noindent\textit{(d) Cross Bispectrum:}
For two signals $x(t)$ and $y(t)$, the cross bispectrum is defined as\cite{kim1979}
\begin{equation}
    B_{c}(f_1,f_2) =
    \frac{1}{M}\sum_{l=1}^M
    X_{f_1}^l \, X_{f_2}^l \, Y_{f_1+f_2}^{*l}.
\end{equation}
The cross bispectrum characterizes quadratic phase coupling between frequency
components of one signal and their nonlinear interaction with another signal.

\noindent\textit{(e) Auto Bicoherence:} To obtain a normalized measure of phase coupling that is independent of signal
amplitude, the bicoherence is defined as\cite{kim1979}
\begin{equation}
    b_a(f_1,f_2) =
    \frac{\left| \frac{1}{M}\sum_{l=1}^M
    X_{f_1}^l X_{f_2}^l X_{f_1+f_2}^{*l} \right|}
    {\left(\frac{1}{M}\sum_{l=1}^M |X_{f_1}^l X_{f_2}^l|^2\right)^{1/2}
     \left(\frac{1}{M}\sum_{l=1}^M |X_{f_1+f_2}^l|^2\right)^{1/2}} .
\label{eq:abicoh}
\end{equation}

The auto bicoherence measures the normalized strength of phase coupling between
$f_1$, $f_2$, and $f_1+f_2$, with values ranging from $0$ (no coupling) to $1$
(perfect phase coupling).

\noindent\textit{(f) Cross Bicoherence:} To obtain a normalized measure of quadratic phase coupling between two different
signals that is independent of signal amplitude, the cross bicoherence is defined as\cite{kim1979}
\begin{equation}
    b_{c}(f_1,f_2) =
    \frac{\left| \frac{1}{M}\sum_{l=1}^M
    X_{f_1}^l X_{f_2}^l Y_{f_1+f_2}^{*l} \right|}
    {\left(\frac{1}{M}\sum_{l=1}^M |X_{f_1}^l X_{f_2}^l|^2\right)^{1/2}
     \left(\frac{1}{M}\sum_{l=1}^M |Y_{f_1+f_2}^l|^2\right)^{1/2}} .
\end{equation}

The cross bicoherence measures the normalized strength of phase coupling between
frequency components $f_1$ and $f_2$ of one signal and their nonlinear interaction
with the component $f_1+f_2$ of another signal. Its values range from $0$
(no phase coupling) to $1$ (perfect quadratic phase coupling).

For random (uncorrelated) signals, the expected bicoherence is not exactly zero due to finite ensemble averaging. 
Therefore, the statistical significance level for the bicoherence is given by\cite{kim1979,Nikias1993},
\begin{equation}
b_{\mathrm{sig}} \approx \frac{1}{\sqrt{M}}
\label{eq:bico_sig}
\end{equation}

The nonlinear coupling is considered statistically significant when $b(f_{1},f_{2}) > \frac{1}{\sqrt{M}}$. Also, for squared bicoherence the statistical significance level is written as: $b_{\mathrm{sig}}^{2} \approx \frac{1}{{M}}$\cite{Nikias1993}.

The spectral and higher-order spectral quantities introduced above provide a framework for characterizing linear and nonlinear interactions among frequency components. Power and cross-power spectra describe energy distribution and linear coupling, while bispectral measures capture triadic phase-coupled interactions responsible for nonlinear energy exchange. These quantities form the basis for constructing the nonlinear energy transfer functions developed in the following section.

\section{Methods for Estimating Nonlinear Energy Transfer Function}
\label{section: methods}
Using the spectral framework developed in the previous section, we describe two approaches for estimating nonlinear energy transfer between spectral components: (i) the \textit{Ritz method}, based on the work of Ritz et al. \cite{ritz1989, ritz1986, Ritz1988}, and (ii) the \textit{Kim method}, based on the work of Kim et al. \cite{kim1996}. The formulation and implementation of these methods are presented in the following subsections.

\subsection{Ritz Method}
\label{subsection: Ritz method}
Based on the nonlinear energy transfer formalism introduced by \textit{Ritz et al.}\cite{ritz1989} and subsequent studies\cite{ritz1986, Ritz1988}, the evolution of turbulent spectral components can be described using a nonlinear wave coupling equation. This formulation separates the dynamics into linear and quadratic interaction terms, providing a natural framework to analyze the direction of nonlinear energy transfer among interacting modes.
\textcolor{black}{Although Ritz's original formulation was developed in wavenumber space, we apply it in the frequency domain using Taylor's frozen-flow hypothesis \cite{Taylor1938,Howes2014}, which enables a direct transformation between wavenumber and frequency ($k \rightarrow f$, $f \rightarrow k$). This hypothesis assumes that the fluctuations are effectively ``frozen'' and are advected past the probe by the background flow without undergoing significant evolution during the measurement interval. This hypothesis, expected to be valid when the advection velocity exceeds the intrinsic phase velocity of the fluctuations. Physically, this implies that the temporal variations measured by a stationary probe arise predominantly from the convection of spatial structures rather than from their intrinsic temporal evolution. The applicability of the hypothesis may be reduced when the intrinsic phase velocity becomes comparable to the advection velocity or when the fluctuations evolve significantly during the measurement interval \cite{Howes2014,Klein2014Violation}. In the present experiment, the measured $E\times B$ flow velocity exceeds the phase velocity of the dominant  modes at the investigated radial locations, thereby supporting the applicability of Taylor's hypothesis for the frequency-to-wavenumber conversion employed in the energy transfer analysis. }This approach simplifies the experimental procedure: wavenumber-space analysis typically requires multi-probe measurements, whereas frequency-space analysis can be performed with a two-probe setup. The governing equation in the frequency domain can then be expressed as\cite{Ritz1988}

\begin{equation}
\frac{\partial \phi(\textit{f},x)}{\partial x} = \Lambda_f^{L} \phi(\textit{f},x) + \frac{1}{2} \sum_{\substack{\textit{f}_1,\, \textit{f}_2 \\ \textit{f}_1 + \textit{f}_2 = \textit{f}}} \Lambda_{\textit{f}}^{Q}(\textit{f}_1, \textit{f}_2)\, \phi(\textit{f}_1,x)\, \phi(\textit{f}_2,x)
\label{eq:ritz1}
\end{equation}

where \( \phi(\textit{f},x) \) is the spatial Fourier spectrum of the fluctuating field \( \varphi(t,x) \) defined by:
\[
\phi(\textit{f},x) = \sum_{x} \varphi(t,x) e^{-i \textit{f}t}
\]


\textcolor{black}{Here, we assume that $\phi(t,x)$ is temporally stationary and has a zero mean value. Temporal stationarity implies that the statistical properties of the signal are invariant in time; in particular, the power associated with each frequency remains unchanged across different time segments.} \textcolor{black}{Eq.~\eqref{eq:ritz1} represents the nonlinear wave-coupling equation governing the evolution of the Fourier spectrum through linear and quadratic nonlinear interactions. These linear and nonlinear processes are characterized by the \textcolor{black}{\textit{linear transfer coefficient}}($\Lambda_f^{L}$) and the quadratic nonlinear coupling coefficient ($\Lambda_{\textit{f}}^{Q}$), respectively. The linear coefficient accounts for physical effects such as linear growth (or damping) and dispersion, whereas the quadratic nonlinear coupling coefficient characterizes the strength of the three-wave interaction. It governs the transfer of energy among coupled modes through three-wave processes, whereby a parent wave ($f$) can decay into two daughter waves ($f_1$) and ($f_2$), or conversely, two waves can merge to form a single mode. The quadratic coupling coefficient arises from convective nonlinearities, including the nonlinear polarization drift and the $\mathbf{E}\times\mathbf{B}$ advection\cite{hasegawa1978}. These nonlinear interactions facilitate the transfer of energy between different modes through wave merging and decay processes, thereby playing a significant role in the spectral evolution of the plasma fluctuations.} The \textcolor{black}{linear transfer coefficient} is defined as \( \Lambda_f^{L} = \gamma_f + ik_f \), with \( \gamma_f \) and \( k_f \) being the linear growth rate and the dispersion relation, respectively. \( \Lambda_{\textit{f}}^{Q} \) is the \textit{nonlinear coupling coefficient}, giving the strength of mode coupling between \( \textit{f}_1 \) and \( \textit{f}_2 \).

The nonlinear wave coupling equation arises from the retention of convective nonlinearities in the governing dynamical equations. For neutral fluids, this coupling originates from the nonlinear advective term $(\mathbf{v}\cdot\nabla)\mathbf{v}$ in the Navier–Stokes equations, whose Fourier representation naturally leads to triadic mode interactions. In plasma systems, an analogous coupling appears in drift-wave turbulence through the nonlinear polarization drift and $\mathbf{E}\times\mathbf{B}$ convection, and is formally described by the Hasegawa–Mima equation\cite{hasegawa1978}. Similar three-wave coupling dynamics have been invoked to model turbulence in diverse physical systems, including shallow-water surface waves\cite{FreilichGuza1984} and atmospheric Rossby waves\cite{HasegawaMaclennanKodama1979}, providing a common framework for describing nonlinear energy transfer across scales.

The method itself consists in computing the fast Fourier transforms, \( \phi(\textit{f},x) \) and \( \phi(\textit{f}, x + \Delta x) \), from the data of two time series measured at a fixed spatial position \( x \) and at another different position \( x + \Delta x \) (i.e., \( \varphi(t,x) \) and \( \varphi(t,x + \Delta x) \)). We can consider the first signal \( \phi(\textit{f},x) \) as the \textit{input} and the space separated signal \( \phi(\textit{f},x + \Delta x) \) as the \textit{output}, such that the output signal can be expressed in terms of the input signal via the linear and quadratic transfer functions, in which the linear and nonlinear coupling coefficients are involved.

We start from Eq.~\eqref{eq:ritz1} to establish the relation between the output and input signals for a change of the spectrum along \( x \). Considering that the Fourier spectrum of \( \phi(f,x) \) is a complex quantity \( \phi(f,x) = |\phi(f,x)|e^{i\Theta(f,x)} \), where \( |\phi(f,x)| \) is the amplitude and \( \Theta(f,x) \) is the phase, we can estimate how it evolves along \( x \) by a difference approach: 


\begin{align}
\frac{\partial \phi(\textit{f},x)}{\partial x}
      &= \frac{\partial\,|\phi(\textit{f},x)|}{\partial x}\,
         e^{i\Theta(\textit{f},x)}
       + |\phi(\textit{f},x)|\,\frac{\partial e^{i\Theta(\textit{f},x)}}{\partial x}
       \notag \\[4pt]
      &= \frac{\partial\,|\phi(\textit{f},x)|}{\partial x}\,
         e^{i\Theta(\textit{f},x)}
       + 
         \frac{\partial\,\Theta(\textit{f},x)}{\partial x}\,
         i\,e^{i\Theta(\textit{f},x)}|\phi(\textit{f},x)|.
\end{align}

Using the forward finite difference formulas, i.e.,
\begin{equation}
\begin{aligned}
\frac{\partial\,|\phi(\textit{f},x)|}{\partial x} 
&\approx \frac{|\phi(\textit{f},x+\Delta x)| - |\phi(\textit{f},x)|}{\Delta x}, \\
\frac{\partial\,\Theta(\textit{f},x)}{\partial x} 
&\approx \frac{\Theta(\textit{f},x+\Delta x) - \Theta(\textit{f},x)}{\Delta x}.
\end{aligned}
\end{equation}

respectively, in Eq.~\eqref{eq:ritz1}, and rearranging the terms, we obtain:

\begin{align}
\phi(\textit{f}, x + \Delta x) 
&= \frac{\Lambda_{\textit{f}}^{L} \Delta x + 1 - i \left[ \Theta(\textit{f}, x + \Delta x) - \Theta(\textit{f}, x) \right]}
         {e^{-i\left[ \Theta(\textit{f}, x + \Delta x) - \Theta(\textit{f}, x) \right]}} \, \phi(\textit{f}, x) \notag \\
&\quad + \frac{1}{2} \sum_{\substack{\textit{f}_1,\, \textit{f}_2 \\ \textit{f}_1 + \textit{f}_2 = \textit{f}}} 
\frac{\Lambda_{\textit{f}}^{Q}(\textit{f}_1, \textit{f}_2) \, \Delta x}
     {e^{-i\left[ \Theta(\textit{f}, x + \Delta x) - \Theta(\textit{f}, x) \right]}} \,
\phi(\textit{f}_1, x)\, \phi(\textit{f}_2, x)
\label{eq:ritz2}
\end{align}

If we define the \textit{input} and \textit{output} Fourier spectra as:

\begin{equation}
\begin{aligned}
\textit{Input: } \quad X &= \phi(\textit{f}, x) \\
\textit{Output: } \quad Y &= \phi(\textit{f}, x + \Delta x)
\end{aligned}
\end{equation}

and the \textit{linear and quadratic transfer functions} as:

\begin{align}
L_{\textit{f}} &=
\frac{\Lambda_{\textit{f}}^{L} \Delta x + 1 - i \left[ \Theta(\textit{f}, x + \Delta x) - \Theta(\textit{f}, x) \right]}
     {e^{-i\left[ \Theta(\textit{f}, x + \Delta x) - \Theta(\textit{f}, x) \right]}}
\label{eq:L_lam} \\[6pt]
Q_{\textit{f}}^{\textit{f}_1 \textit{f}_2} &=
\frac{\Lambda_{\textit{f}}^{Q}(\textit{f}_1, \textit{f}_2)\, \Delta x}
     {e^{-i\left[ \Theta(\textit{f}, x + \Delta x) - \Theta(\textit{f}, x) \right]}}
\label{eq:Q_lam}
\end{align}

then, Eq.~\eqref{eq:ritz2} can be written concisely as:

\begin{equation}
Y_{\textit{f}} = L_{\textit{f}} X_{\textit{f}}  + \sum_{\substack{\textit{f}_1 \geq \textit{f}_2 \\ \textit{f} = \textit{f}_1 + \textit{f}_2}} 
Q_{\textit{f}}^{\textit{f}_1 \textit{f}_2} X_{\textit{f}_1}\, X_{\textit{f}_2}
\label{eq:Y_X}
\end{equation}

\subsubsection{Estimation of Linear and Quadratic Transfer Functions}
\label{subsubsection: estimation of L and Q}
We now proceed to determine $L_f$ and $Q_f^{f_1 f_2}$, and apply them within the framework of the previously derived expressions.dispersion relations derived above. To obtain the expressions of \( L_{\textit{f}} \) and \( Q_{\textit{f}}^{\textit{f}_1 \textit{f}_2} \), one can multiply Eq.~\eqref{eq:Y_X} with the complex-conjugate of \(X_{\textit{f}}\), and \( X_{\textit{f}_1} X_{\textit{f}_2}\ \), respectively, and average over many statistically similar realizations. The results are as follows:

\begin{equation}
\langle Y_{\textit{f}} X_{\textit{f}}^* \rangle = L_{\textit{f}} \langle X_{\textit{f}} X_{\textit{f}}^* \rangle + 
\sum_{\substack{\textit{f}_1 \geq \textit{f}_2 \\ \textit{f} = \textit{f}_1 + \textit{f}_2}}  
Q_{\textit{f}}^{\textit{f}_1, \textit{f}_2} \langle X_{\textit{f}_1} X_{\textit{f}_2} X_{\textit{f}}^* \rangle
\end{equation}

\begin{equation}
\langle Y_{\textit{f}} X_{\textit{f}_1^{\prime}}^{*} X_{\textit{f}_2^{\prime}}^{*} \rangle = 
L_{\textit{f}} \langle X_{\textit{f}} X_{\textit{f}_1^{\prime}}^{*} X_{\textit{f}_2^{\prime}}^{*} \rangle + 
\sum_{\substack{\textit{f}_1 \geq \textit{f}_2 \\ \textit{f} = \textit{f}_1 + \textit{f}_2}} 
Q_{\textit{f}}^{\textit{f}_1, \textit{f}_2} \langle X_{\textit{f}_1} X_{\textit{f}_2} X_{\textit{f}_1^{\prime}}^{*} X_{\textit{f}_2^{\prime}}^{*} \rangle
\label{eq:ritz4}
\end{equation}

where,  \( \textit{f} = \textit{f}_1 + \textit{f}_2 = \textit{f}_1^{\prime} + \textit{f}_2^{\prime}\) ; The notation $\langle \cdot \rangle$ denotes the ensemble average over realizations.\\

The evaluation of the fourth-order moment 
$\langle X_{f_1} X_{f_2} X_{f_1'}^{*} X_{f_2'}^{*} \rangle$ 
is computationally demanding. To reduce the complexity, the Ritz method employs Millionshchikov's approximation~\cite{ritz1986, Millionshchikov1941}, whereby contributions from terms satisfying 
$(f_1, f_2) \neq (f_1', f_2')$ 
are neglected. A similar approach has been employed in various theoretical models of strong turbulence\cite{TerryDiamond1985, TerryDiamondShaing1986}. Consequently, Eq.~\eqref{eq:ritz4} reduces to:

\begin{equation}
\langle Y_{\textit{f}} X_{\textit{f}_1}^{*} X_{\textit{f}_2}^{*} \rangle = 
L_{\textit{f}} \langle X_{\textit{f}} X_{\textit{f}_1}^{*} X_{\textit{f}_2}^{*} \rangle + 
Q_{\textit{f}}^{\textit{f}_1 \textit{f}_2} \langle |X_{\textit{f}_1} X_{\textit{f}_2}|^2 \rangle
\label{eq:ritz5}
\end{equation}

Operating on this equation one gets:

\begin{equation}
Q_{\textit{f}}^{\textit{f}_1 \textit{f}_2} = 
\frac{\langle Y_{\textit{f}} X_{\textit{f}_1}^{*} X_{\textit{f}_2}^{*} \rangle - 
L_{\textit{f}} \langle X_{\textit{f}} X_{\textit{f}_1}^{*} X_{\textit{f}_2}^{*} \rangle}
     {\langle |X_{\textit{f}_1} X_{\textit{f}_2}|^2 \rangle}
\label{eq:ritz6}
\end{equation}

Substituting this expression of \( Q_{\textit{f}}^{\textit{f}_1 \textit{f}_2} \) into Eq.~\eqref{eq:ritz5} finally, we get:

\begin{equation}
L_{\textit{f}} =
\frac{
\langle Y_{\textit{f}} X_{\textit{f}}^{*}\rangle - 
\sum_{\substack{\textit{f}_1 \geq \textit{f}_2 \\ \textit{f} = \textit{f}_1 + \textit{f}_2}} 
\frac{
\langle X_{\textit{f}}^{*} X_{\textit{f}_1} X_{\textit{f}_2} \rangle
\langle Y_{\textit{f}} X_{\textit{f}_1}^{*} X_{\textit{f}_2}^{*} \rangle
}
{\langle |X_{\textit{f}_1} X_{\textit{f}_2}|^2 \rangle}
}
{
\langle X_{\textit{f}} X_{\textit{f}}^{*} \rangle - 
\sum_{\substack{\textit{f}_1 \geq \textit{f}_2 \\ \textit{f} = \textit{f}_1 + \textit{f}_2}} 
\frac{
|\langle X_{\textit{f}}^{*} X_{\textit{f}_1} X_{\textit{f}_2} \rangle|^2
}
{\langle |X_{\textit{f}_1} X_{\textit{f}_2}|^2 \rangle}
}
\label{eq:ritz7}
\end{equation}

which is needed to calculate the \textit{linear growth rate}(\( \gamma_{\textit{f}} \)) and dispersion relation$(k_f)$. The value of the \textit{quadratic transfer function} \( Q_{\textit{f}}^{\textit{f}_1 \textit{f}_2} \) can be obtained by substituting \( L_{\textit{f}}\) into Eq.~\eqref{eq:ritz6} together with the “averaged products” of Fourier transforms of the measured signals.
In our notation of spectral quantities described in section-\ref{section: sepectral framework}, we can rewrite Eq.~\eqref{eq:ritz6} and Eq.~\eqref{eq:ritz7} as:

\begin{equation}
L_{\textit{f}} =
\frac{
P_c(f) - 
\sum_{\substack{\textit{f}_1 \geq \textit{f}_2 \\ \textit{f} = \textit{f}_1 + \textit{f}_2}} 
\frac{
B_a(f_1, f_2) B_c^*(f_1, f_2)
}
{E_{12}}
}
{
P_a(f) - 
\sum_{\substack{\textit{f}_1 \geq \textit{f}_2 \\ \textit{f} = \textit{f}_1 + \textit{f}_2}} 
\frac{
|B_a(f_1, f_2)|^2
}
{E_{12}}
}
\label{eq:L_f}
\end{equation}

\begin{equation}
Q_{\textit{f}}^{\textit{f}_1 \textit{f}_2} = 
\frac{B_c^*(f_1, f_2) - 
L_{\textit{f}} B_a^*(f_1, f_2)}
     {E_{12}}
\label{eq:Q_f}
\end{equation}

Here, we denoted $E_{12} = \langle |X_{\textit{f}_1} X_{\textit{f}_2}|^2 \rangle$. These transfer function definitions are used to estimate the corresponding quantities in the following sections.

\subsubsection{Estimation of Power and Energy Transfer Functions}
\label{subsubsection: estimation of T and W}
One can write a wave kinetic equation for the power spectrum 
\( P_{\textit{f}} = \langle X_{\textit{f}} X_{\textit{f}}^* \rangle \) 
in terms of the coupling coefficients \( \Lambda_{\textit{f}}^L \) and \( \Lambda_{\textit{f}}^Q \), 
by multiplying Eq.~\eqref{eq:ritz1} with \( \phi^*(\textit{f}, x) \)~\cite{ritz1989}.

\begin{equation}
\frac{\partial \left[ \phi(\textit{f}, x)\, \phi^*(\textit{f}, x) \right]}{\partial x} 
= \frac{\partial \phi(\textit{f}, x)}{\partial x} \, \phi^*(\textit{f}, x) 
+ \frac{\partial \phi^*(\textit{f}, x)}{\partial x} \, \phi(\textit{f}, x)
\end{equation}

Using Eq.~\eqref{eq:ritz1} in the above equation and ensemble-averaging this equation takes the form of wave kinetic equation \cite{kim1996}:

\begin{equation}
\frac{\partial P_{\textit{f}}}{\partial x} = 2\gamma_{\textit{f}} P_{\textit{f}} + 
T_f, \quad \text{where} \quad 
T_f = \sum_{\substack{\textit{f}_1 , \textit{f}_2 \\ \textit{f} = \textit{f}_1 + \textit{f}_2}} T_{\textit{f}}(\textit{f}_1, \textit{f}_2)
\label{eq:wave_kinetic_eq}
\end{equation}

Where, the power transfer function is written as: 

\begin{equation}
T_{\textit{f}}(\textit{f}_1, \textit{f}_2) 
= \mathrm{Re} \left[ \Lambda_{\textit{f}}^{Q}(\textit{f}_1, \textit{f}_2)\, 
\langle X_{\textit{f}}^{*} X_{\textit{f}_1} X_{\textit{f}_2} \rangle \right] = \mathrm{Re} \left[ \Lambda_{\textit{f}}^{Q}(\textit{f}_1, \textit{f}_2)\, 
B_a(f_1, f_2) \right]
\label{eq:Tf}
\end{equation}

Where, $\Lambda_{\textit{f}}^{Q}(\textit{f}_1, \textit{f}_2)$ can be obtained from Eq.~\eqref{eq:Q_lam}.

Once the power transfer function is calculated then the energy transfer function can be easily calculated as : 
\begin{equation}
W^{f}_{\mathrm{NL}} = T_f = \sum_{\substack{f_1, f_2 \\ f = f_1 + f_2}} T_{f}(f_1, f_2)
\label{eq:W_nl}
\end{equation}

To check the reliability of the estimated nonlinear energy transfer, we use spectral energy conservation as a physical consistency condition, i.e.,
\(\sum_f W_{\mathrm{NL}}^{f} = 0\).
Any deviation from this relation indicates an imbalance in the evaluated transfer. To quantify this imbalance, we define the \emph{energy mismatch parameter} \( W_{\mathrm{mis}} \) \cite{xia2004} as
\begin{equation}
W_{\mathrm{mis}} = \frac{\sum_f W_{\mathrm{NL}}^{f}}{\sum_f \lvert W_{\mathrm{NL}}^{f} \rvert}.
\label{eq:w_mis}
\end{equation}
This parameter measures how far the computed nonlinear energy transfer departs from exact conservation. For a perfectly balanced transfer, \( W_{\mathrm{mis}} = 0 \), while nonzero values indicate residual imbalance in the spectral energy budget.

\subsubsection{Computational Algorithm for Ritz method}
\label{subsubsection: computational algoritm Ritz}
In this subsection, we outline the computational procedure employed to estimate the linear and quadratic transfer functions using the Ritz method. The algorithm is implemented in the frequency domain and is based on ensemble-averaged spectral quantities obtained from multiple realizations of the input–output data. The steps of the algorithm as follows:

\begin{itemize}

\item \textit{Step 1: Original time-series data}  

The input and output signals are given as time-domain sequences,
\( X(t) \) and \( Y(t) \), respectively.

\item \textit{Step 2: Pre-processing}
\begin{itemize}

    \item \textit{Step 2.1: Ensemble and data points}  

    Define the number of ensemble realizations (M) and the number of data points per realization (N).

    \item \textit{Step 2.2: Mean removal and detrending}  

    Apply mean removal or detrending to the time-series data as appropriate. Mean removal is performed when the data exhibit a DC offset, whereas detrending is applied in the presence of low-frequency trends.

    \item \textit{Step 2.3: Overlapping and window function}  

    Not applying overlapping and window function to the time series data. Because highly correlated samples that do not significantly improve statistical convergence when sufficient independent realizations are available. And windowing can distort phase relationships and bias higher-order statistics.

    \item \textit{Step 2.4: FFT}  

    Apply the Fast Fourier transform (FFT) to the input and output time-series data to obtain their frequency-domain representations. The resulting spectral components of the input and output signals are denoted by \( X_i(f) \) and \( Y_i(f) \), respectively, for \( i = 1, 2, \ldots, M \).

\end{itemize}

\item \textit{Step 3: Computation of valid frequency pairs and second and third order cumulants}  

For each frequency index \( f = -\frac{N}{2}, \ldots, \frac{N}{2}-1 \), all frequency pairs \( (f_1, f_2) \) satisfying the resonance condition \( f_1 + f_2 = f \) are calculated. Using these valid triads, the relevant statistical quantities are evaluated, including the power spectral densities \( P_a(f) \) and \( P_c(f) \), the bispectra \( B_a(f_1, f_2) \) and \( B_c(f_1, f_2) \), and the corresponding bicoherences \( b_a(f_1, f_2) \) and \( b_c(f_1, f_2) \).

\item \textit{Step 4: Computation of Transfer Functions}  

Calculate the linear and quadratic transfer functions, i.e., \( L_f \) and \( Q_f^{f_1, f_2} \).

\item \textit{Step 5: Computation of Coupling Coefficients}  

Calculate the linear and quadratic coupling coefficients, i.e., \( \Lambda_f^L \) and \( \Lambda_f^Q(f_1, f_2) \), respectively, by multiplying the phase difference between input and output data to the transfer functions.

\item \textit{Step 6: Computation of Power and Energy transfer functions}  

Calculate the Power transfer function \( (T_f(f_1, f_2)) \) and Energy transfer function (\( W_{NL}^f \)) by summing \( T_f(f_1, f_2) \) along the \( f = f_1 + f_2 \) axis.

\item \textit{Step 7: Outputting Results}  

Output the following quantities: \(L_f, Q_f^{f_1, f_2}, \Lambda_f^L, \Lambda_f^Q(f_1, f_2), T_f(f_1, f_2), W_{NL}^f\).
\end{itemize}

We use the above algorithm for Ritz method in case of both simulated data and experimental data. However, using Ritz method to experimental data requires a data selection process which is discussed in the following sections.

\subsubsection{Simulation Test and Analysis}
\label{subsubsection: simulation test ritz}
To verify the validity of the method described in the previous section, we have performed a computer simulation. First we define the analytical form of linear and quadratic transfer functions, then we generate output signal from a given Gaussian input. Next, we estimate the linear and quadratic transfer functions using the method described in Section~(\ref{subsubsection: estimation of L and Q}), and then we compare the results of the estimated transfer function with the analytical ones.

\paragraph{I. Simulation Data Generation:} To carry out the simulation test we have defined a ``\textit{nonlinear black-box model}" in which we can generate a non-Gaussian output from a Gaussian input signal following Eq.~\eqref{eq:Y_X}. This model consists of a linear transfer function($L_f$) which makes the input and output spectra similar in shape as expected from a stationary state and a quadratic transfer function($Q_{f}^{f_1,f_2}$) which incorporates the nonlinear coupling among the spectral components of the input signal. The schematic diagram of this nonlinear black box model is shown in Fig.~\ref{fig:block_diagram} below.
\begin{figure}[H]
    \centering
    \includegraphics[width=0.65\linewidth]{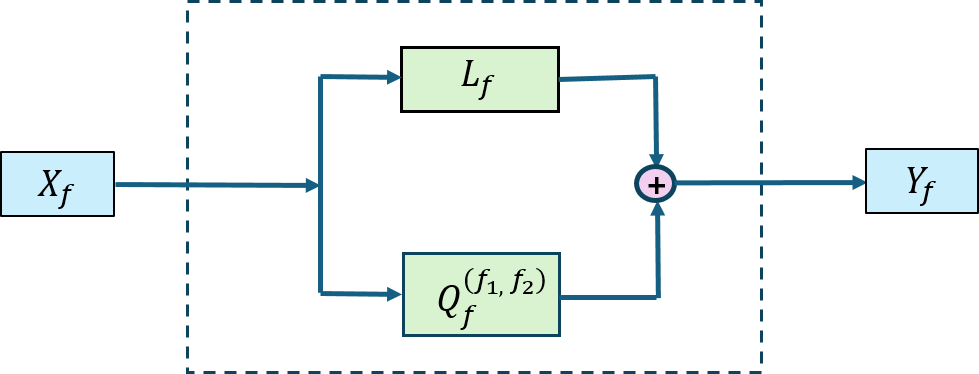}
    \caption{Non-Linear Black Box Model}
    \label{fig:block_diagram}
\end{figure}

The analytical linear and quadratic transfer functions defined in the following way \cite{ritz1986}:  

\begin{equation}
L_f = 1.0 - 0.4\frac{f^2}{f_{\text{Nyq}}^2} + i \, 0.8 \frac{f}{f_{\text{Nyq}}},
\label{ana_Ltrans_fn}
\end{equation}

\begin{equation}
Q_{f}^{f_1,f_2} = \frac{i}{5 f_{\text{Nyq}}^4} 
\frac{ f_1 f_2 \left( f_2^2 - f_1^2 \right)}{1 + \dfrac{f^2}{f_{\text{Nyq}}^2}} .
\label{ana_Qtrans_fn}
\end{equation}
where, $f= f_1 + f_2$ and $f_{Nyq}$ : Nyquist frequency.\\

The choice of these functions is arbitrary, but their values are of the same order of magnitude as those predicted by the Hasegawa–Mima equation \cite{hasegawa1978}. \textcolor{black}{These analytical transfer functions closely resemble those reported by Shen et al.\cite{Shen2020ImprovedBispectrum} for a nonlinear developing turbulence model, in which the input and output power spectra differ, a feature also observed in shallow-water ocean waves.} Now, to carry out simulation, we need to generate non-linearity in our input and output signal as we are looking for the nonlinear energy transfer function. A white Gaussian noise as an input signal shows zero bispectrum indicating there is no nonlinearity in the signal. The non-linearity can be introduced to the signal by passing it through a ``non-linear black box" of type shown in Fig.~\ref{fig:block_diagram}. Here we have taken zero mean white Gaussian noise as input signal and passed it  through a series of five identical ``non-linear black boxes" which are connected in series. So, in this case a Gaussian white noise signal is fed into the first black-box. Due to the nonlinear behavior of this system, the output is no longer Gaussian and is passed as the input to the next black-box. This process is repeated sequentially, and for the simulation we consider the input–output pair of the fifth black box as our final input($X_f$) and output($Y_f$) in frequency domain. We can always do an inverse fourier transform to get input and output time series signals.\\

As we have time series signals, now we can study power spectrum, Bi-spectrum and Bi-coherence which are already defined in section-\ref{section: sepectral framework}. The auto power spectra $(P_a(f))$ of the input and output signals after the fifth iteration (i.e., after the fifth nonlinear black box) are shown in Fig.~\ref{fig:psd5} below.
\begin{figure}[H]
    \centering
    \includegraphics[width=0.55\linewidth]{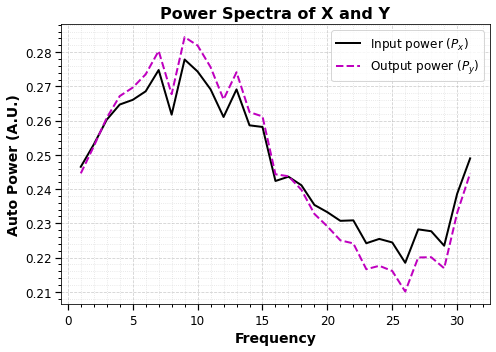}
    \caption{Power Spectrum of Input and Output signals after $5^{th}$ iteration}
    \label{fig:psd5}
\end{figure}

From the power spectra shown in Fig.~\ref{fig:psd5}, it is evident that the input and output spectra do not completely coincide. This mismatch is more pronounced at higher frequencies, a feature commonly observed in shallow-water surface waves in the ocean\cite{FreilichGuza1984}, and it is indicative of a developing turbulence regime\cite{Shen2020ImprovedBispectrum}.

The auto bispectrum $(B_a(f_1, f_2))$ and cross bispectrum $(B_c(f_1, f_2))$  for the input signal after $5^{th}$ iteration is shown in Fig.~\ref{fig:bis5} below.

\begin{figure}[H]
    \centering
    \includegraphics[width=0.99\linewidth]{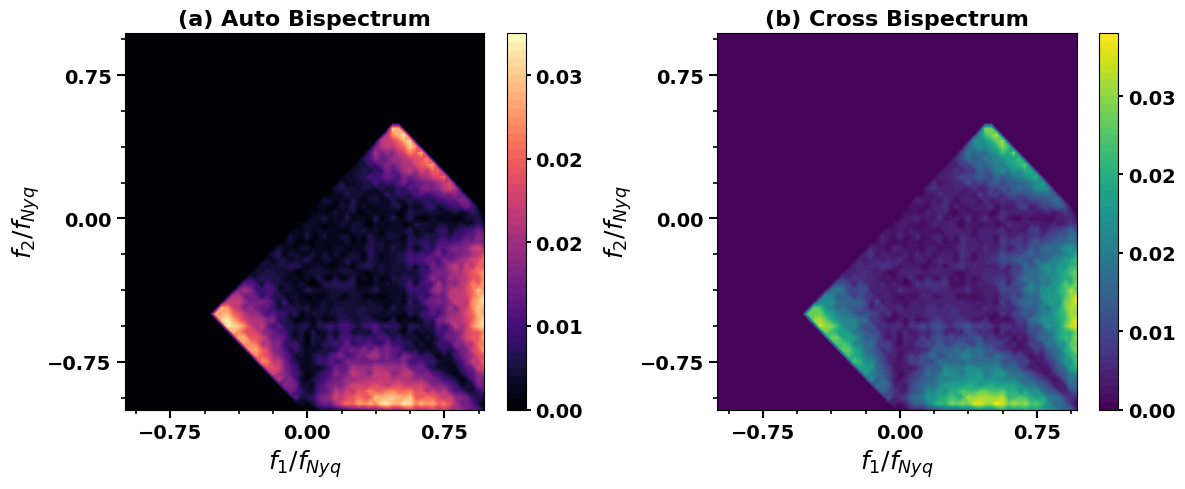}
    \caption{(a) Auto Bispectrum and (b) Cross Bispectrum after $5^{th}$ iteration.}
    \label{fig:bis5}
\end{figure}

\begin{figure}[H]
    \centering
    \includegraphics[width=0.98\linewidth]{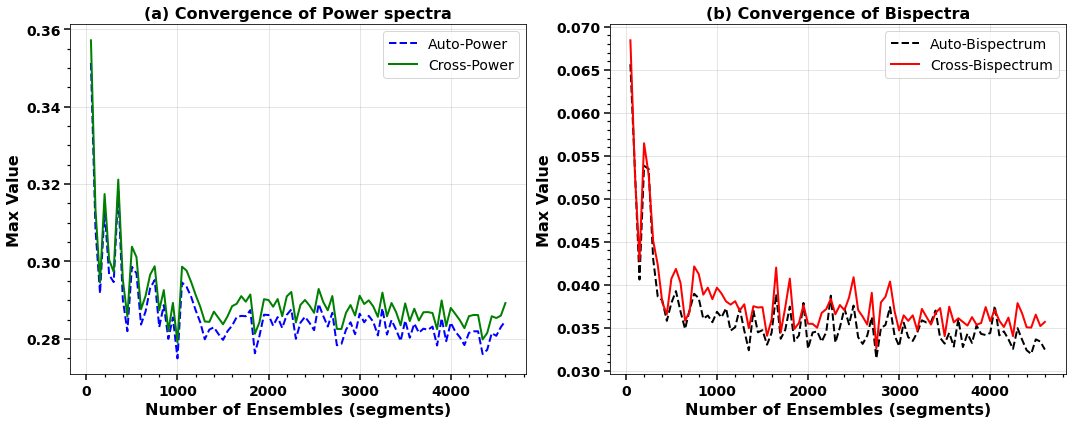}
    \caption{Convergence plot of (a) Auto and Cross Power Spectra (b) Auto and Cross Bispectra.}
    \label{fig:bis_con}
\end{figure}

\paragraph{II. Convergence Test:}
In the Ritz method, obtaining reliable estimates of the linear and quadratic transfer functions requires a sufficiently large number of ensembles. These quantities are constructed from auto- and cross-power spectra as well as higher-order cumulants, namely the auto- and cross-bispectra, which are known to converge slowly with respect to the number of realizations. When only a small number of ensembles is used, the estimates are dominated by statistical fluctuations, resulting in large variance and poor accuracy. Increasing the ensemble size reduces these fluctuations through averaging, allowing the transfer functions to converge toward their true statistical values. To quantify convergence, we examine the evolution of the auto- and cross-power spectra and the auto- and cross-bispectra as functions of the number of ensembles. The results are considered to be convergent when the relative change of these quantities between successive ensemble sizes falls below 5\%. This criterion ensures that both second- and third-order statistics are sufficiently stable for subsequent analysis of the linear and quadratic transfer functions. The corresponding convergence behavior is illustrated in Fig.~\ref{fig:bis_con}.

From the above plots, we observe that the amplitudes of the auto- and cross-power spectra as well as the auto- and cross-bispectra exhibit pronounced oscillatory behavior for small ensemble sizes \(M\), indicating strong statistical fluctuations when only a limited number of realizations are available. As \(M\) increases, these fluctuations are progressively suppressed and the estimates approach stable values. Beyond \(M \approx 2000\), the relative change in these quantities reduces to about \(0.5\%\), which is already well below the adopted convergence threshold of \(5\%\). Furthermore, increasing the ensemble size improves the statistical accuracy even more. As seen in Fig.~\ref{fig:bis_con}, the relative change falls below \(0.1\%\) once \(M \gtrsim 3500\), indicating near-complete convergence of both second- and third-order statistics. For the present simulations, we therefore use \(M = 2500\) realizations of a Gaussian signal, each consisting of \(N = 64\) points, which is sufficient to obtain reliable estimates of the linear and quadratic transfer functions. However, for experimental data, where additional noise and non-idealities are present, we employ ensemble sizes larger than \(3500\) to ensure higher statistical robustness and more reliable transfer-function estimation.

\begin{figure}[H]
    \centering
    \includegraphics[width=0.90\linewidth]{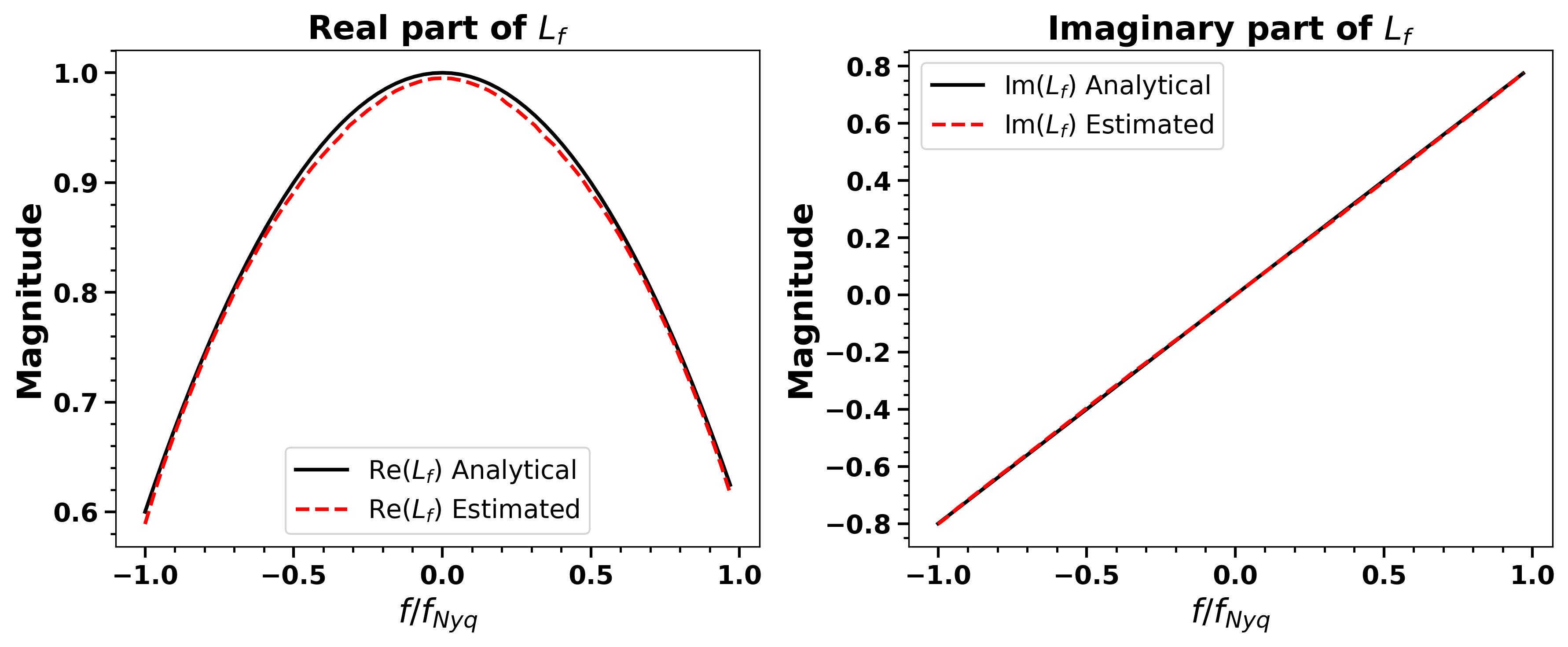}
    \caption{Benchmarking of $L_f$ with Analytical Function.}
    \label{fig:ben_L}
\end{figure}

\begin{figure}[H]
    \centering
    \includegraphics[width=1\linewidth]{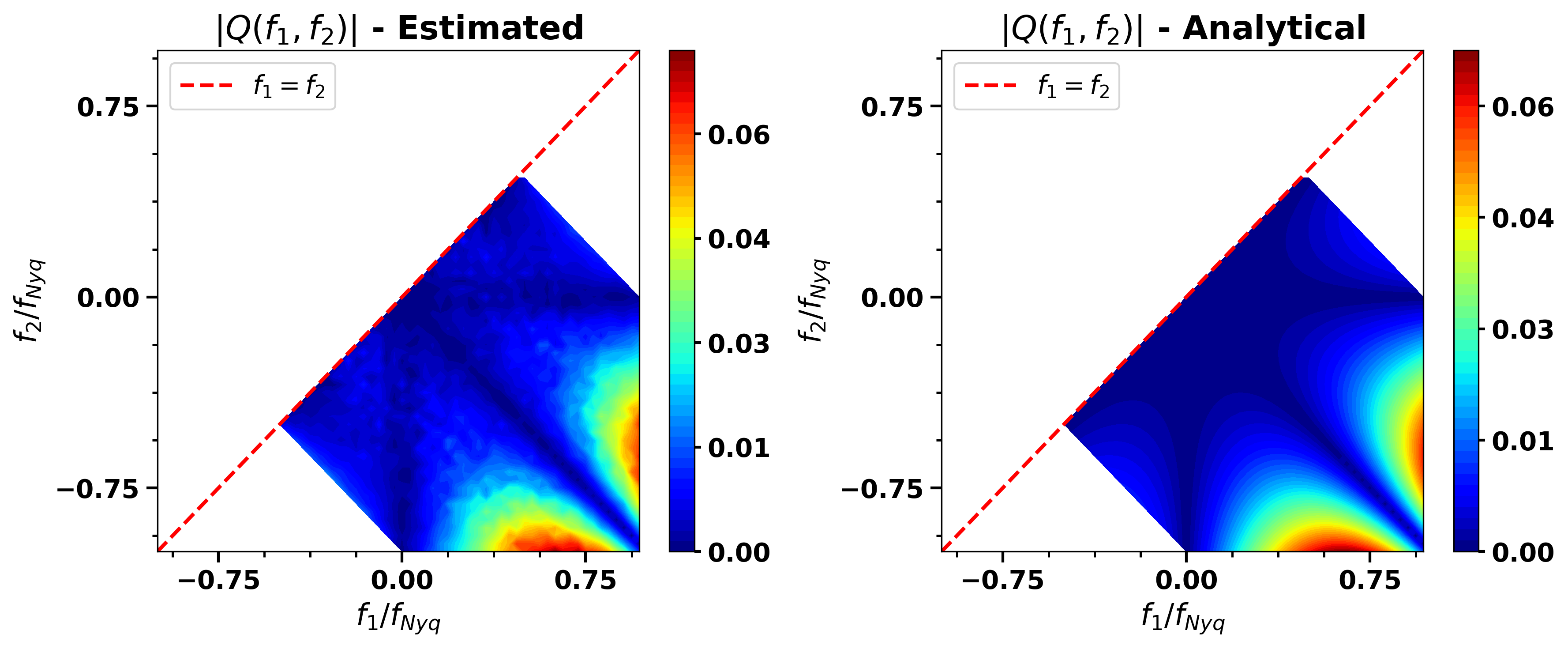}
    \caption{Benchmarking of $Q_f^{f_1,f_2}$ with Analytical Function.}
    \label{fig:ben_Q}
\end{figure}

\paragraph{III. Simulation Validation: } To validate our formulation, we benchmark the estimated linear and quadratic transfer functions against their corresponding analytical expressions. Fig.~\ref{fig:ben_L} and Fig.~\ref{fig:ben_Q} shows the comparison for $L_f$ and $Q^{f_1,f_2}_f$ with their respective analytical functions.

As discussed earlier, we use the input and output data at iteration \(=5\), for which the estimated transfer functions show good agreement with the analytical results. A small deviation is observed in the linear transfer function near zero frequency, and the contour plot of the estimated quadratic transfer function appears slightly noisier than its analytical counterpart. This behavior primarily arises from the use of the Millionshchikov's approximation in the present method, which affects the estimation of higher-order statistics. Such limitations do not arise in Kim’s method, where this approximation is not used, as will be discussed in the subsequent sections.
  
\subsubsection{Applicability of Ritz Method}
\label{subsubsection: applicability of Ritz}
In the previous section, we have discussed that if we feed a white Gaussian noise at the input of the nonlinear black box then it will gain some nonlinearity at the output due to the nonlinear nature of the black box. This signature of the nonlinearity can be identified by the non-zero bispectrum of the signal.
On the other hand we can identify the nonlinearity by looking at the distribution function of the signal, more specifically the nonlinearity can be identified by Skewness and Kurtosis of the signal's distribution function. 
The \textit{skewness} of a signal $\{x_i\}_{i=1}^N$ is defined as \cite{Bendat2010RandomData, Spiegel2018Statistics},  
\begin{equation}
\text{Skewness} = \frac{1}{N}\sum_{i=1}^N \left( \frac{x_i - \mu}{\sigma} \right)^3,
\end{equation}
where $\mu = \tfrac{1}{N}\sum_{i=1}^N x_i$ is the mean and $\sigma$ is the standard deviation of the signal.  
Skewness corresponds to the \emph{third-order central moment}, and it measures the \emph{asymmetry} of the probability distribution. A positive skewness indicates a distribution with a longer tail on the right, while a negative skewness indicates a longer tail on the left.  
For a Gaussian distribution, $\text{Skewness} = 0$\cite{Bendat2010RandomData}.

The \textit{kurtosis} of the signal is defined as \cite{Bendat2010RandomData, Spiegel2018Statistics}, 
\begin{equation}
\text{Kurtosis} = \frac{1}{N}\sum_{i=1}^N \left( \frac{x_i - \mu}{\sigma} \right)^4,
\end{equation}
which is the \emph{fourth-order central moment}, measuring the \emph{peakedness} and tail heaviness of the distribution. A higher kurtosis indicates heavier tails and sharper peak compared to a Gaussian distribution, while lower kurtosis indicates lighter tails and a flatter peak.
For a Gaussian distribution, $\text{Kurtosis} = 3$.The excess kurtosis is defined as the kurtosis measured relative to its Gaussian benchmark value of 3 \cite{Bendat2010RandomData}.

Skewness, being a third-order statistic, is directly related to the bispectrum \cite{ritz1986}, which is the Fourier transform of the third-order cumulant. On the other hand, kurtosis, as a fourth-order statistic, is connected to fourth-order cumulants, thereby providing insight into higher-order correlations. For purely Gaussian signals, all higher-order cumulants beyond the second order vanish; hence, any non-zero skewness or excess kurtosis serves as a clear indicator of non-Gaussianity and nonlinearity in the signal. Thus, skewness and kurtosis serve as compact statistical measures of nonlinearity and deviations from Gaussian behavior in signal data, complementing the more detailed spectral tools such as bispectrum and the fourth order cumulants.\\

In the previous section we have discussed about the non-linear black box model in which a Gaussian signal was passed through a number of black boxes which are connected in series. As an effect of this we obtained some amount of nonlinearity in the output signal at every iteration. Here, we can observe the signature of nonlinearity by calculating skewness and kurtosis of the output signal at every iteration. The distribution functions before the first iteration (i.e, $1^{st}$ black box) and after the fifth iteration(i.e, $5^{th}$ black box) along with the skewness and kurtosis of the distribution are shown in the Fig.~\ref{fig:prob_dis_1} below.
\begin{figure}[H]
    \centering
    \includegraphics[width=0.8\linewidth]{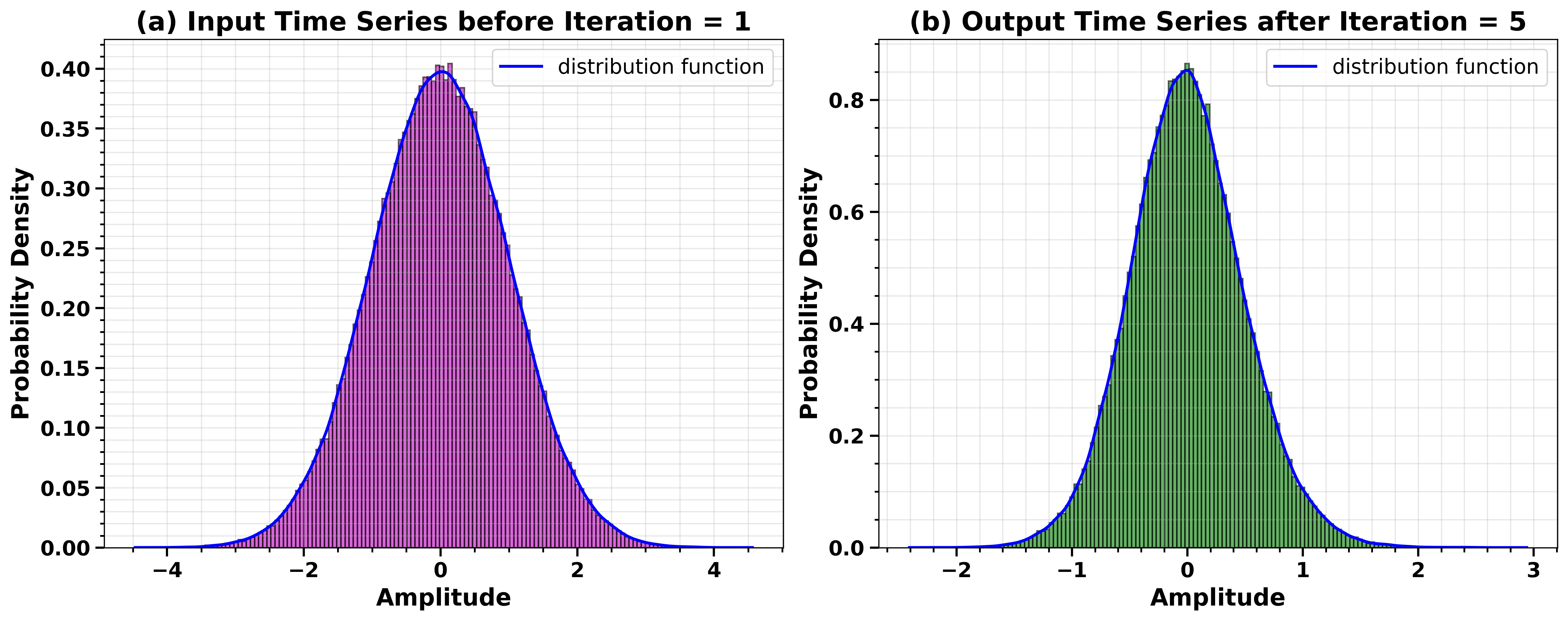}
    \caption{(a)Probability distribution of input time series having $\mu = 0.001$, $\sigma = 1.00$, Skewness $=0.001$, Excess Kurtosis $=0.01$. (b) Probability distribution of output time series having $\mu = 0.002$, $\sigma = 0.52$, Skewness $= 0.15$, Excess Kurtosis $=0.40$ at fifth iteration. }
    \label{fig:prob_dis_1}
\end{figure}

\begin{figure}[H]
\centering

\begin{subfigure}[b]{0.49\textwidth}
    \centering
    \includegraphics[width=\textwidth]{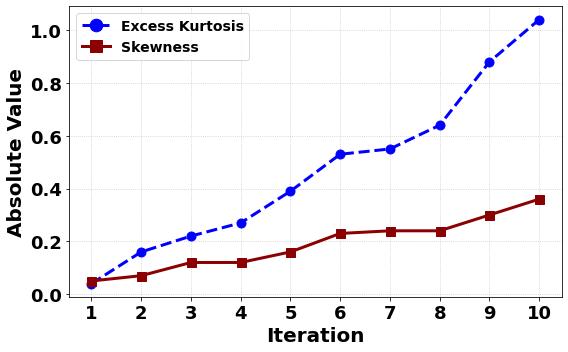}
    \caption{Evolution of skewness and excess kurtosis with iteration number.}
    \label{fig:skw_kurt}
\end{subfigure}
\hfill
\begin{subfigure}[b]{0.49\textwidth}
    \centering
    \includegraphics[width=\textwidth]{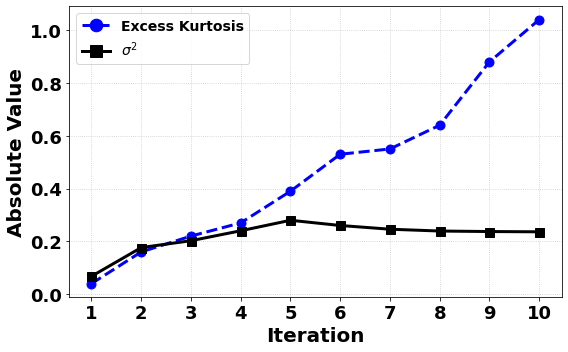}
    \caption{Evolution of square of standard deviation($\sigma^2$) and excess kurtosis with iteration number.}
    \label{fig:var_kurt}
\end{subfigure}

\caption{Evolution of statistical parameters with nonlinear iterations to examine the validity of Millionshchikov’s approximation.}
\label{fig:sk}
\end{figure}

From the above figure we observe that before the first iteration (i.e, $1^{st}$ black box) the distribution of input time series (Gaussian white noise) gives the expected mean and standard deviation but we find small non-zero skewness and kurtosis values are still there which are nothing but the consequence of not taking long signal length. On the other hand, the distribution at the output of the fifth iteration (i.e., the $5^{\mathrm{th}}$ black box) exhibits a noticeable increase in skewness and excess kurtosis, indicating that substantial nonlinearity has been introduced into the signal after passing through the fifth nonlinear black box (i.e., iteration $=5$). By observing the output across successive iterations, we find that both skewness and excess kurtosis increase with the number of iterations. This trend is illustrated in Figure~\ref{fig:sk}(a), where results up to 10 iterations are shown.


From the Fig.~\ref{fig:sk}(a), it is clear that kurtosis increases more rapidly than skewness with iteration. So, we see that with the increase of nonlinearity, skewness and kurtosis of the distribution increases - here comes the main issue of applicability of Ritz method. In the Ritz method, a key assumption is that the fourth-order statistical moments of the system can be approximated by the square of the second-order moments. This assumption is known as Millionshchikov’s approximation \cite{ritz1986, Millionshchikov1941}. However, this approximation has a limitation. The quantity called kurtosis is a direct statistical measure of the fourth-order central moment of a distribution. When kurtosis remains close to its Gaussian value, the approximation works reasonably well. But as kurtosis increases significantly, the higher-order statistics deviate strongly from Gaussian behavior. In such cases, the fourth-order moment can no longer be represented simply as the square of the second-order moment. \textcolor{black}{Further, the inapplicability of Millionshchikov’s approximation is also supported by Fig.~\ref{fig:sk}(b), where the square of the standard deviation ($\sigma^2$) and the excess kurtosis are plotted as a function of iteration number. Since the standard deviation ($\sigma$) is related to the second-order central moment, the fourth-order moments can be approximated by the square of the second-order moments only when $\sigma^2$ remains comparable to the excess kurtosis, i.e., when Millionshchikov’s approximation is valid. From Fig.~\ref{fig:sk}(b), it can be seen that up to iteration $=5$, the excess kurtosis and $\sigma^2$ remain relatively close. However, beyond this point, the excess kurtosis increases to significantly higher values, resulting in a substantial deviation between the two quantities and indicating the breakdown of Millionshchikov’s approximation.} Therefore, when the kurtosis becomes large, the fundamental assumption behind the Ritz method breaks down, and the method becomes invalid. This marks the key limitation in the applicability of the Ritz method. Hence, the Ritz method cannot be applied to highly nonlinear signals. 

\begin{figure}[H]
    \centering
    \includegraphics[width=0.9\linewidth]{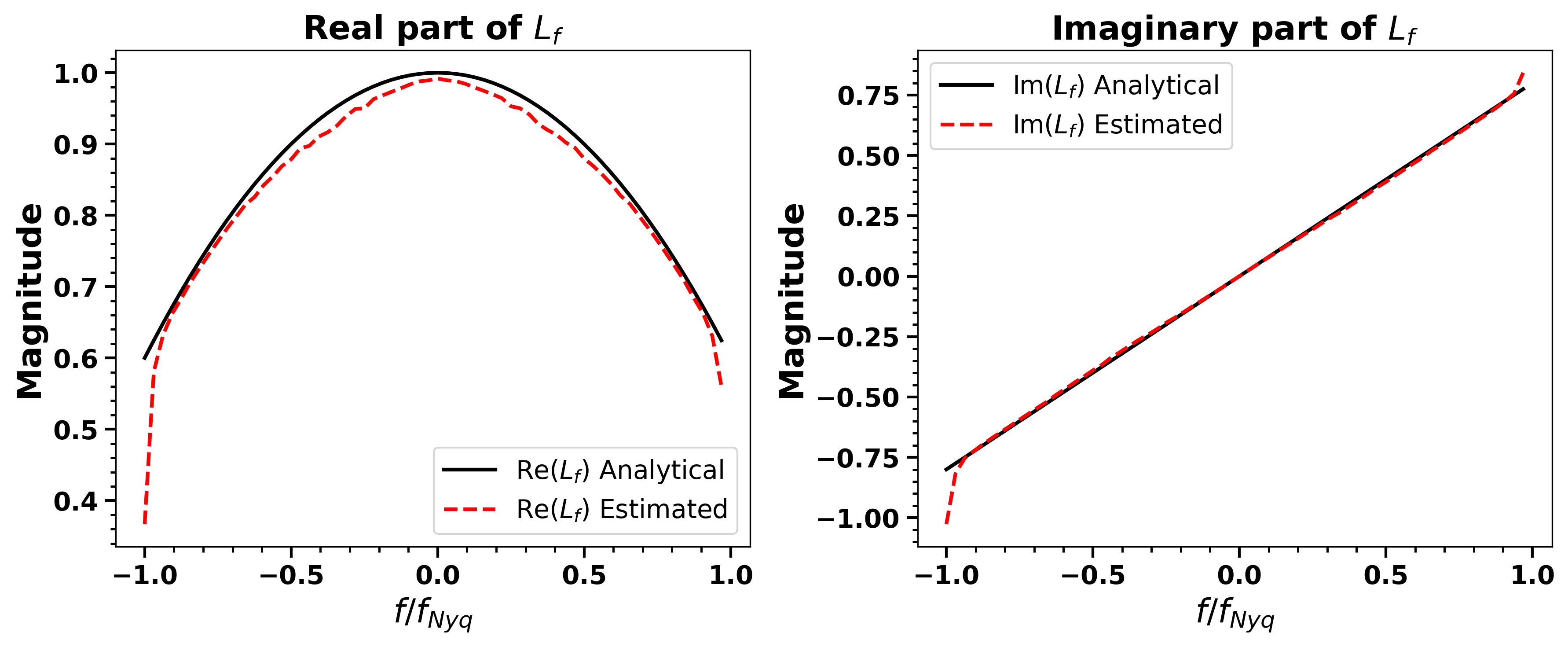}
    \caption{Plot of $L_f$ with Analytical Function at iteration = 7.}
    \label{fig:L7}
\end{figure}

\begin{figure}[H]
    \centering
    \includegraphics[width=1\linewidth]{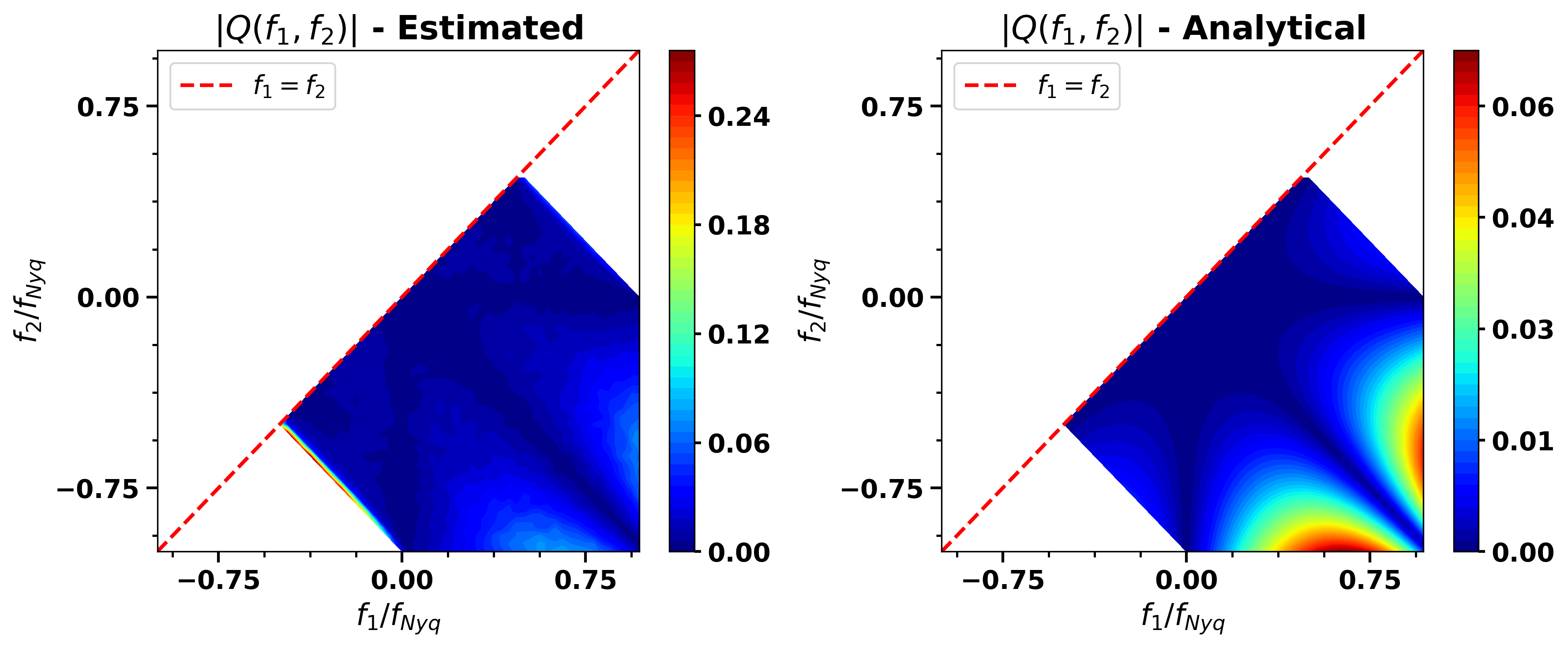}
    \caption{Plot of $Q_f^{f_1,f_2}$ with Analytical Function at iteration = 7.}
    \label{fig:Q7}
\end{figure}

In this study, using artificial data, we observed that the Ritz method fails to correctly estimate the transfer functions once the excess kurtosis exceeds approximately \textit{0.40}, which occurs after iteration $=5$. As shown earlier, both the linear and quadratic transfer functions exhibit good agreement with the analytical functions at iteration $=5$ in Fig.~\ref{fig:ben_L} and Fig.~\ref{fig:ben_Q}. However, at iteration $=7$, where the kurtosis reaches approximately \textit{0.55}, the transfer function plots exhibit clear deviations from the analytical functions, as shown in Fig.~\ref{fig:L7} and Fig.~\ref{fig:Q7}. The above plots in Fig.~\ref{fig:L7} and Fig.~{\ref{fig:Q7} show a significant deviation of the estimated transfer functions from the analytical ones. \textcolor{black}{These results are also consistent with Fig.~\ref{fig:sk}(b), which shows that Millionshchikov’s approximation breaks down beyond iteration $=5$. Consequently, the assumptions underlying the Ritz method are no longer satisfied, indicating that the Ritz method is not applicable for iterations greater than 5.} In the next section, we discuss the Kim method and validate it using simulation data, as done for the Ritz method, and also examine its performance at higher iterations.

\subsection{Kim Method}
\label{subsection: Kim method}
As discussed earlier, the Ritz method suffers from several inherent limitations. One major source of inaccuracy arises from the use of the Millionshchikov approximation\cite{Millionshchikov1941}, in which fourth-order moments are approximated by the square of second-order moments. As a consequence, the Ritz method loses accuracy for fluctuation data exhibiting significant non-Gaussianity or high kurtosis, as demonstrated in the previous section. In addition, the application of the Ritz method to experimentally measured fluctuations often results in unrealistically large growth or damping coefficients across the entire spectrum.
According to Kim et al.\cite{kim1996}, these unphysical results originate from the fact that the Ritz method implicitly assumes that all measured fluctuations satisfy the nonlinear wave coupling equation [Eq.~\eqref{eq:ritz1}]. In practice, experimental signals contain non-ideal fluctuations that do not obey this equation, such as contributions from systematic errors, higher-order nonlinear interactions beyond quadratic coupling, and instrumental or environmental noise. To overcome these limitations, Kim et al.\cite{kim1996} proposed a modified formalism in which the measured spectra $(X_f, Y_f)$ are decomposed into an ideal component $(\beta_f, \alpha_f)$ that participates in the linear and quadratic interaction processes described by Eq.~\eqref{eq:ritz1}, and a nonideal component $(X_f^{ni}, Y_f^{ni})$ that does not contribute to these processes. That is, 

\begin{equation}
X_f = \beta_f + X_f^{\mathrm{ni}}, \qquad
Y_f = \alpha_f + Y_f^{\mathrm{ni}}
\label{nideal}
\end{equation}

Furthermore, unlike the Ritz method, Kim et al. avoided the Millionshchikov's approximation and explicitly retained the fourth-order moments in the calculation of the nonlinear transfer functions, leading to a more accurate and physically consistent description of nonlinear energy transfer in experimental turbulence data. Using Eq.~\ref{nideal} in Eq.~\ref{eq:Y_X}, and expanding the wave-coupling equation into a hierarchy of moment equations by multiplying by $X_f^{}$, $Y_f^{}$, and $X_{f_1'}^{} X_{f_2'}^{}$, we obtain

\begin{equation}
\left.
\begin{aligned}
\langle Y_f X_f^{*} \rangle 
&= L_f \langle \beta_f \beta_f^{*} \rangle 
+ \sum_{f_1 \ge f_2 \atop f = f_1 + f_2}
Q_{f}^{f_1,f_2}\,
\langle X_{f_1} X_{f_2} X_f^{*} \rangle ,
\\[6pt]
\langle \alpha_f \alpha_f^{*} \rangle 
&= L_f \langle X_f Y_f^{*} \rangle 
+ \sum_{f_1 \ge f_2 \atop f = f_1 + f_2}
Q_{f}^{f_1,f_2}\,
\langle X_{f_1} X_{f_2} Y_f^{*} \rangle ,
\\[6pt]
\langle Y_f X_{f_1}^{*} X_{f_2}^{*} \rangle 
&= L_f \langle X_f X_{f_1}^{*} X_{f_2}^{*} \rangle
+ \sum_{f_1' \ge f_2' \atop f = f_1' + f_2' = f_1 + f_2}
Q_{f}^{f_1',f_2'}\,
\langle X_{f_1'} X_{f_2'} X_{f_1}^{*} X_{f_2}^{*} \rangle
\end{aligned}
\right\}
\label{eq:moment_eqs}
\end{equation}

Here, $\langle \cdot \rangle$ denotes ensemble averaging. In contrast to the Millionshchikov's approximation, the fourth-order cumulant $\langle X_{f_1'} X_{f_2'} X_{f_1}^{} X_{f_2}^{} \rangle$ is explicitly retained in the moment equation [Eq.~\eqref{eq:moment_eqs}]. This renders the nonlinear energy transfer formalism more general. Moreover, retaining the fourth-order term improves the accuracy of the energy transfer estimates, albeit at the cost of increased computational effort. Inspection of the resulting moment equations shows that there are four unknown quantities, namely the $L_f$, $Q_{f}^{f_1,f_2}$, $\langle \beta_f \beta_f^{} \rangle$ and $\langle \alpha_f \alpha_f^{} \rangle$, while only three independent moment equations are available.

To close the system, Kim et al.\cite{kim1996} imposed the condition that the power spectrum is fully saturated by linear and three-wave coupling processes. This assumption corresponds to spatial stationarity of the turbulence and leads to the constraint

\begin{equation}
\frac{\partial P_f}{\partial x}
\approx \frac{\langle \alpha_f \alpha_f^{*} \rangle - \langle \beta_f \beta_f^{*} \rangle}{\Delta x}
= 2\gamma_f P_f + T_f
= 0 ,
\label{eq:stationarity}
\end{equation}

where $P_f$ is the ideal power spectrum, $\gamma_f$ is the linear growth (or damping) rate, and $T_f$ denotes the nonlinear energy transfer term. From Eq.~\ref{eq:stationarity}, the fourth closure relation immediately follows as

\begin{equation}
\langle \alpha_f \alpha_f^{*} \rangle =
\langle \beta_f \beta_f^{*} \rangle .
\label{eq:4th_mom_eq}
\end{equation}

Consequently the set of monemts equations[Eq.~\eqref{eq:moment_eqs}] can be written in matrix form, and from that we can easily calculate the linear and quadratic transfer function as: 

\begin{equation}
L_f =
\frac{
\langle Y_f X_f^{*} \rangle
-
(\mathbf{B}^{*})^{T}\!\cdot\!\mathbf{F}^{-1}\!\cdot\!\mathbf{A}
}{
\langle \beta_f \beta_f^{*} \rangle
-
(\mathbf{A}^{*})^{T}\!\cdot\!\mathbf{F}^{-1}\!\cdot\!\mathbf{A}
}
\label{linear_trns}
\end{equation}

\begin{equation}
\mathbf{Q}_f
=
\bigl[
(\mathbf{B}^{*})^{T}
-
L_f (\mathbf{A}^{*})^{T}
\bigr]
\cdot \mathbf{F}^{-1}
\label{quad_trans}
\end{equation}
Here, $\mathbf{A}$ and $\mathbf{B}$ denote vectors composed of third-order moments, corresponding to the auto- and cross-bispectra, respectively, while $\mathbf{F}$ represents the matrix of fourth-order moments\cite{kim1996}. Here, the expression for the power transfer function $T_f(f_1,f_2)$ is identical to that used in the Ritz method, since both approaches are based on the same wave kinetic equation [Eq.~\eqref{eq:wave_kinetic_eq}]. Consequently, the power transfer function in Kim’s method is given by,

\begin{equation}
T_{\textit{f}}(\textit{f}_1, \textit{f}_2) 
= \mathrm{Re} \left[ \Lambda_{\textit{f}}^{Q}(\textit{f}_1, \textit{f}_2)\, 
\langle X_{\textit{f}}^{*} X_{\textit{f}_1} X_{\textit{f}_2} \rangle \right] = \mathrm{Re} \left[ \Lambda_{\textit{f}}^{Q}(\textit{f}_1, \textit{f}_2)\,\cdot A \right]
\label{eq:Tf}
\end{equation}

Here, the expression for $\Lambda_f^{Q}(f_1,f_2)$ follows from Eq.(\ref{eq:Q_lam}), provided that Eq.(\ref{quad_trans}) is used for the quadratic transfer function estimation. Also, from the above definition of power transfer function, we can easily estimate the nonlinear energy transfer function($W_{NL}^f$) from Eq.(\ref{eq:W_nl}).

Furthermore, the Kim method allows for a decomposition of the measured fluctuations into ideal and nonideal components. The nonideal component accounts for contributions arising from experimental noise, measurement uncertainties, and interactions with physical processes that lie outside the applicability of the underlying model. In the present work, this decomposition is not performed, and all observed fluctuations are treated as ideal. With this prescription, we use $ X_f$ and $Y_f$ instead of $\beta$ and $\alpha$ respectively in Eq.(\ref{eq:moment_eqs}) and Eq.(\ref{eq:4th_mom_eq}) and subsequently obtained the linear transfer function as

\begin{equation}
L_f =
\frac{
\langle Y_f X_f^{*} \rangle
-
(\mathbf{B}^{*})^{T}\!\cdot\!\mathbf{F}^{-1}\!\cdot\!\mathbf{A}
}{
\langle X_f X_f^{*} \rangle
-
(\mathbf{A}^{*})^{T}\!\cdot\!\mathbf{F}^{-1}\!\cdot\!\mathbf{A}
}
\label{linear_trns_mod}
\end{equation}
and the quadratic transfer function follows accordingly from Eq.(\ref{quad_trans}). In this treatment, we can write the Eq.~\eqref{eq:stationarity} for the spatial stationarity $(i.e, \frac{\partial P_f}{\partial x}=0)$ as 
\begin{equation}
\frac{\langle Y_f Y_f^{*} \rangle - \langle X_f X_f^{*} \rangle}{\Delta x}
= 2\gamma_f P_f + T_f
= 0 ,
\label{eq:stationarity2}
\end{equation}

Here, $P_f$ represents the total power associated with the frequency $f$. And the closure relation reduced to,
\begin{equation}
    \langle X_f X_f^{*} \rangle = \langle Y_f Y_f^{*} \rangle
    \label{eq: closure_eqn}
\end{equation}

In this formulation, the linear growth rate is expressed as 
\(\gamma_f = \Re(\Lambda_f^L)\), with the corresponding \textcolor{black}{linear transfer coefficient}
\((\Lambda_f^L)\) extracted from the linear transfer function$(L_f)$ defined in Eq.~\eqref{eq:L_lam}. However, the power transfer function and the energy transfer function follow the same form as in Eq.(\ref{eq:Tf}) and Eq.(\ref{eq:W_nl}).The formulation developed above is applicable to experimental fluctuation data; however, its reliability must first be established. Accordingly, we validate the method using numerical simulations with analytically prescribed transfer functions, following an approach similar to that employed for the Ritz method. The details of this validation procedure and the corresponding results are presented in the subsequent section.



\subsubsection{Computational Algorithm for Kim Method}
\label{subsubsection: computational algo kim}

For the Kim method, the initial computational process is the same as that of the Ritz method, i.e., we follow the same procedures from \textit{Step 1} to \textit{Step 2}. From \textit{Step 3} onward, the procedure is modified and presented as follows:

\textit{Step 3: For $f = -\frac{N}{2}$ to $\frac{N}{2}-1$, execute the following steps:}

\begin{itemize}

    \item \textit{Step 3.1:} Identify all frequency pairs $(f_1, f_2)$ that satisfy $f_1 + f_2 = f$, and store the number of such pairs in \textit{Npairs}.
    
    \item \textit{Step 3.2:} Compute the arrays $A_f$ and $B_f$, and construct the matrix $F_f$.
    
    \item \textit{Step 3.3:} Evaluate the linear and quadratic transfer functions, $L_f$ and $Q_f$.
    
    \item \textit{Step 3.4:} Calculate the \textcolor{black}{linear transfer coefficient} and store it in the array $\Lambda_f^L(f)$. Compute the quadratic coupling coefficients for all pairs $(1 \ldots \textit{Npairs})$ and assign them to the array $\Lambda_f^Q$.

\end{itemize}

Subsequently,  to compute the power transfer function $T_f(f_1, f_2)$ and the nonlinear energy transfer function $W_{NL}^f$, steps are carried out in the same manner as in the Ritz method, from \textit{Step 6} to \textit{Step 7}. We use the above algorithm for Ritz method in case of both simulated data and experimental data which is discussed in the subsequent sections.

\subsubsection{Simulation Test and Analysis}
\label{subsubsection: simulation test Kim}
To validate the method, we employ the same analytical linear and quadratic transfer functions introduced in Sec.~\ref{subsubsection: simulation test ritz} [Eqs.~(\ref{ana_Ltrans_fn}) and (\ref{ana_Qtrans_fn})]. The input and output data are generated using the same nonlinear black-box model adopted in the Ritz-method validation. For the present simulations, the input–output pair from the fifth black box (i.e., iteration $=5$) is taken as the final input $X_f$ and output $Y_f$ in the frequency domain. The simulations consist of $M = 2500$ realizations, each containing $N = 64$ data points. 

To benchmark the implementation, the estimated linear coefficient $L_f$ and nonlinear coupling coefficient $Q_f^{f_1,f_2}$ are compared with their corresponding analytical expressions, and the results are presented in the following figures. From the Fig.~\ref{fig:kim_simulation_validation}, it is evident that the estimated transfer functions closely match their corresponding analytical expressions, thereby confirming the correct implementation of the Kim method. Furthermore, the results obtained using the Kim method are noticeably smoother and more accurate than those obtained from the Ritz method, as demonstrated by the benchmarking plots. This improvement arises from the explicit retention of fourth-order moments in the Kim formulation. In the following section, we therefore employ the Kim method to address the limitations encountered in the Ritz approach.

\begin{figure}[H]
    \centering
    \includegraphics[width=0.85\linewidth]{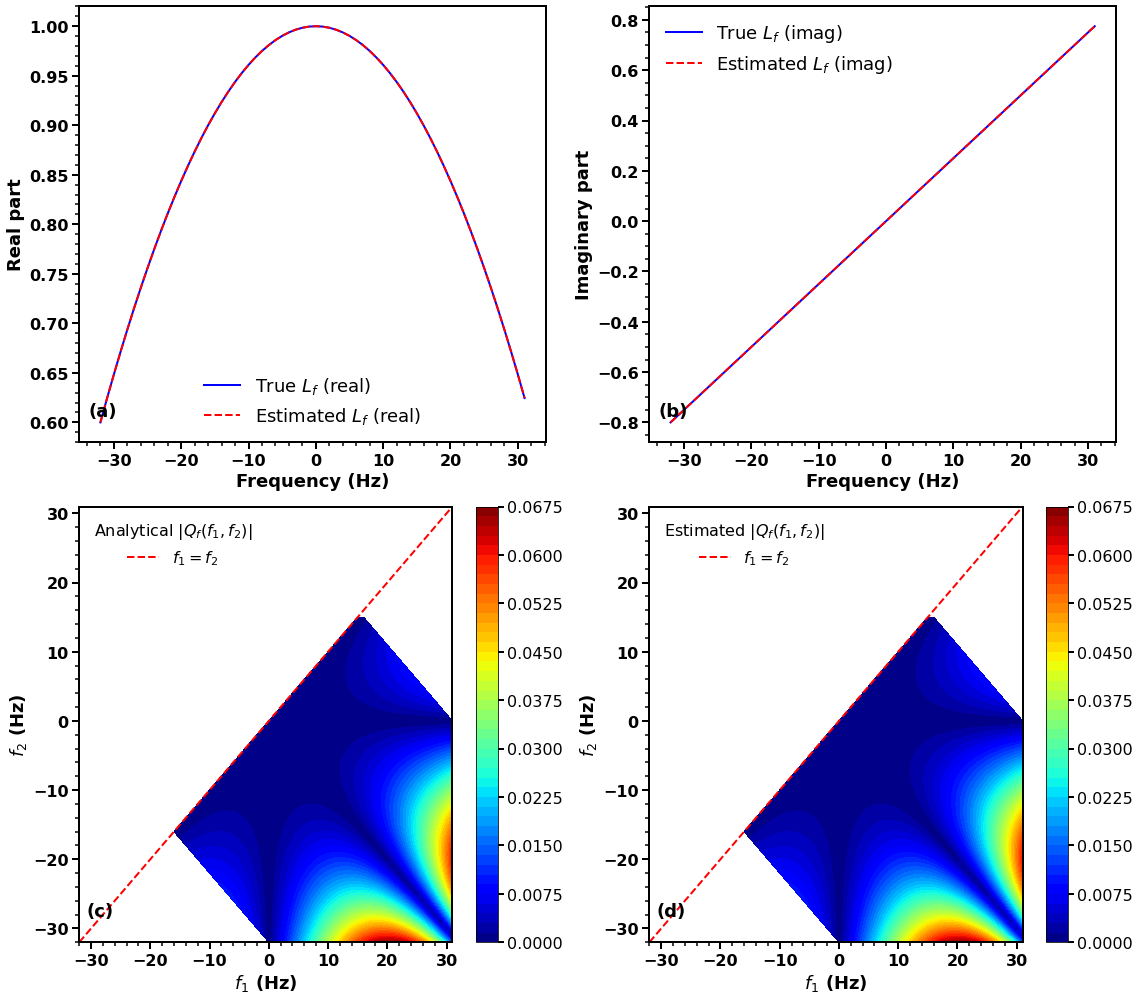}
    \caption{
    Simulation validation of the Kim method.
    Panels (a) and (b) show the real and imaginary parts of the linear transfer function $L_f$.
    Panels (c) and (d) compare the analytical and estimated magnitudes of the quadratic coupling coefficient
    $\lvert Q_f(f_1,f_2) \rvert$.
    }
    \label{fig:kim_simulation_validation}
\end{figure}

\subsubsection{Addressing the Limitations of the Ritz Method through Kim Method}
\label{subsubsection: addresing issue of Ritz by kim}
The primary limitation of the Ritz method is its inability to adequately handle data exhibiting high kurtosis. As discussed in Sec.~\ref{subsubsection: applicability of Ritz}, kurtosis is a direct statistical measure of the fourth-order central moment of a distribution. However, the Ritz method relies on Millionshchikov’s approximation\cite{Millionshchikov1941}, in which the fourth-order cumulant is replaced by the square of second-order cumulants. This approximation becomes invalid for strongly non-Gaussian or highly nonlinear data. 

In contrast, the Kim method does not invoke this approximation and explicitly retains the fourth-order moments, enabling it to more accurately capture nonlinear interactions and to remain robust for data with high kurtosis. This can be easily justified from the benchmarking plots of $L_f$ and $Q_f^{f_1,f_2}$ at higher iterations. The plots for iteration = 10 is shown in the Fig.~\ref{fig:kim_simulation_iter10}.

From the Fig.~\ref{fig:kim_simulation_iter10}, it is evident that the Kim method continues to yield results in good agreement with the analytical solutions even at iteration 10, corresponding to a kurtosis value of 1.04. In contrast, the Ritz method shows noticeable deviations beginning at iteration 7, where the kurtosis reaches approximately 0.55. These observations indicate that the Kim method is more robust to increasing non-Gaussianity and is therefore more suitable than the Ritz method for the analysis of general nonlinear data. In the analysis of experimental fluctuation data, we primarily employ the Kim method due to its robustness in handling strong nonlinearities. However, in the present work, both the Kim and Ritz methods are applied to the same experimental dataset in order to quantitatively assess the deviation between the two approaches when applied to real measurements. This comparison and its implications are discussed in detail in the next section.

\begin{figure}[H]
    \centering
    \includegraphics[width=0.85\linewidth]{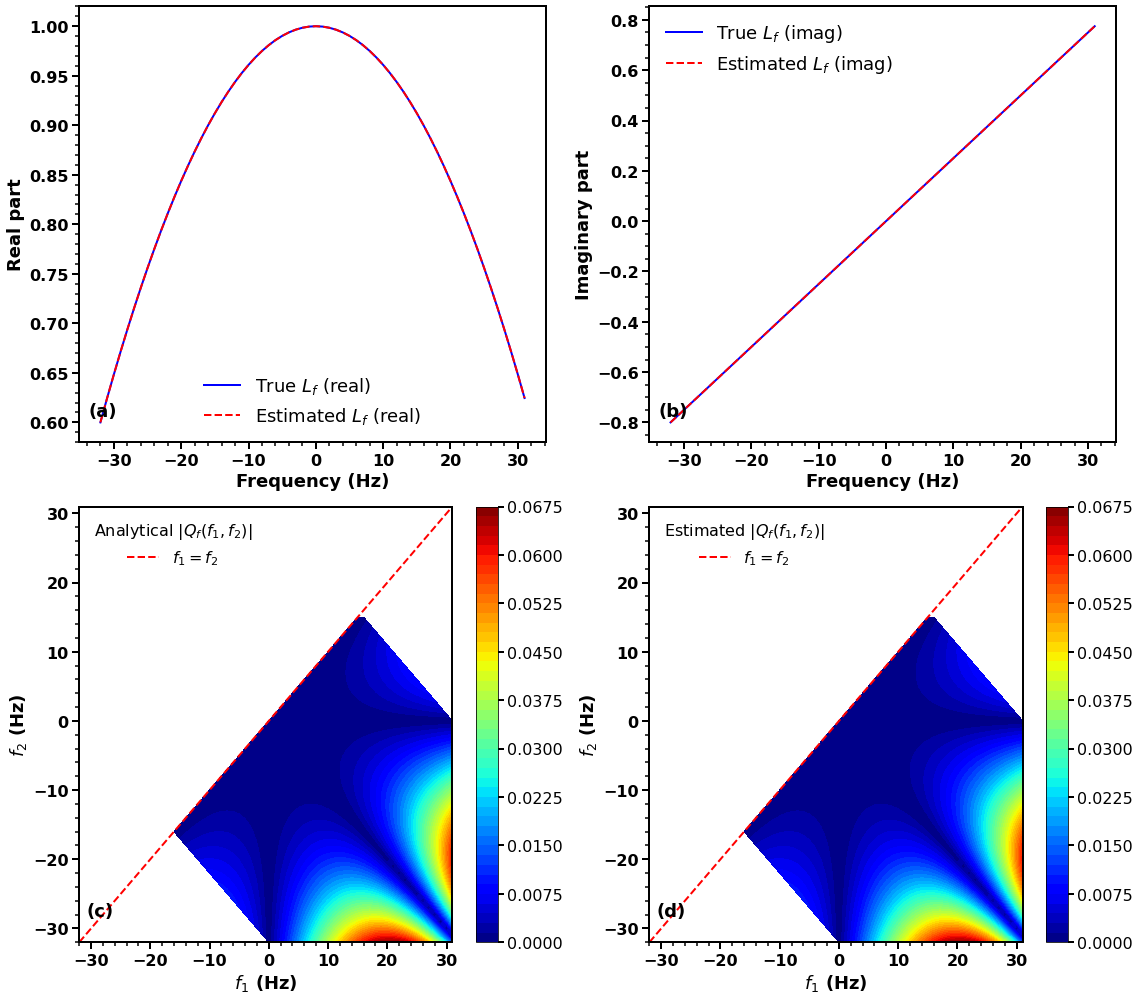}
    \caption{
    Panels (a) and (b) show the real and imaginary parts of the linear transfer function $L_f$.
    Panels (c) and (d) compare the analytical and estimated magnitudes of the quadratic coupling coefficient
    $\lvert Q_f(f_1,f_2) \rvert$ at iteration = 10.
    }
    \label{fig:kim_simulation_iter10}
\end{figure}


\subsubsection{Discussion on Growth Rate and Energy Transfer Function}
\label{subsubsection: growth rate and ETF simuation}

As discussed earlier, we can estimate the linear growth rate $\gamma_f$ and the nonlinear energy transfer function $T_f$ by using Kim method. These two quantities describe different but complementary aspects of fluctuation dynamics in frequency space. The growth rate $\gamma_f$ indicates whether the fluctuation at frequency $f$ grows or decays in time. A positive value $\gamma_f>0$ means that the mode at frequency $f$ is unstable and its amplitude increases with time. In this case, the mode extracts energy from the system and acts as a source of energy. A negative value $\gamma_f<0$ means that the mode is stable and damped, so its amplitude decreases and the mode loses energy unless it is supported by nonlinear interactions with other modes. Whereas, the energy transfer function $T_f$ describes how energy is redistributed among different frequencies through nonlinear three-wave (triad) interactions satisfying $f=f_1+f_2$. With the present sign convention, $T_f>0$ represents a net nonlinear transfer of energy into the mode at frequency $f$ from other frequencies, while $T_f<0$ represents a net transfer of energy out of the mode at $f$ toward other frequencies.

The relation between these two quantities is given by Eq.~\eqref{eq:stationarity2}, which shows that growth and transfer play opposite roles. If $\gamma_f>0$, then $T_f<0$, meaning that an unstable mode generates energy and redistributes it to other frequencies through nonlinear coupling. If $\gamma_f<0$, then $T_f>0$, meaning that a damped mode can persist only by receiving energy from other, typically dominant, modes. Therefore, the growth rate identifies where energy is created or removed in the spectrum, while the energy transfer function determines how that energy is transported across frequencies. Their combined behavior controls the spectral evolution and saturation of the fluctuations.

\begin{figure}[H]
    \centering
    \includegraphics[width=0.9\textwidth]{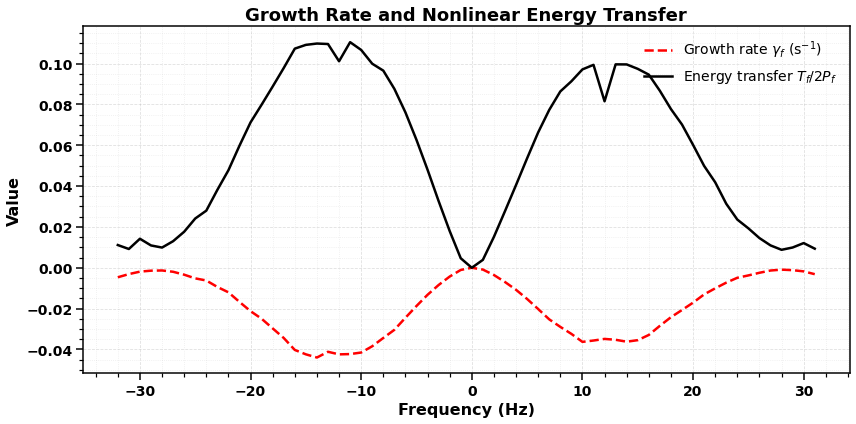}
    \caption{Growth rate $\gamma_f$ and nonlinear energy transfer $T_f/2P_f$ as a function of frequency.}
    \label{fig:growth_rate_etf}
\end{figure}
The growth rate and energy transfer function are plotted in Fig.~\ref{fig:growth_rate_etf}. It can be observed that \(\gamma_f\) and \(T_f/2P_f\) do not sum to zero, indicating that the spatial stationarity condition given in Eq.~\eqref{eq:stationarity2} is not satisfied within this method.
From Fig.~\ref{fig:psd5}, it is evident that the power spectra of the input and output data do not perfectly coincide, which leads to a mismatch in the complementary behavior of \(\gamma_f\) and \(T_f/2P_f\). This deviation is not unphysical; rather, it arises from the choice of the analytical linear and quadratic transfer functions adopted from \cite{ritz1986}. As discussed in Sec.~\ref{subsubsection: simulation test ritz}, these transfer functions correspond to a developing turbulence regime, in which the growth rate and nonlinear transfer function are not balanced and the input and output power spectra are unequal. A different choice of transfer functions, corresponding to fully developed turbulence \cite{ritz1989} and more consistent with the simulated dynamics, may improve the spatial stationarity of the data. This would better satisfy Eq.~\eqref{eq:stationarity2} and yield a closer complementary behavior between the nonlinear transfer function and the linear growth rate.

Here, we introduce a parameter called the \textit{energy imbalance parameter} \((\Delta_{\mathrm{im}})\), defined as the frequency-averaged spectral energy balance between growth rate and nonlinear energy transfer function,
\begin{equation}
\Delta_{\mathrm{im}} = \left\langle 2\gamma_f P_f + T_f \right\rangle_f .
\label{delta}
\end{equation}
For an ideally spatially stationary system, the local balance is satisfied at each frequency, yielding \(\Delta_{\mathrm{im}} = 0\). A finite value of \(\Delta_{\mathrm{im}}\) therefore measures the net excess or deficit of energy when linear growth and nonlinear redistribution do not fully compensate, indicating that the spectrum is not perfectly stationary. Equivalently, \(\Delta_{\mathrm{im}} = 0\) corresponds to a fully developed turbulent regime, whereas deviations from zero indicate a developing turbulent state.

In our case, for the simulated data analyzed using Kim method, we obtain \(\Delta_{\mathrm{im}} = 0.007\), which is very small but indicating developing turbulent state. This allows Eq.~\eqref{eq:stationarity2} to be written approximately as
\(2\gamma_f P_f + T_f \approx 0\), or \(\gamma_f \approx -T_f/(2P_f)\), consistent with the behavior observed in Fig.~\ref{fig:growth_rate_etf}. Consequently, Eq.~\eqref{eq:stationarity2} implies \(\langle Y_f Y_f^{*} \rangle \approx \langle X_f X_f^{*} \rangle\) which supports the closure equation i.e, Eq.~\eqref{eq: closure_eqn}. Therefore, although the turbulence is still developing, the spatial stationarity condition is approximately satisfied, and Kim method can be applied with reasonable accuracy.

\section{Application on Experimental Data and 
Analysis }
\label{section: application to expt data}
In the present study, experiments are carried out in a cylindrical plasma device known as the Inverse Mirror Plasma Experimental Device (IMPED \cite{bose2015inverse, sayak2015inverse}). The plasma is axially uniform over a length of approximately 2~m in the main chamber. \textcolor{black}{The radius of the plasma column in the IMPED device is \(8\,\mathrm{cm}\), and the corresponding operating conditions and plasma parameters are summarized in Table~\ref{tab:simParameters}.} IMPED has an additional control parameter, \(R_m\), defined as the ratio of the magnetic field in the main chamber to that in the source chamber. Previously, IMPED has been operated in different experimental regimes to investigate nonlinear interactions among primary instabilities such as drift wave (DW), Rayleigh--Taylor (RT), and Kelvin--Helmholtz (KH) modes. At a magnetic field of 550~G and a neutral pressure of \(1 \times 10^{-4}\,\text{mbar}\), it has been demonstrated that variations in the density scale length and velocity shear scale length can lead to a transition from drift wave instability (DWI) to Kelvin--Helmholtz instability (KHI), along with their nonlinear interactions \cite{roy2025experimental}. With a further reduction in the magnetic field, reduced magnetic tension facilitate the onset of Rayleigh--Taylor instability (RTI) along by different KHI modes \cite{Rosh2025experimental}. An increase in ion--neutral collisions introduces nonlinearities in both density and velocity fields, rendering the system more susceptible to nonlinear mode coupling. Recently, a study conducted by our group at 350~G and a pressure of \(4 \times 10^{-4}\,\text{mbar}\) provided a detailed identification of the primary modes, namely DWI and RTI, along with their nonlinear interactions and energy transfer mechanisms, which is submitted for publication. In the present work, for the same parameter, we perform a comprehensive analysis of nonlinear energy transfer between drift-wave (DW) and Rayleigh–Taylor (RT) modes using both the Ritz and Kim methods at two radial locations exhibiting distinct statistical properties. These locations are selected to assess the validity and applicability of each method with respect to the underlying statistical character and the spatial stationarity of the measured fluctuations, as discussed in the following sections. Long time-series measurements of plasma density fluctuations ($\tilde{n}$) and floating potential fluctuations ($\tilde{\phi}_f$), each of 10~s duration, are obtained using a three-tip Langmuir probe to have large number of statistical realization such that the error in the higher order moments can be minimized as discussed at the section~\ref{subsubsection: simulation test ritz}. The fluctuations $\tilde{n}$ and $\tilde{\phi}_f$ are used as proxies for ion saturation current fluctuations ($\tilde{I}_{sat}$) and plasma potential fluctuations ($\tilde{\phi}_p$), respectively, assuming electron temperature fluctuations ($\tilde{T}_e$) are negligible \cite{roy2025experimental, Rosh2025experimental,burin2005transition}. Each probe tip ($C_1$, $C_2$, and $C_3$) has a length of 4~mm and a diameter of 0.5~mm. \textcolor{black}{The \(C_1\) and \(C_2\) probe tips are used to measure density fluctuations (\(\tilde{n}\)), whereas the \(C_3\) tip measures floating potential fluctuations (\(\tilde{\phi}_f\)). A detailed description of the probe configuration can be found in Fig.~2 in Ref.~\cite{Rosh2025experimental}}.
\textcolor{black}{The data are acquired at a sampling frequency of 1~MHz, corresponding to a sampling rate of 1~MS/s. For a data acquisition duration of 10~s, the total number of acquired samples is therefore $10^7$.} The cross-phase between density and potential fluctuations measured using probes $C_1$ and $C_3$, along with the normalized fluctuation level of the corresponding modes, is used to identify the underlying instability mechanisms. For the estimation of the nonlinear energy transfer function, the ``two-probe method" is employed, in which two Langmuir probes are spatially separated by a distance $\Delta x$ to measure fluctuations at $x$, $\phi(x,t)$, and at $x+\Delta x$, $\phi(x+\Delta x,t)$. In the present experiment, density fluctuation data ($\tilde{n}$) are obtained using two poloidally separated Langmuir probes with a fixed separation of $\Delta x = 6.4$~mm. In this framework, the fluctuation signal measured at position $x$, $\phi(x,t)$, is treated as the input signal ($X$), while the signal measured at $x+\Delta x$, $\phi(x+\Delta x,t)$, is considered as the output signal ($Y$), as discussed earlier. In the experiment, only two spatially separated signals are available. To identify which signal should be treated as the input and which as the output, a cross-correlation function (CCF) analysis\cite{Bendat2010RandomData} is performed. The CCF provides information on the direction of flow or wave propagation, thereby allowing the identification of propagation from $X$ (input signal) to $Y$ (output signal) or $Y$ to $X$. The details of this analysis are discussed in the next section.

\begin{table}[H]
    \centering
    \caption{Typical plasma and operational conditions of IMPED device}
    \label{tab:simParameters}
    \begin{tabular}{|l|c|l|}
        \hline
        \textbf{Parameter} & \textbf{Range} & \textbf{Unit} \\
        \hline
        Magnetic Field \( B \) & 100-1200 & G \\
        Operating Pressure (Argon) & \(2\times10^{-5}-1\times10^{-3}\) & mBar \\
        Peak Electron Density & \(8 \times 10^{11}\) & \(\text{cm}^{-3}\) \\
        Electron Temperature & (2 - 5) & eV \\
        Ion Temperature & 0.025 & eV \\
        Density Scale Length & (2-5) & cm \\
        Electron Plasma Frequency & (0.3 - 9) & GHz \\
        Electron Gyro Frequency & (0.3 - 3.4) & GHz \\
        Ion Gyro Frequency & (3.8 - 45.6) & KHz \\
        Electron Gyro Radius & (0.04 - 0.3) & mm \\
        Ion Gyro Radius & (1 - 13) & mm \\
        Electron-Neutral Collision Frequency & (6.6 - 213) & KHz \\
        Ion-Neutral Collision Frequency & (1.18 - 23.7) & KHz \\
        Electron-ion Collision Frequency & (0.6-800) & MHz \\
        Ion-Neutral Mean Free Path & (2.9 - 58) & cm \\
        Plasma Radius & 8 & cm \\
        \hline
    \end{tabular}
\end{table}

\subsection{Estimation of direction of flow propagation}
\label{subsection: direction of flow}

To determine the direction of propagation, the cross-correlation function (CCF)\cite{Bendat2010RandomData} is evaluated and is defined as
\begin{equation}
C_{XY}(\tau) = \sum_{\tau=-\infty}^{\infty} X(t)\,Y(t+\tau) d\tau
\label{eq:CCF}
\end{equation}

where $X(t)$ and $Y(t)$ represent the time series of the two measured potential signals, and $\tau$ denotes the time lag. 
If a finite time delay exists between the two signals, the CCF peak is shifted away from zero lag. A positive lag indicates propagation from $X$ to $Y$, validating the initial assumption, whereas a shift toward negative lag implies that the propagation direction is reversed. Fig.~\ref{fig:ccf_5.76cm} shows the cross-correlation function (CCF) at a radial position of 5.76~cm. The cross-correlation between $X$ and $Y$ exhibits a distinct peak at a positive time lag, indicating a finite propagation delay. This confirms that the wave propagates from $X$ to $Y$, justifying the identification of $X$ as the input signal and $Y$ as the output signal. The same analysis is performed in other radial locations before undergoing the non-linear energy transfer analysis.

\begin{figure}[H]
    \centering
    \includegraphics[width=0.6\textwidth]{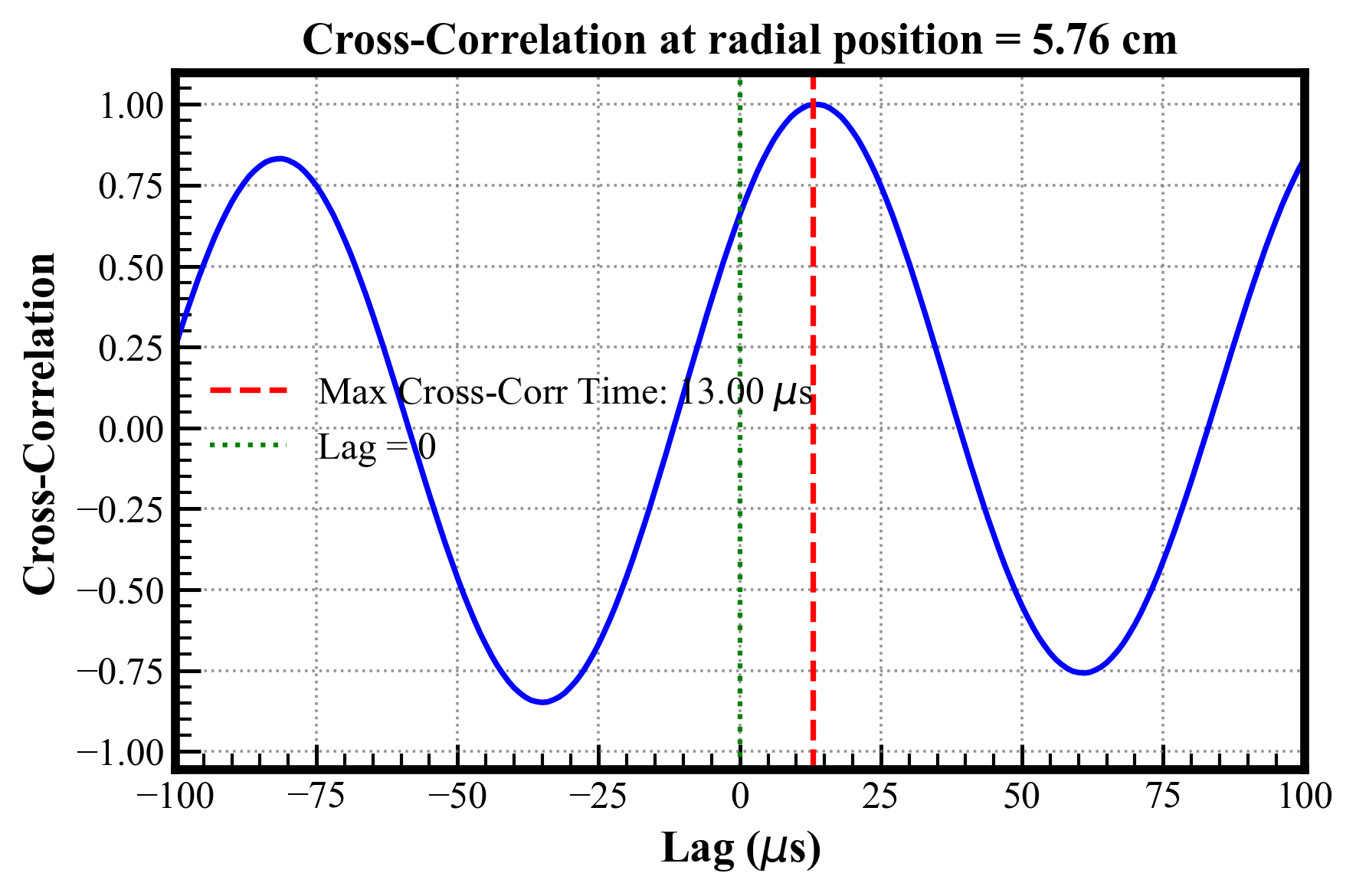}
    \caption{Cross-correlation function (CCF) at a radial position of 5.76~cm showing a positive time delay between the input ($X$) and output ($Y$) signals.}
    \label{fig:ccf_5.76cm}
\end{figure}

\subsection{Experimental Data Selection}
\label{subsection: data selection} 
To ensure the validity and applicability of the Ritz and Kim methods on experimental data, density fluctuations($\tilde{n}$) are selected at two distinct radial locations based on their statistical parameters(mainly skewness and kurtosis) and spatial stationarity. These criteria and their implications are discussed in detail in the following paragraphs.

\paragraph{I. Statistical Parameters:} From the discussion in the preceding sections, it is evident that the Ritz method cannot be applied to arbitrary time-series data. The applicability of the Ritz method is restricted to signals that are approximately Gaussian in nature, which implies that the data should exhibit low skewness and kurtosis. In contrast, the Kim method does not impose such constraints on the underlying data distribution. The statistical properties of the fluctuations, including skewness and kurtosis, are evaluated over a range of radial positions from 0 to 7.04~cm to guide the appropriate data selection for the application of the Ritz and Kim methods, as shown in Fig.~\ref{fig:stat_par}.


\begin{figure}[H]
    \centering

    \begin{subfigure}[b]{0.49\textwidth}
        \centering
        \includegraphics[width=\textwidth]{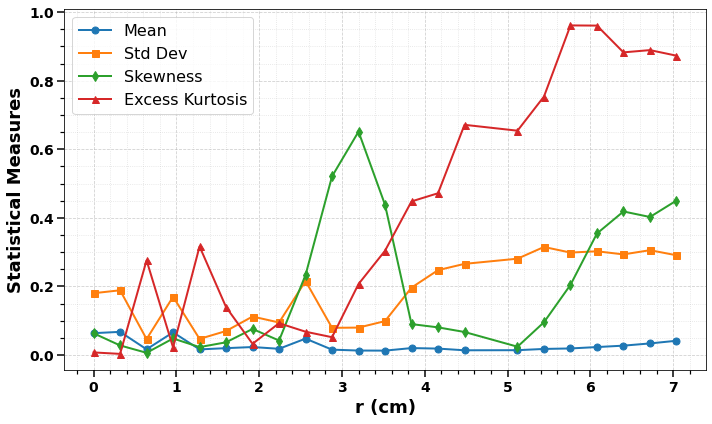}
        \caption{Variation of statistical parameters with radial position.}
        \label{fig:stat_par}
    \end{subfigure}
    \hfill
    \begin{subfigure}[b]{0.49\textwidth}
        \centering
        \includegraphics[width=\textwidth]{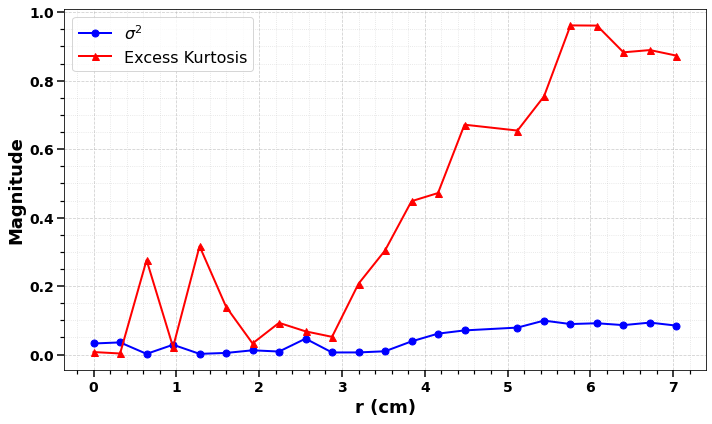}
        \caption{Comparison of the variance ($\sigma^2$) and excess kurtosis as a function of radial position.}
        \label{fig:var_kurt}
    \end{subfigure}

    \caption{Statistical properties of the fluctuation signal at different radial positions. Panel (a) shows Radial variation of statistical parameters (mean, standard deviation, skewness, and excess kurtosis) of the measured density fluctuations($\tilde{n}$), while panel (b) compares the ($\sigma^2$) and excess kurtosis to assess the validity of Millionshchikov's approximation.}
    \label{fig:stats_IoS}
\end{figure}

From Fig.~\ref{fig:stat_par}, it can be observed that the data for the Ritz method can be selected below a radial position of 4~cm to maintain the excess kurtosis within $0.40$, consistent with the simulation results. \textcolor{black}{Furthermore, Fig.~\ref{fig:var_kurt} shows that, for radial positions up to approximately $r = 4$~cm, the values of $\sigma^2$ and excess kurtosis remain relatively close. This suggests that Millionshchikov's approximation may be applicable in this region, and consequently, the Ritz method is expected to provide reliable estimates. Moreover, at certain radial locations within $r = 4$~cm, the agreement between $\sigma^2$ and excess kurtosis is particularly good, indicating conditions under which the Ritz method is most likely to be valid.} In the present analysis, the absolute values of the statistical parameters are used, as their signs do not influence the interpretation of the results. Based on the above analysis, the energy transfer-function analysis is performed at two distinct radial locations: one closer to the plasma center and the other farther away. In this study, these locations correspond to $r = 2.24$~cm and $r = 5.76$~cm, respectively. The probability distribution functions of the input data at these radial locations are shown in Fig.~\ref{fig:pdfs}. \textcolor{black}{The smooth curves represent the Gaussian Kernel Density Estimation (KDE), which provides a continuous estimate of the probability density function by smoothing the discrete data using Gaussian kernels.} From the Fig.~\ref{fig:pdf_2.24}, it is observed that at $r = 2.24$~cm the distribution is nearly Gaussian, with skewness $= 0.042$ and excess kurtosis $= 0.093$. In contrast, at $r = 5.76$~cm the distribution deviates significantly from Gaussian behavior, exhibiting skewness $= 0.205$ and excess kurtosis $= 0.961$ as shown in Fig.~\ref{fig:pdf_5.76}. \textcolor{black}{Accordingly, Ritz’s method is expected to yield reliable results at $r = 2.24$~cm, but to fail at $r = 5.76$~cm. This is consistent with Fig.~\ref{fig:var_kurt}, where $\sigma^2$ remains close to the excess kurtosis at $r = 2.24$~cm, whereas a significant deviation between the two quantities is observed at $r = 5.76$~cm.
.} Because Kim’s method does not depend explicitly on the skewness or kurtosis of the data, it is expected to provide consistent results at both radial locations. A detailed comparison of both approaches is presented in the following sections.

\begin{figure}[H]
    \centering
    \begin{subfigure}{0.49\textwidth}
        \centering
        \includegraphics[width=\linewidth]{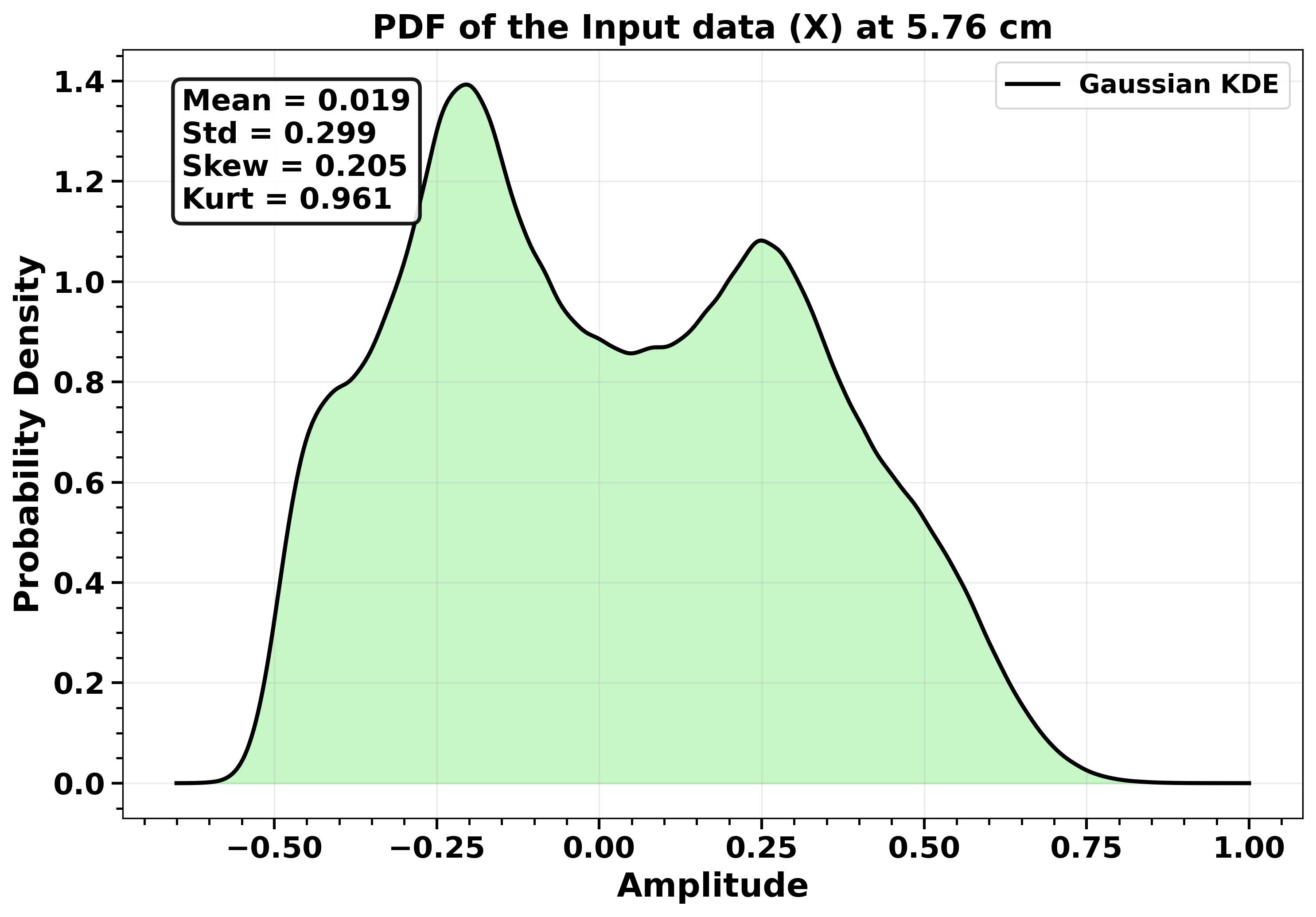}
        \caption{PDF at $r = 5.76$~cm.}
        \label{fig:pdf_5.76}
    \end{subfigure}
    \hfill
    \begin{subfigure}{0.49\textwidth}
        \centering
        \includegraphics[width=\linewidth]{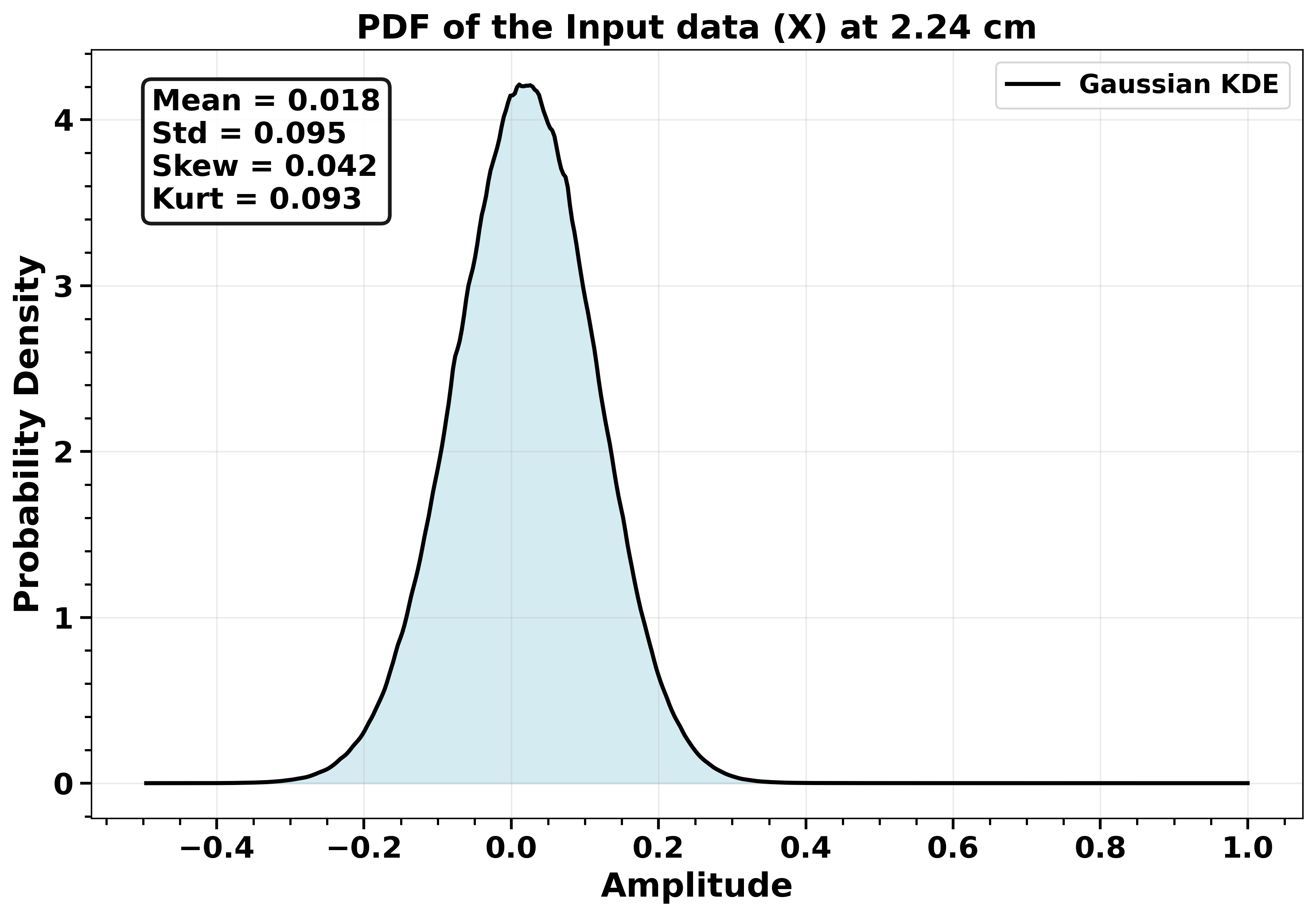}
        \caption{PDF at $r = 2.24$~cm.}
        \label{fig:pdf_2.24}
    \end{subfigure}
    \caption{Probability distribution functions of the measured density fluctuations ($\tilde{n}$) at two radial locations. \textcolor{black}{Gaussian KDE: Gaussian Kernel Density Estimation of the probability density function of the measured fluctuations.}}
    \label{fig:pdfs}
\end{figure}

\textcolor{black}{The PDF at $r=5.76$ cm exhibits a multimodal structure with more than one prominent peak, indicating a significant departure from a single Gaussian distribution. This behavior is consistent with the enhanced nonlinear activity at this radial location, as evidenced by the higher-order statistical moments and the comparatively higher bicoherence values presented in the following sections. The multimodal nature of the PDF may arise from nonlinear intermittency and active energy transfer between coupled modes, leading to multiple preferred fluctuation-amplitude states.}

\paragraph{II. Spatial Stationarity:} Spatial stationarity is defined as the condition in which the power associated with a given frequency component of a signal remains invariant in space. This implies that the power spectrum preserves both its shape and magnitude across different spatial locations. In this study, the fluctuation signals are considered spatially stationary when the power spectra of the input (\(X\)) and output (\(Y\)), separated by \(\Delta x\), are equal and exhibit the same spectral shape as indicated by Eq.~\eqref{eq:stationarity2}.

Previous studies \cite{ritz1989,Shen2020ImprovedBispectrum} have shown that Ritz’s method yields accurate results when applied to developing plasma turbulence, but fails for fully developed turbulence. Although Kim’s method was originally formulated for fully developed turbulence, it has also been shown to provide reliable results for developing turbulence models \cite{Shen2020ImprovedBispectrum}. \textcolor{black}{In fully developed turbulence, the input and output power spectra typically coincide}, indicating good spatial stationarity and broadband characteristics, \textcolor{black}{whereas in developing turbulence the spectra do not fully coincide, suggesting poorer spatial stationarity and the turbulent cascade has not yet reached a fully developed state}. However, Kim’s method becomes inapplicable when the input and output spectra do not follow similar trends \cite{kim1987ocean}. Consequently, Kim’s method remains suitable for developing turbulence cases in which the input and output spectra exhibit similar trends, even if they are not completely \textcolor{black}{coincidence}.

In the experimental data, coherent structures appear at low frequencies superimposed on a broadband turbulent background, indicating a developing turbulence regime. Although the input and output spectra do not fully coincide, they exhibit similar spectral trends, under which Kim’s method performs well. The Ritz method is also applicable under these developing turbulence conditions. As described earlier, the data are analyzed at two radial locations, $r = 2.24$~cm and $r = 5.76$~cm. \textcolor{black}{The power spectra of the input (\(X\)) and output (\(Y\)) signals at these two positions are shown in Fig.~\ref{fig:psd_compare}. Here, \(P_X\) and \(P_Y\) denote the power spectra of the density fluctuations corresponding to the input signal \(X(t)\) and output signal \(Y(t)\), respectively. The input and output signals are obtained from the \(C_2\) and \(C_1\) probe tips, which measure density fluctuations at different poloidal positions.} At $r = 5.76$~cm, the power spectra of $X$ and $Y$ nearly coincide, indicating good spatial stationarity. In contrast, at $r = 2.24$~cm the spectra do not coincide exactly, although in both cases the input and output spectra exhibit similar trends. Since the power spectra of the input and output at both radial locations exhibit similar spectral trends, we argue that the Kim method is applicable at both positions. These arguments are verified through nonlinear energy transfer function analysis, which is discussed in the following sections.



\begin{figure}[H]
    \centering
    \begin{subfigure}{0.49\textwidth}
        \centering
        \includegraphics[width=\linewidth]{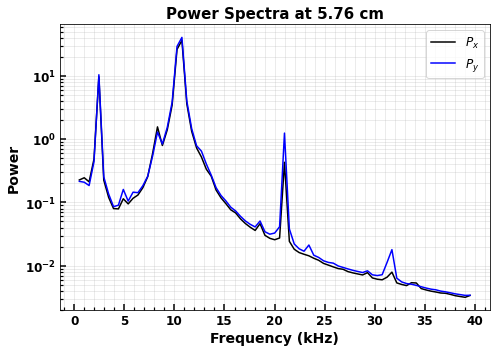}
        \caption{Power spectra at $r = 5.76$ cm.}
        \label{fig: PSD_5.76}
    \end{subfigure}
    \hfill
    \begin{subfigure}{0.49\textwidth}
        \centering
        \includegraphics[width=\linewidth]{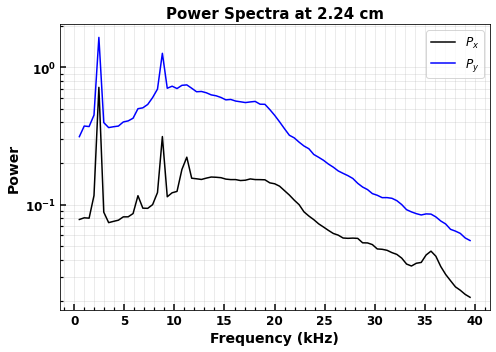}
        \caption{Power spectra at $r = 2.24$ cm.}
        \label{fig:PSD_2.24}
    \end{subfigure}
    \caption{\textcolor{black}{Comparison of the input power spectrum (\(P_X\)) and output power spectrum (\(P_Y\)) of density fluctuations corresponding to the input (\(X(t)\)) and output (\(Y(t)\)) signals, measured using the \(C_2\) and \(C_1\) probe tips, respectively, at \(r = 5.76\) cm and \(r = 2.24\) cm. The probe tips are located at different poloidal positions.}}
    \label{fig:psd_compare}
\end{figure}

\subsection{Analysis and Results}
\label{section: Analysis and results}
We now present the nonlinear transfer-function analysis of the experimental data using both the Kim and Ritz methods. As discussed earlier, the selected radial locations in this study correspond to \(r = 2.24~\mathrm{cm}\) and \(r = 5.76~\mathrm{cm}\), respectively. As discussed in the previous section [see Fig.~\ref{fig:stats_IoS}], the data at $r = 2.24$~cm exhibit low kurtosis, making them suitable for application of Ritz’s method. In contrast, Kim’s method does not depend explicitly on the skewness or kurtosis of the data and is therefore expected to provide consistent results at both radial locations. Moreover, from the stationarity viewpoint, the input and output spectra at both positions follow similar trends, supporting the applicability of Kim’s method at each location. To assess the validity of Ritz’s method, its results are systematically compared with those obtained using Kim’s method. Agreement between the two approaches indicates the applicability of Ritz’s method, whereas discrepancies signify its breakdown at a given radial position. Detailed analyses using both methods are presented in the following sections.

In present analysis, the frequency range is restricted to $40~\mathrm{kHz}$ because the dominant spectral modes lie within this band. Although modes inside this range can exchange energy with modes outside the measured spectrum, such interactions cannot be directly resolved. As a result, energy transferred from the measured frequencies to unobserved higher-frequency modes may appear as linear damping, even though it actually originates from nonlinear spectral redistribution. Accordingly, the computed $W_{\mathrm{mis}}$ does not represent the energy mismatch over the entire spectrum, but only the mismatch within the limited $40~\mathrm{kHz}$ band considered in the analysis.

For the estimation of the power spectra, bicoherence, and energy transfer functions, the number of data points per segment is chosen as $N = 2048$, and the number of ensemble realizations as $M = 4882$, corresponding to a total data length of $10~\mathrm{s}$ sampled at $1~\mathrm{MHz}$. This choice yields a frequency resolution of approximately $244~\mathrm{Hz}$. A large ensemble size is employed to ensure accurate estimation of higher-order moments, at the expense of spectral resolution. However, the resulting resolution of $244~\mathrm{Hz}$ is adequate for the present analysis, since the dominant modes are well resolved within this bandwidth.

\subsubsection{Nonlinear Energy Transfer Analysis at Radial Position $r = 5.76$~cm}
\label{subsubsection: analysis at 5.76}
In this section, the nonlinear energy transfer analysis at the radial location $r = 5.76$~cm is presented using both the Kim and Ritz methods, as described in the following paragraphs. Before discussing the energy transfer results, we first present the power spectrum and auto-bicoherence of the density fluctuations ($\tilde{n}$), shown in Fig.~\ref{fig:exp_data_5.76_PSD_Bx}.

\begin{figure}[H]
    \centering
    \includegraphics[width=1\linewidth]{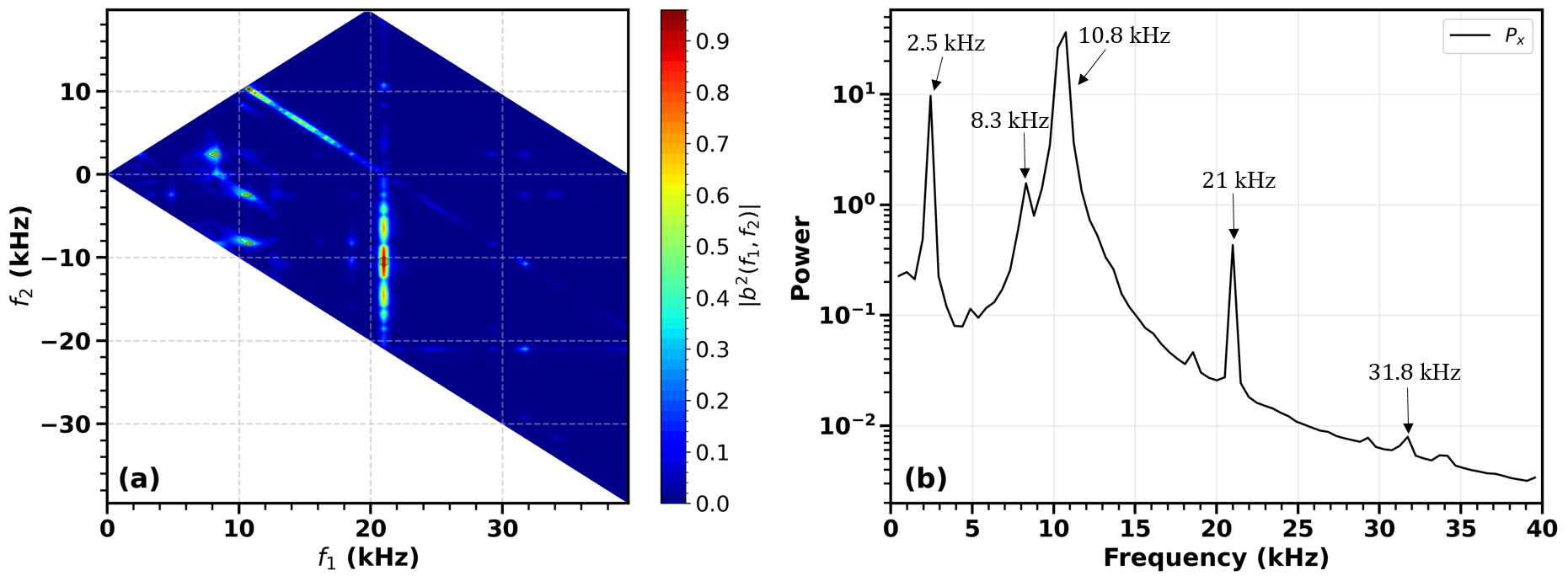}
    \caption{(a) Auto bicoherence analysis. (b) Density fluctuation spectra for $\tilde{n}$ of DW and RT modes at $r = 5.76$ cm.}
    \label{fig:exp_data_5.76_PSD_Bx}
\end{figure}

The auto–power spectrum and auto–bicoherence are computed using Eqs.~(\ref{eq:aPSD}) and (\ref{eq:abicoh}), respectively, as described in Section~\ref{section: sepectral framework}. The auto–bicoherence is evaluated over both positive and negative frequency regions by exploiting the symmetry relations discussed by Kim \textit{et al.}~\cite{kim1979}. From Fig.~\ref{fig:exp_data_5.76_PSD_Bx}(b), three dominant frequency modes are observed, which are listed in the Table--\ref{tab:5.76}, The instability types, poloidal wave numbers corresponding to these modes are also listed in the Table--\ref{tab:5.76}. From Fig.~\ref{fig:exp_data_5.76_PSD_Bx}(a), the frequency components at 
\( f_1 = 8.3~\mathrm{kHz} \) , \( f_2 = 2.5~\mathrm{kHz} \) and \( f = 10.8~\mathrm{kHz} \) are observed to be nonlinearly coupled, as indicated in the positive bicoherence region. This interaction satisfies the frequency resonance condition 
\( f_1 + f_2 = f \) as well as the corresponding wavenumber matching condition 
$k_{\theta,1} + k_{\theta,2} = k_{\theta}$, (see Table--\ref{tab:5.76}). Since the wave propagation is predominantly in the poloidal direction, the resonance condition is evaluated using only the poloidal wavenumber component. However, a small deviation from the exact wavenumber matching condition can arise due to finite spectral width of turbulent modes, plasma inhomogeneity, experimental uncertainty in wavenumber estimation, and resonance broadening associated with collisional and nonlinear effects. Therefore, nonlinear coupling is considered satisfied within the experimental uncertainty range. In the negative region, additional interactions consistent with the relations \( f - f_1 = f_2 \) and \( f - f_2 = f_1 \) are also observed. The observed finite bicoherence confirms coherent phase coupling among these modes, with all identified interactions showing bicoherence values above the statistical significance level [Eq.~\eqref{eq:bico_sig}], 
$b_{\mathrm{sig}}^{2} =\frac{1}{M} =  \frac{1}{4882} \approx 2 \times 10^{-4}$.  Furthermore, the frequency mode at $21~\mathrm{kHz}$ (where the 21~kHz mode is identified as the second harmonic of the 10.8~kHz mode, within the spectral resolution of 244~Hz.) is observed to interact with a broad band extending from the low-frequency region up to $21~\mathrm{kHz}$ in the negative bicoherence domain. \textcolor{black}{However, the bicoherence values indicate that the nonlinear interaction is strongest for the coupling between the 21~kHz mode and the 8--12.5~kHz frequency band in the negative-bicoherence region.} 
In the positive bicoherence region, frequency modes spanning from the low-frequency range up to $10.8~\mathrm{kHz}$ interact with the band between $10.8~\mathrm{kHz}$ and $21~\mathrm{kHz}$, as illustrated in Fig.~\ref{fig:exp_data_5.76_PSD_Bx}(a).

\begin{table}[H]
\centering
\setlength{\abovecaptionskip}{0pt}
\caption{Measured mode frequencies with corresponding poloidal ($k_{\theta}$) and radial ($k_r$) wavenumbers(absolute values), and identified instability type.}
\label{tab:5.76}
\vspace{1mm}
\begin{tabular}{cccc}
\toprule
Frequency (kHz) & $k_{\theta}$ (cm$^{-1}$) & Mode Type \\
\midrule
2.5  & 0.36 & Drift Wave (DW) \\
8.3  & 1.12 & \textcolor{black}{Mix mode of DW and RT} \\
10.8 & 1.49 & Rayleigh--Taylor (RT) \\
\bottomrule
\end{tabular}
\end{table}

While these results clearly indicate the presence of three-wave coupling among the modes, bicoherence alone does not provide information about the direction or magnitude of the associated energy exchange. To address this limitation, we employ nonlinear energy transfer analysis to quantify the direction and rate of energy flow among the interacting modes, as discussed in the following paragraphs.

\paragraph{I. Analysis using Kim Method:}

Figure~\ref{fig:ETF_Kim_5.76}(a--b) shows the power transfer function $(T_f(f_1,f_2))$ and the corresponding nonlinear energy transfer function $(W^{f}_{\mathrm{NL}})$ computed from two density fluctuation signals separated by 6.4~mm in the poloidal direction at the radial position $r=5.76$~cm. In the two-dimensional contour map of \( T_f(f_1,f_2) \) shown in Fig.~\ref{fig:ETF_Kim_5.76}(a), the positive (red) contour regions  correspond to energy gain at the coupled frequency \( f = f_1 + f_2 \) due to contributions from \( f_1 \) and \( f_2 \), whereas negative (blue) contour regions  indicate energy transfer from \( f \) back to its coupling frequencies. Figure~\ref{fig:ETF_Kim_5.76}(b) presents the nonlinear energy transfer function \( W^{f}_{\mathrm{NL}} \), obtained by integrating the power transfer function along the resonance line \( f = f_1 + f_2 \), where positive peaks indicate the energy gain and negative peaks indicate energy loss from the individual modes.

\begin{figure}[H]
    \centering
    \includegraphics[width=1\linewidth]{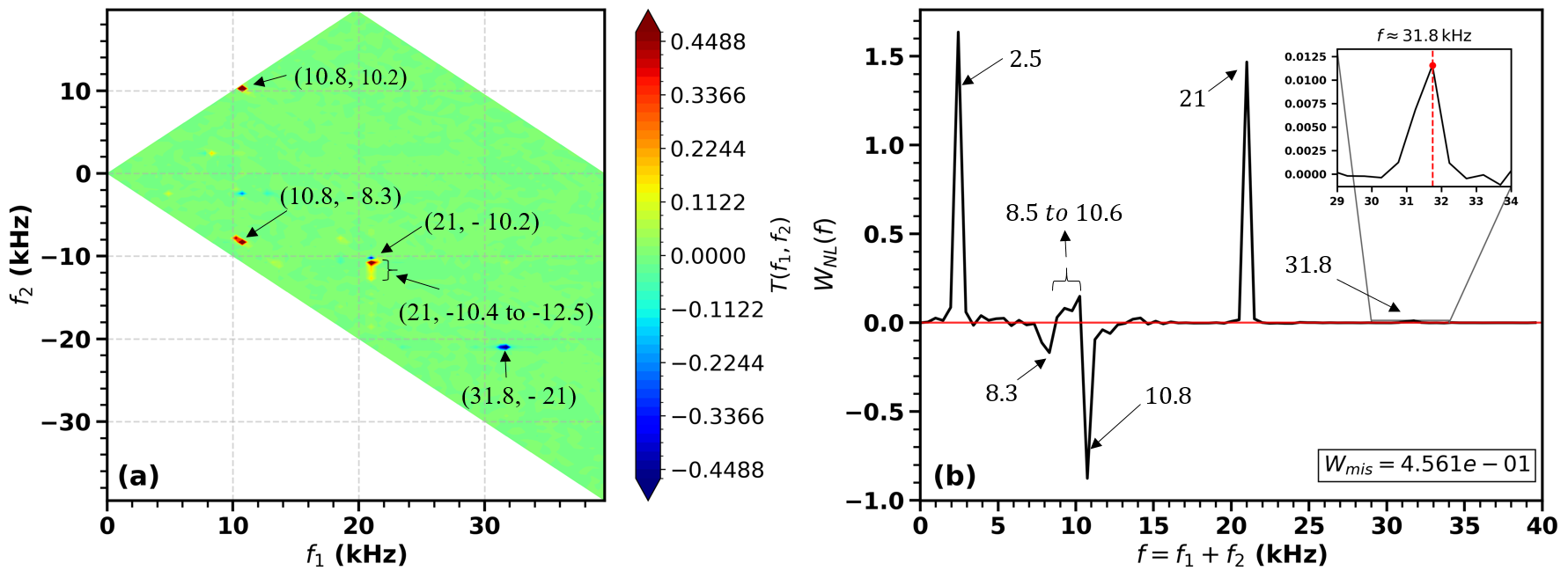}
    \caption{\textcolor{black}{Results obtained using the Kim method at $r = 5.76~cm$: (a) Power transfer function $T_f(f_1,f_2)$.
    (b) Energy transfer function $W^{f}_{\mathrm{NL}}(f)$ for $\tilde{n}$
    fluctuations associated with DW and RT modes. The coordinates
    $(f_1,f_2)$ (in kHz) corresponding to the dominant nonlinear interaction regions are marked in panel (a), while the corresponding frequencies
    $f=f_1+f_2$ are indicated in panel (b). Positive (red) regions in
    $T_f(f_1,f_2)$ denote triads for which energy is transferred from the
    interacting modes $(f_1,f_2)$ to the mode at $f=f_1+f_2$, producing
    positive peaks in $W^{f}_{\mathrm{NL}}(f)$. Negative (blue) regions denote
    triads for which energy is transferred from the mode at $f=f_1+f_2$ to the
    interacting modes $(f_1,f_2)$, resulting in negative peaks (dips) in
    $W^{f}_{\mathrm{NL}}(f)$. 
    For better visualization, the color scale in panel (a) is limited to the
    $99.9^{\mathrm{th}}$ percentile of $T_f(f_1,f_2)$; values exceeding this
    threshold are saturated in the colormap. The contribution at $f = 31.8$ kHz is relatively weak but remains above the noise floor; therefore, a zoomed inset is included in panel (b).}}
    \label{fig:ETF_Kim_5.76}
\end{figure}

From the two-dimensional contour plot of \( T_f(f_1,f_2) \) as shown in Fig.~\ref{fig:ETF_Kim_5.76}(a), \textcolor{black}{it is observed that the drift-wave (DW) mode at 2.5~kHz is gaining energy from the mix mode and Rayleigh--Taylor (RT) mode at 8.3~kHz and 10.8~kHz respectively}, following the interaction \((f_1 = 10.8~\text{kHz}) + (f_2 = -8.3~\text{kHz})\rightarrow (f = 2.5~\text{kHz})\). The one-dimensional spectrum \( W_{\mathrm{NL}}^{f} \) in Fig.~\ref{fig:ETF_Kim_5.76}(b) further supports this observations showing a positive peak near 2.5~kHz and negative peaks near 8.3~kHz and 10.8~kHz, indicating a net transfer of energy to the 2.5~kHz~(DW) mode and from the 8.3~kHz~(mix mode) and 10.8~kHz~(RT) modes through nonlinear coupling. Furthermore, Fig.~\ref{fig:ETF_Kim_5.76}(a) shows that the RT mode at \(10.8~\mathrm{kHz}\) loses energy by transferring it to modes at \(21~\mathrm{kHz}\), \(10.2~\mathrm{kHz}\), and \(31.8~\mathrm{kHz}\). These transfers occur through the triad interactions $(f = 10.8~\mathrm{kHz}) \rightarrow (f_1 = 21~\mathrm{kHz}) + (f_2 = -10.2~\mathrm{kHz})$ and $(f = 10.8~\mathrm{kHz}) \rightarrow (f_1 = 31.8~\mathrm{kHz}) + (f_2 = -21~\mathrm{kHz})$, which are identified by the negative contour (blue) regions in the two-dimensional power transfer function. \textcolor{black}{Further, energy is transferred from the 10.8~kHz and 10.2~kHz modes to the 21~kHz mode via the triad interaction
$
(f_1 = 10.8~\mathrm{kHz}) + (f_2 = 10.2~\mathrm{kHz})
\rightarrow
(f = 21~\mathrm{kHz}),
$
as indicated by the positive (red) contour region in Fig.~\ref{fig:ETF_Kim_5.76}(a). The $W_{\mathrm{NL}}^{f}$ spectrum in Fig.~\ref{fig:ETF_Kim_5.76}(b) further supports these observations, exhibiting a negative peak at $10.8~\mathrm{kHz}$ and positive peaks at $21~\mathrm{kHz}$. The positive peaks at $31.8~\mathrm{kHz}$ and $10.2~\mathrm{kHz}$ are comparatively weak, indicating that a net transfer of energy occurs away from the $10.8~\mathrm{kHz}$ mode toward the $21~\mathrm{kHz}$ mode. In addition, the $21~\mathrm{kHz}$ mode interacts with a band of frequencies in the range $10.4$--$12.5~\mathrm{kHz}$ and transfers energy to the $8.5$--$10.6~\mathrm{kHz}$ band, as indicated by the positive contour regions (red and yellow) in Fig.~\ref{fig:ETF_Kim_5.76}(a). This observation is consistent with Fig.~\ref{fig:ETF_Kim_5.76}(b), which exhibits a broad positive peak spanning approximately $8.5$--$10.6~\mathrm{kHz}$.} The remaining peaks in the $W_{\mathrm{NL}}^{f}$ spectrum correspond to relatively weak nonlinear interactions.

\paragraph{II. Analysis using Ritz Method:}
Here, the nonlinear energy transfer analysis of the same data at the radial location \( r = 5.76~\mathrm{cm} \) is performed using the Ritz method. Figure~\ref{fig:ETF_Ritz_5.76}(a--b) presents the power transfer function \( T_f(f_1,f_2) \) and the corresponding nonlinear energy transfer rate \( W^{f}_{\mathrm{NL}} \), computed from the same density fluctuation data used for the Kim-method analysis in the previous section. From the two-dimensional contour plot of $T_f(f_1,f_2)$ shown in Fig.~\ref{fig:ETF_Ritz_5.76}(a), the RT mode at $10.8~\mathrm{kHz}$ is observed to gain energy from the DW mode at $2.5~\mathrm{kHz}$ and the mix mode at $8.3~\mathrm{kHz}$ through the triad interaction $(f_1 = 2.5~\mathrm{kHz}) + (f_2 = 8.3~\mathrm{kHz}) \rightarrow (f = 10.8~\mathrm{kHz})$ indicated by positive contour(red) region. In addition, the mode at $21~\mathrm{kHz}$ interacts with a frequency band from $9.5~\mathrm{kHz}$ to $13.5~\mathrm{kHz}$, producing positive contour (red) regions in the power transfer function. The mode at $12.7~\mathrm{kHz}$ transfers energy to $21~\mathrm{kHz}$ and $8.3~\mathrm{kHz}$ via the triad interaction $(f = 12.7~\mathrm{kHz}) \rightarrow (f_1 = 21~\mathrm{kHz}) + (f_2 = -8.3~\mathrm{kHz})$, which appears as negative contours (blue) in the transfer map. Furthermore, the RT mode at $10.8~\mathrm{kHz}$ loses energy to it's harmonics at $21~\mathrm{kHz}$ and $31.8~\mathrm{kHz}$ modes through the triad interaction $(f = 10.8~\mathrm{kHz}) \rightarrow (f_1 = 31.8~\mathrm{kHz}) + (f_2 = -21~\mathrm{kHz})$, also indicated by negative contour (blue) regions.  The one-dimensional spectrum $W_{\mathrm{NL}}^{f}$ in Fig.~\ref{fig:ETF_Ritz_5.76}(b) supports these trends, showing negative peaks near $2.5~\mathrm{kHz}$ and $12.7~\mathrm{kHz}$, and positive peaks in the $7.5$--$11.5~\mathrm{kHz}$ band and near $31.8~\mathrm{kHz}$. The remaining peaks correspond to relatively weak nonlinear interactions.

\begin{figure}[H]
    \centering    \includegraphics[width=1\linewidth]{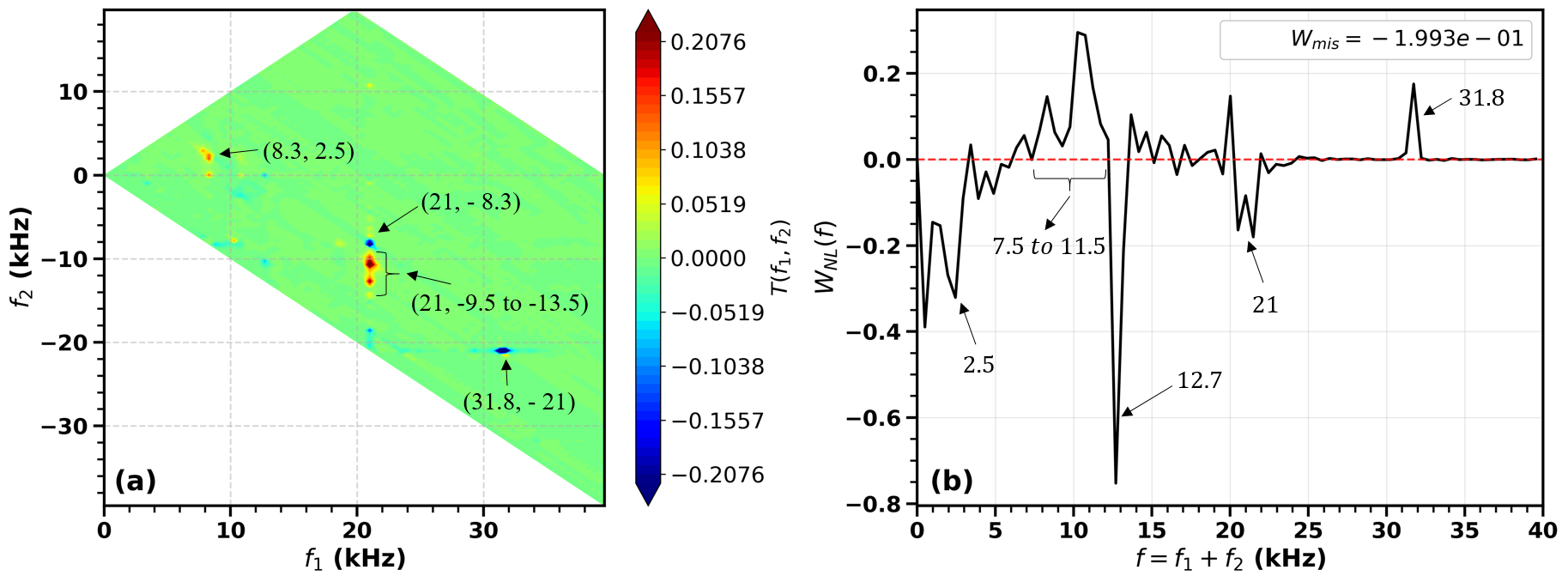}
    \caption{\textcolor{black}{Results obtained using the Ritz method at $r=5.76~\mathrm{cm}$:
    (a) Power transfer function $T_f(f_1,f_2)$.
    (b) Energy transfer function $W^{f}_{\mathrm{NL}}(f)$ for $\tilde{n}$
    fluctuations associated with DW and RT modes. The coordinates
    $(f_1,f_2)$ (in kHz) corresponding to the dominant nonlinear interaction regions are marked in panel (a), while the corresponding frequencies
    $f=f_1+f_2$ are indicated in panel (b). Positive (red) regions in
    $T_f(f_1,f_2)$ denote triads for which energy is transferred from the
    interacting modes $(f_1,f_2)$ to the mode at $f=f_1+f_2$, producing
    positive peaks in $W^{f}_{\mathrm{NL}}(f)$. Negative (blue) regions denote
    triads for which energy is transferred from the mode at $f=f_1+f_2$ to the
    interacting modes $(f_1,f_2)$, resulting in negative peaks (dips) in
    $W^{f}_{\mathrm{NL}}(f)$. 
    For better visualization, the color scale in panel (a) is limited to the
    $99.9^{\mathrm{th}}$ percentile of $T_f(f_1,f_2)$; values exceeding this
    threshold are saturated in the colormap.}}
    \label{fig:ETF_Ritz_5.76}
\end{figure}

The accuracy of the nonlinear energy transfer estimation is assessed using the energy mismatch parameter \( W_{\mathrm{mis}} \) (shown in the legends of Fig.~\ref{fig:ETF_Kim_5.76}(b) and Fig.~\ref{fig:ETF_Ritz_5.76}(b)). For the density fluctuation \( \tilde{n} \) at the radial position \( r = 5.76~\mathrm{cm} \), the mismatch is \( W_{\mathrm{mis}} = 0.456 \approx 0.46 \) for the Kim method, whereas for the Ritz method it is \( W_{\mathrm{mis}} = -0.199 \approx -0.20 \). This implies that about 46\% of the net transfer is unbalanced for the Kim method and nearly 20\% for the Ritz method. In the Kim method, $W_{\mathrm{mis}}$ is positive, indicating that the net nonlinear transfer into the resolved modes slightly exceeds the transfer out of them. This imbalance likely arises from additional energy input from the background plasma or from couplings with other fluctuations (e.g., potential fluctuations) that are not included in the present analysis. In contrast, for the Ritz method, $W_{\mathrm{mis}}$ is negative, implying that the estimated loss marginally exceeds the corresponding gain. This suggests that part of the transferred energy is exchanged with the background plasma or with unaccounted modes outside the scope of the present analysis.

From the nonlinear energy transfer analysis using both methods at the radial location 5.76~cm, it is evident that the results differ significantly and, in some cases, even show opposite trends. \textcolor{black}{From Fig.~\ref{fig:ETF_Ritz_5.76}(a--b), it is observed that the contour plots obtained using the Ritz method are comparatively noisier than those obtained using the Kim method. Consistent with this observation, the corresponding 1D profile of $W_{NL}$ exhibits broader peaks around the 2.5~kHz and 21~kHz modes, along with several spurious oscillations that are not expected from the physical analysis.} This discrepancy can be understood from the data selection discussion in Sec.~\ref{subsection: data selection}, where the kurtosis is found to be \(0.96\) (Fig.~\ref{fig:stats_IoS}), indicating a strong deviation from Gaussian statistics. From the simulation study, it was also observed that the Ritz method becomes unreliable for kurtosis values beyond \(\sim 0.40\). Therefore, at the radial location \( r = 5.76~\mathrm{cm} \), the Ritz method is not valid and its results are not physically reliable. In contrast, the Kim method does not rely on such statistical approximations and remains applicable for non-Gaussian data. Moreover, at this radial location we observe good spatial stationarity (Fig.~\ref{fig:psd_compare}(a)), and the input and output power spectra exhibit similar spectral trends; despite the developing turbulence regime indicated by the spectra, the Kim method provides a more reliable description of nonlinear energy transfer at this radial position. In summary, at this radial location we find, the RT mode at \(10.8~\mathrm{kHz}\) transfers energy to the DW mode at \(2.5~\mathrm{kHz}\) and mix mode at  \(8.3~\mathrm{kHz}\) and also to other modes at \(10.2~\mathrm{kHz}\), and \(21~\mathrm{kHz}\). In the next section, we present the nonlinear energy transfer analysis using both the Ritz and Kim methods at a radial location close to the plasma center \((r = 2.24~\text{cm})\), where the data exhibit low kurtosis, in order to examine whether the results obtained from the Ritz method agree with those from the Kim method.

\subsubsection{Nonlinear Energy Transfer Analysis at Radial Position $r = 2.24$~cm}
\label{subsubsection: analysis at 2.24}
We analyze the nonlinear energy transfer at the radial position \( r = 2.24~\mathrm{cm} \) using both the Kim and Ritz approaches. As a first step, the spectral properties of the density fluctuations \( \tilde{n} \) are examined through the power spectrum and auto-bicoherence, presented in Fig.~\ref{fig:exp_data_2.24_PSD_Bx}, before discussing the corresponding energy transfer results.

\begin{figure}[H]
    \centering
    \includegraphics[width=1\linewidth]{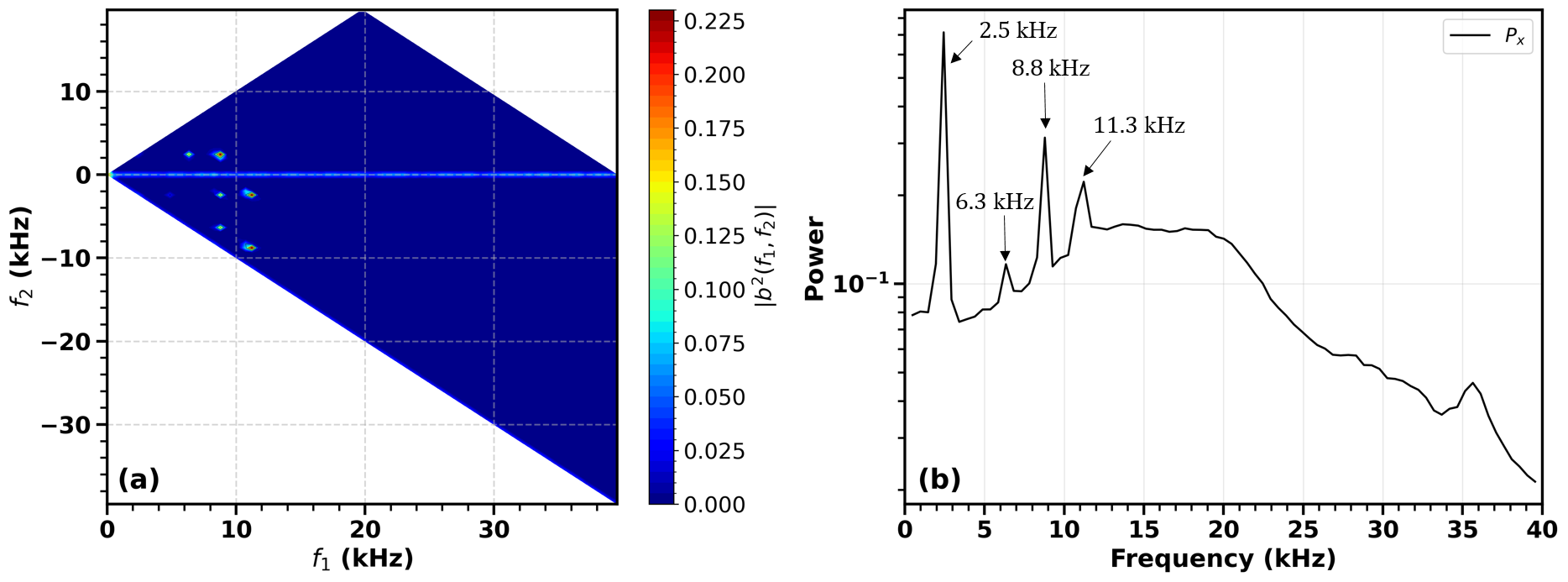}
    \caption{(a) Auto bicoherence analysis. (b) Density fluctuation spectra for $\tilde{n}$ of DW and RT modes at $r = 2.24$ cm.}
    \label{fig:exp_data_2.24_PSD_Bx}
\end{figure}

The spectral resolution and data length kept same as in case of the radial position 5.76~cm. From auto-power spectrum plot Fig.~\ref{fig:exp_data_2.24_PSD_Bx}(b), three dominant frequency modes are observed those are listed in the Table--\ref{tab:2.24}. The instability types, radial and poloidal wave numbers
corresponding to these modes are also listed in the Table -\ref{tab:2.24}. From Fig.~\ref{fig:exp_data_2.24_PSD_Bx}(a), it is evident that the frequency components at
\( f_1 = 8.8~\text{kHz} \), \( f_2 = 2.5~\text{kHz} \) and \( f = 11.3~\text{kHz} \) are nonlinearly interact as indicated in the positive bicoherence region. This interaction satisfies the frequency resonance conditions \( f_1 + f_2 = f \). In the negative region, additional interactions consistent with the relations \( f - f_1 = f_2 \) and \( f - f_2 = f_1 \) are also observed. Similarly, we observe another relatively weak interaction: \( f_1 = 6.3~\text{kHz} \), \( f_2 = 2.5~\text{kHz} \) and \( f = 8.8~\text{kHz} \) nonlinearly interact, satisfying the resonance condition \( f_1 + f_2 = f \), as evident in the positive bicoherence region. Similarly, additional interactions are observed in the negative region as discussed earlier. The observed finite bicoherence confirms coherent phase coupling among these modes, with all identified interactions showing bicoherence values above the statistical significance level, 
$b_{\mathrm{sig}}^{2} = \frac{1}{4882} \approx 2 \times 10^{-4}$. However, the wavenumber resonance condition between the poloidal wavenumbers $(k_{\theta})$ (see Table--\ref{tab:2.24}) is not exactly satisfied, which may be attributed to relatively weak nonlinear three-wave coupling (indicated by low bicoherence magnitude) at this radial location and uncertainties in wavenumber estimation. Since the wave propagation is predominantly poloidal, the resonance condition is evaluated using only the poloidal wavenumber component. Small deviations from exact matching can arise due to the finite spectral width of turbulent fluctuations, plasma inhomogeneity, and nonlinear or collisional resonance broadening. Therefore, the nonlinear coupling condition is considered satisfied within experimental uncertainty.

\begin{table}[H]
\centering
\setlength{\abovecaptionskip}{0pt}
\caption{Measured mode frequencies with corresponding poloidal ($k_{\theta}$) and radial ($k_r$) wavenumbers(absolute values), and identified instability type.}
\label{tab:2.24}
\vspace{1mm}
\begin{tabular}{cccc}
\toprule
Frequency (kHz) & $k_{\theta}$ (cm$^{-1}$) & Mode Type \\
\midrule
2.5  & 0.95 & Drift Wave (DW) \\
8.8 & 1.94 & \textcolor{black}{ Mix mode of DW and RT} \\
11.3 & 3.15 & Rayleigh--Taylor (RT) \\
\bottomrule
\end{tabular}
\end{table}

We now determine the direction of energy transfer among these frequency components using nonlinear energy transfer analysis based on both the Kim and Ritz methods, as discussed below.

\paragraph{I. Analysis using Kim Method:}
Figure~\ref{fig:ETF_Kim_2.24}(a--b) presents the power transfer function \( T_f(f_1,f_2) \) and the associated nonlinear energy transfer function \( W^{f}_{\mathrm{NL}} \), evaluated from two density fluctuation signals separated by 6.4~mm in the poloidal direction at the radial location \( r = 2.24 \)~cm. The computation uses the same frequency resolution as adopted in the bicoherence analysis.

\begin{figure}[H]
    \centering
    \includegraphics[width=1\linewidth]{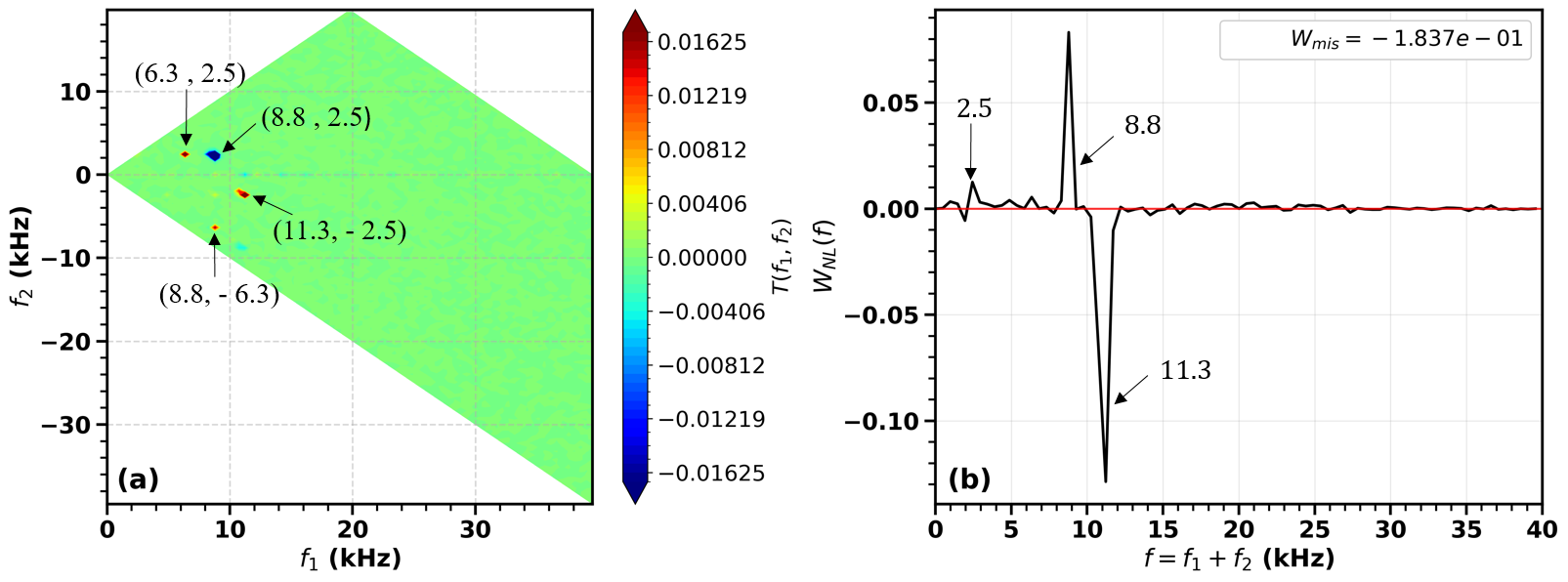}
    \caption{\textcolor{black}{Results obtained using the Kim method at $r = 2.24~cm$: (a) Power transfer function $T_f(f_1,f_2)$.
    (b) Energy transfer function $W^{f}_{\mathrm{NL}}(f)$ for $\tilde{n}$
    fluctuations associated with DW and RT modes. The coordinates
    $(f_1,f_2)$ (in kHz) corresponding to the dominant nonlinear interaction regions are marked in panel (a), while the corresponding frequencies
    $f=f_1+f_2$ are indicated in panel (b). Positive (red) regions in
    $T_f(f_1,f_2)$ denote triads for which energy is transferred from the
    interacting modes $(f_1,f_2)$ to the mode at $f=f_1+f_2$, producing
    positive peaks in $W^{f}_{\mathrm{NL}}(f)$. Negative (blue) regions denote
    triads for which energy is transferred from the mode at $f=f_1+f_2$ to the
    interacting modes $(f_1,f_2)$, resulting in negative peaks (dips) in
    $W^{f}_{\mathrm{NL}}(f)$. 
    For better visualization, the color scale in panel (a) is limited to the
    $99.9^{\mathrm{th}}$ percentile of $T_f(f_1,f_2)$; values exceeding this
    threshold are saturated in the colormap.}}
    \label{fig:ETF_Kim_2.24}
\end{figure}

From the two-dimensional contour plot of \( T_f(f_1,f_2) \) shown in Fig.~\ref{fig:ETF_Kim_2.24}(a), it is observed that the Rayleigh--Taylor (RT) mode at 11.3~kHz loses energy through nonlinear coupling to the mix mode at 8.8~kHz and the drift-wave (DW) mode at 2.5~kHz, satisfying the interaction \((f = 11.3~\text{kHz}) \rightarrow (f_1 = 8.8~\text{kHz}) + (f_2 = 2.5~\text{kHz})\). Another relatively weak interaction is also evident, \((f_1 = 6.3~\text{kHz}) + (f_2 = 2.5~\text{kHz}) \rightarrow (f = 8.8~\text{kHz})\), indicating that the mix mode at 8.8~kHz gains energy through nonlinear coupling with the 6.3~kHz mode and DW mode at 2.5~kHz. \textcolor{black}{Another triad interaction \((f_1 = 8.8~\text{kHz}) + (f_2 = -6.3~\text{kHz}) \rightarrow (f = 2.5~\text{kHz})\) shows that the DW mode at 2.5~kHz gains energy from the modes at 6.3~kHz and 8.8~kHz.} The one-dimensional spectrum \( W_{\mathrm{NL}}^{f} \) in Fig.~\ref{fig:ETF_Kim_2.24}(b) supports these observations, showing a negative peak near 11.3~kHz and prominent positive peaks near 8.8~kHz and 2.5~kHz. The weak positive peak at 2.5~kHz arises because this mode gains energy in one interaction and loses energy in another, leading to a small net gain. \textcolor{black}{In the case of the 8.8~kHz mode, energy is gained through two nonlinear interactions, while energy is lost through only a single interaction, as discussed earlier. Consequently, this mode exhibits a prominent net energy gain, which appears as a pronounced positive peak in the $W_{NL}^f$ spectrum.} The results indicate a net nonlinear energy transfer from the 11.3~kHz RT mode toward the 8.8~kHz mix mode and the 2.5~kHz DW mode, while the faint contribution near 6.3~kHz reflects comparatively weak nonlinear coupling, as evidenced by the low bicoherence associated with this mode. Other peaks in \( W_{\mathrm{NL}}^{f} \) correspond to minor interactions. Now, we analyse the same data with Ritz method which is presented in the next section.

\paragraph{II. Analysis using Ritz Method:}
Here, the nonlinear energy transfer analysis of the same density fluctuation data at the radial location \( r = 2.24~\mathrm{cm} \) is performed using the Ritz method. The two-dimensional contour plot of \( T_f(f_1,f_2) \) shown in Fig.~\ref{fig:ETF_Ritz_2.24}(a) is qualitatively identical to that obtained using the Kim method in Fig.~\ref{fig:ETF_Kim_2.24}(a), indicating consistent nonlinear interactions. The dominant coupling processes are the same as those discussed for the Kim method. The one-dimensional spectrum \( W_{\mathrm{NL}}^{f} \) in Fig.~\ref{fig:ETF_Ritz_2.24}(b) further supports these observations and shows the same trend as in the Kim method, leading to the same conclusion: a net nonlinear energy transfer from the 11.3~kHz RT mode to the 8.8~kHz mix mode and the 2.5~kHz DW mode. The faint contribution near 6.3~kHz indicates relatively small coupling as evidenced by the low bicoherence associated with this mode, while the remaining peaks in \( W_{\mathrm{NL}}^{f} \) correspond to minor interactions.

\begin{figure}[htbp]
    \centering
    \includegraphics[width=1\linewidth]{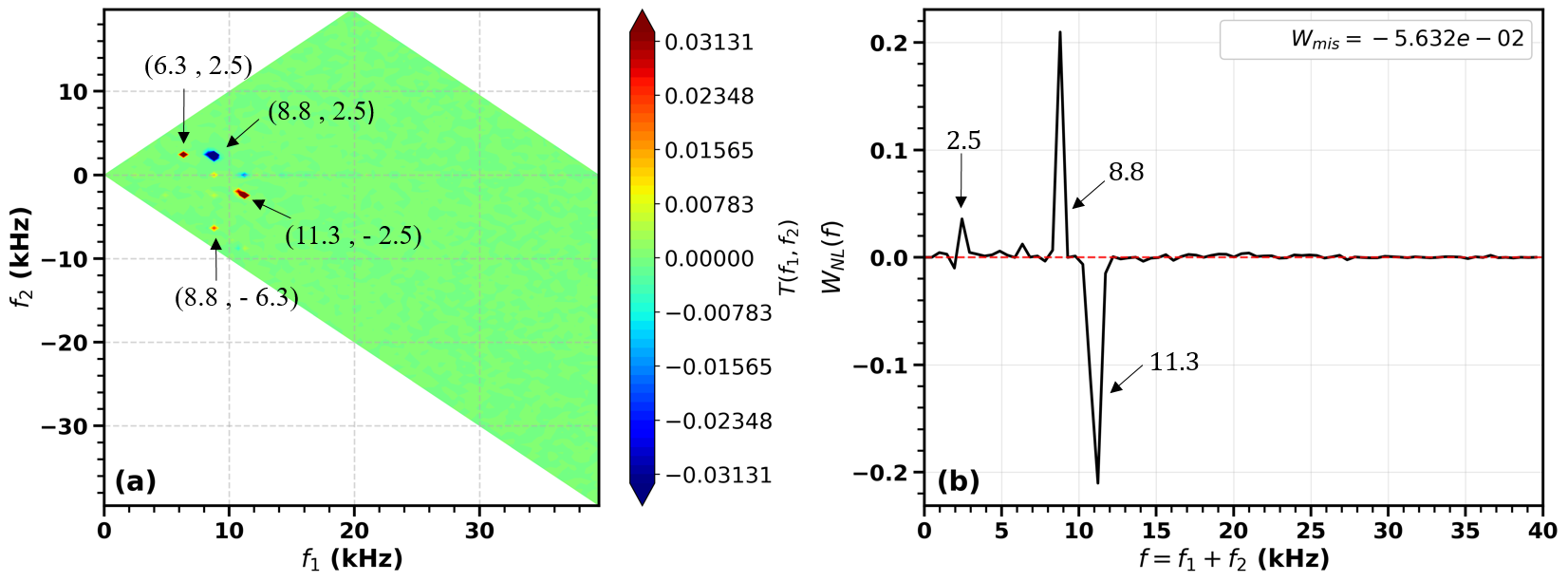}
    \caption{\textcolor{black}{Results obtained using the Ritz method at $r = 2.24~cm$: (a) Power transfer function $T_f(f_1,f_2)$.
    (b) Energy transfer function $W^{f}_{\mathrm{NL}}(f)$ for $\tilde{n}$
    fluctuations associated with DW and RT modes. The coordinates
    $(f_1,f_2)$ (in kHz) corresponding to the dominant nonlinear interaction regions are marked in panel (a), while the corresponding frequencies
    $f=f_1+f_2$ are indicated in panel (b). Positive (red) regions in
    $T_f(f_1,f_2)$ denote triads for which energy is transferred from the
    interacting modes $(f_1,f_2)$ to the mode at $f=f_1+f_2$, producing
    positive peaks in $W^{f}_{\mathrm{NL}}(f)$. Negative (blue) regions denote
    triads for which energy is transferred from the mode at $f=f_1+f_2$ to the
    interacting modes $(f_1,f_2)$, resulting in negative peaks (dips) in
    $W^{f}_{\mathrm{NL}}(f)$. 
    For better visualization, the color scale in panel (a) is limited to the
    $99.9^{\mathrm{th}}$ percentile of $T_f(f_1,f_2)$; values exceeding this
    threshold are saturated in the colormap.}}
    \label{fig:ETF_Ritz_2.24}
\end{figure}

The accuracy of the nonlinear energy transfer estimation is evaluated using the energy mismatch parameter \( W_{\mathrm{mis}} \), shown in the legends of Fig.~\ref{fig:ETF_Kim_2.24}(b) and Fig.~\ref{fig:ETF_Ritz_2.24}(b). For the density fluctuation \( \tilde{n} \) at \( r = 2.24~\mathrm{cm} \), the mismatch is \( W_{\mathrm{mis}} = -0.1837 \approx -0.18 \) for the Kim method and \( W_{\mathrm{mis}} = -0.056 \approx -0.06 \) for the Ritz method. This indicates that about 18\% and 6\% of the net transfer remain unbalanced for the Kim and Ritz methods, respectively. In both cases, the negative sign of \( W_{\mathrm{mis}} \) implies that the estimated loss slightly exceeds the corresponding gain, suggesting that part of the transferred energy is exchanged with background plasma or other unaccounted modes not included in the present analysis.

From the nonlinear energy transfer analysis at the radial location \( r = 2.24~\mathrm{cm} \) using both the Kim and Ritz methods, it is evident that the results obtained from the two approaches agree closely. In this region, the Ritz method performs reliably because the kurtosis of the data is low (\(  \approx 0.09 \)), which is well below the critical limit (\(\ \approx 0.40 \)) identified from the simulation study for the validity of the Ritz method. This good agreement with the Kim method indicates that the Ritz method can be safely applied in regions characterized by low-kurtosis statistics. However, from Fig.~\ref{fig:psd_compare}(b) we find that the spatial stationarity at this location is poor, although the input and output power spectra exhibit similar spectral shapes, which supports the validity of the Kim method at this location. \textcolor{black}{We also find that at this location, the Ritz method yields a smaller value of the mismatch parameter $W_{mis}$ than the Kim method, indicating a smaller residual energy imbalance in the estimated nonlinear energy budget. It should be noted that $W_{mis}$ is not used as a direct criterion for method applicability, since its value can be influenced by factors such as finite frequency-band truncation, transfer to unmeasured fields, transfer to mean profiles, and statistical uncertainties.} In summary, at this radial location the dominant interaction shows that the RT mode at \(11.3~\mathrm{kHz}\) transfers energy to the DW mode at \(2.5~\mathrm{kHz}\) as well as to the mix mode at \(8.8~\mathrm{kHz}\). In the next section, we present the growth-rate analysis at the radial location \(r = 2.24~\mathrm{cm}\), where both the Kim and Ritz methods perform reliably.

\subsubsection{Discussion on Growth Rate}
\label{subsubsection: growth rate expt}
In the previous section, we identified which modes transfer energy to other modes. In this section, we present a growth-rate analysis, which allows us to examine the growth or damping of individual modes and how these behaviors are connected to the energy-transfer processes. Specifically, we analyze the growth rates at the radial position \( r = 2.24~\mathrm{cm} \) (since at this location both methods are applicable), can be shown as shown in Fig.~\ref{fig:growth_rate_spectrum}.

\begin{figure}[htbp]
    \centering
    \includegraphics[width=0.75\textwidth]{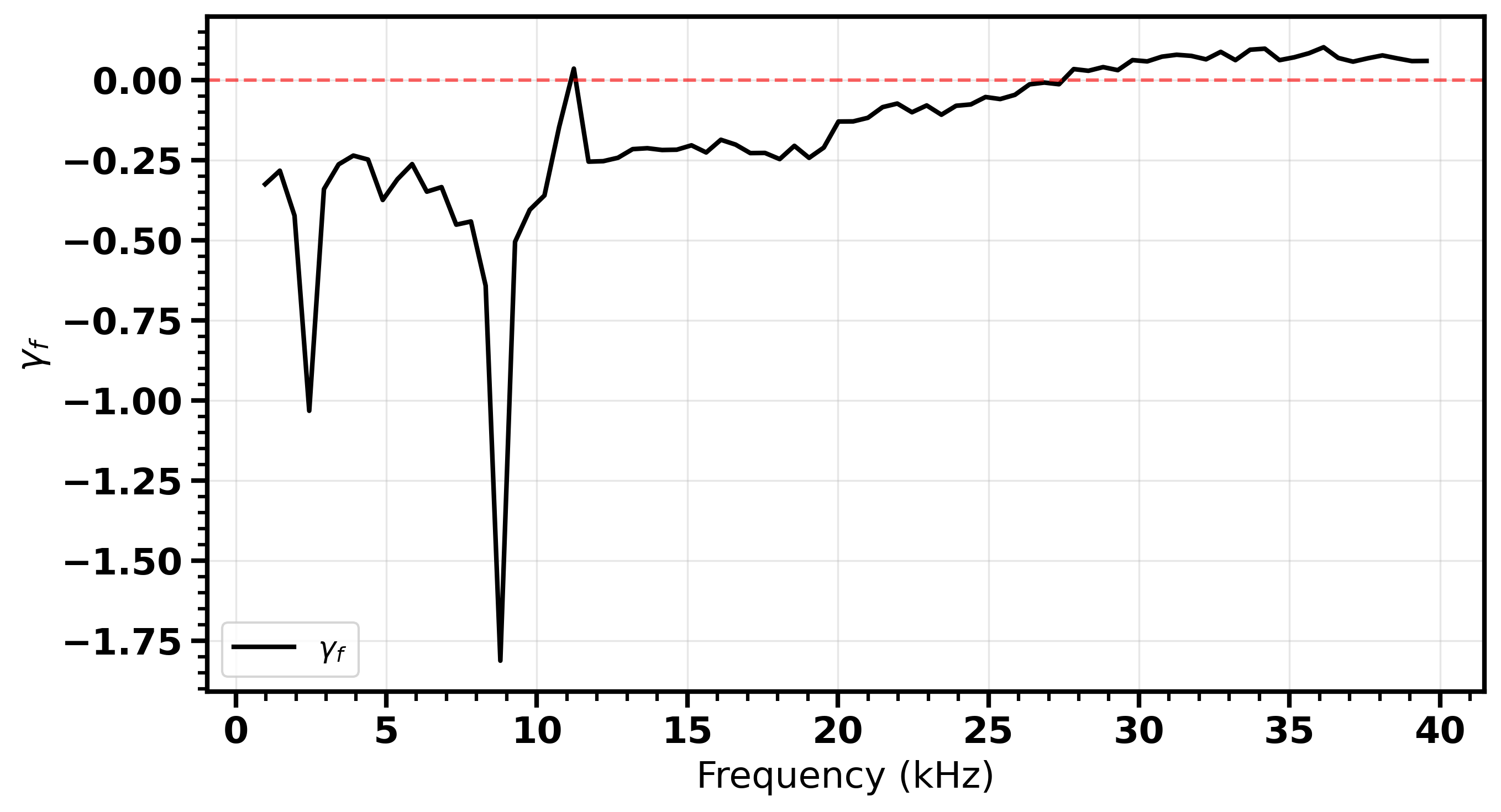}
    \caption{Growth rate spectrum $\gamma_f$ as a function of frequency. The dashed line indicates $\gamma_f = 0$.}
    \label{fig:growth_rate_spectrum}
\end{figure}
Here, we provide a qualitative discussion of the growth rate plot rather than a quantitative analysis. This approach allows us to focus on the overall trends, such as which modes are growing or damping and how their behavior correlates with the energy transfer between modes, without delving into exact numerical values or rates. From Fig.~\ref{fig:growth_rate_spectrum}, the mode near \( f = 10.8~\mathrm{kHz} \) has a small positive growth rate and is therefore unstable, whereas the modes near \( f = 2.5~\mathrm{kHz} \) and \( f = 8.3~\mathrm{kHz} \) are strongly damped. This implies that the \(10.8~\mathrm{kHz}\) mode grows linearly and the energy gained by this mode is redistributed to other modes or to the background equilibrium due to nonlinear interaction, while the lower-frequency modes are damped and receive energy through nonlinear interactions following the spatial stationarity condition Eq.~\eqref{eq:stationarity2}. The energy transfer analysis reveals the nonlinear redistribution of energy, demonstrating that the \(10.8~\mathrm{kHz}\) mode acts as a donor, transferring energy to the \(8.3~\mathrm{kHz}\) and \(2.5~\mathrm{kHz}\) modes which is clearly reflected in the energy transfer function shown in Fig.~\ref{fig:ETF_Kim_2.24}. Thus, although the growth-rate spectrum identifies which modes are linearly driven or damped, the direction of spectral energy flow is governed by nonlinear interactions captured by the energy transfer analysis. Furthermore, the positive growth rates in the higher-frequency band indicate that these modes are unstable, and the energy contained in these frequencies is transported to lower-frequency bands or to the background equilibrium, consistent with the damping observed in the low-frequency modes. In addition, we evaluate the energy imbalance parameter \(\Delta_{\mathrm{im}}\) from Eq.~\eqref{delta} at this location and obtain a value of \(0.06\). Although small, this finite value indicates a developing turbulent state. Nevertheless, the small magnitude of \(\Delta_{\mathrm{im}}\) suggests that Kim method remains valid and applicable at this position.

\section{Discussion}
\label{section: discussion}
The behavior of nonlinear interactions among various instability modes, namely Rayleigh--Taylor and drift modes, observed in the measured plasma density fluctuations in the Inverse Mirror Plasma Experimental Device (IMPED), are investigated in detail. Prior to applying the methods to experimental data, the computational implementations of both the Ritz and Kim methods are benchmarked against analytical transfer functions obtained from a developing plasma turbulence model. The existing computational methods are further explored with suitable improvisations to overcome their inherent limitations and constraints. It is observed that the Ritz method performs well for fluctuations that are close to Gaussian, with low skewness and kurtosis. However, for cases with large kurtosis, the method deteriorates because its formulation approximates fourth-order moments, thereby introducing errors in the computation of the energy transfer functions. In contrast, Kim’s method performs much better for fluctuation data with larger kurtosis and yields consistent estimates of the energy transfer functions, because the method does not rely on the \textit{Millionshchikov approximation} and instead retains the fourth-order moments.

Both Ritz and Kim methods are tested to compute the energy transfer among the modes observed in the experimental fluctuation data of IMPED. The applicability of these methods with respect to spatial stationarity is also examined. We find that Kim's method yields reliable estimates of the energy transfer function even under weak spatial stationarity in developing turbulence, provided the spectral shapes of the power spectra are identical. These trends are clearly observed in the experimental data at the two radial locations, \(r = 5.76~\mathrm{cm}\) and \(r = 2.24~\mathrm{cm}\). In this work, we observe that the Kim method performs well at both radial locations, whereas the Ritz method yields reliable results only at \(r = 2.24~\mathrm{cm}\). This behavior can be explained by the statistical nature of the data and the degree of spatial stationarity. \textcolor{black}{In particular, the mismatch parameter \(W_{mis}\) for the Ritz method is relatively small at \(r = 2.24~\mathrm{cm}\) (about \(0.06\)), indicating a relatively small residual energy mismatch at this location. The applicability of the Ritz method is assessed primarily from the statistical properties of the fluctuations, particularly the validity of the Millionshchikov approximation and the low excess kurtosis observed at this radial position.}

\section{Conclusion}
\label{section: conclusion}
The nonlinear energy transfer among drift-wave and Rayleigh--Taylor modes in IMPED plasma has been investigated using the Ritz and Kim methods. Benchmarking against a developing plasma turbulence model showed that the Ritz method provides reliable estimates for near-Gaussian fluctuations, whereas its performance deteriorates with increasing kurtosis. In contrast, the Kim method remains robust for both Gaussian and non-Gaussian fluctuations. Consistent trends are observed in the experimental data. The Kim method yields reliable estimates at both \(r = 5.76~\mathrm{cm}\) and \(r = 2.24~\mathrm{cm}\), whereas the Ritz method performs reliably only at \(r = 2.24~\mathrm{cm}\). These results highlight the importance of fluctuation statistics, particularly excess kurtosis, in assessing the suitability of energy-transfer estimation methods. The energy transfer analysis further reveals that, at \(r = 5.76~\mathrm{cm}\), the RT mode at \(10.8~\mathrm{kHz}\) and the mixed mode at \(8.3~\mathrm{kHz}\) primarily transfer energy to the DW mode at \(2.5~\mathrm{kHz}\). In contrast, at \(r = 2.24~\mathrm{cm}\), the RT mode at \(11.3~\mathrm{kHz}\) transfers energy to both the DW mode at \(2.5~\mathrm{kHz}\) and the mixed mode at \(8.8~\mathrm{kHz}\). Thus, the analysis indicates a net transfer of energy from RT-dominated fluctuations toward a comparatively lower-frequency DW mode at both radial locations. \textcolor{black}{However, since the nonlinear interaction is significantly stronger at \(r = 5.76~\mathrm{cm}\), the overall direction of energy transfer is inferred primarily from the results at this radial location. Therefore, the dominant nonlinear energy transfer process can be interpreted as the transfer of energy from an RT mode and a mixed DW--RT mode to a comparatively low-frequency DW mode.}

\section{Outlook}
\label{section: outlook}
In the present study, the energy transfer analysis is restricted to a single-field framework, where the transfer functions are estimated using the Kim and Ritz methods. Although single-field analysis provides useful insight into nonlinear energy transfer within an individual fluctuating quantity, it cannot capture the full complexity of plasma turbulence, in which multiple coupled fields such as density, potential, temperature, and velocity interact nonlinearly \cite{Xu2009PoP,Xu2010PoP}. In drift-wave turbulence experiments, where the density–potential cross-correlation is well below unity, in such systems, key processes such as drift-wave–zonal-flow coupling, the exchange between internal and kinetic energy channels cannot be fully described within a single-field formulation\cite{Xu2009PoP}. A multi-field turbulence approach is therefore required to quantify nonlinear coupling and energy transfer among different fluctuating fields, as demonstrated in experimental studies and theoretical treatments of plasma turbulence \cite{Xu2009PoP,Manz2008PPCF,Itoh2017PFR}. Our future work will involve extending the current methodology to a multi-field framework, building upon the validated single-field results achieved here through the Kim and Ritz methods.

\section*{Acknowledgments}
The authors acknowledge Jignesh Patel, Kalpesh Doshi, Kirti Mahajan, Manisha Bhandankar, Praveenlal, Minsha Shah, and Vijay Patel for their experimental support. 

\section*{Declaration of Interest}

The authors declare that they have no conflict of interest.

\section*{References}
\begingroup
\renewcommand{\section}[2]{}%
\bibliographystyle{unsrt}   
\bibliography{refs}

@article{smith1974fft,
  author       = {Smith, D. E. and Powers, E. J. and Caldwell, G. S.},
  title        = {Fast-Fourier-Transform Spectral-Analysis Techniques as a Plasma Fluctuation Diagnostic Tool},
  journal      = {IEEE Transactions on Plasma Science},
  volume       = {PS-2},
  pages        = {261--272},
  year         = {1974},
  publisher    = {IEEE}
}

@article{smith1974nonlinear,
  author       = {Smith, D. E. and Powers, E. J. and Caldwell, G. S.},
  title        = {Nonlinear Three-Wave Interactions in Plasma Turbulence},
  journal      = {IEEE Transactions on Plasma Science},
  volume       = {PS-2},
  pages        = {261--267},
  year         = {1974},
  publisher    = {IEEE}
}

@book{Bendat2010RandomData,
  author    = {Julius S. Bendat and Allan G. Piersol},
  title     = {Random Data: Analysis and Measurement Procedures},
  edition   = {4th},
  publisher = {Wiley},
  year      = {2010}
}

@book{Spiegel2018Statistics,
  author    = {Murray R. Spiegel and Larry J. Stephens},
  title     = {Schaum's Outline of Statistics},
  edition   = {6th},
  publisher = {McGraw-Hill Education},
  year      = {2018},
  chapter   = {Moments, Skewness, and Kurtosis}
}

@article{smith1973power,
  author  = {Smith, D. E. and Powers, E. J.},
  title   = {Experimental Investigation of Nonlinear Wave Interactions in a Turbulent Plasma},
  journal = {Physics of Fluids},
  volume  = {16},
  number  = {},
  pages   = {1373--1380},
  year    = {1973},
}

@article{elgar1985,
  author  = {Elgar, S. and Guza, R. T.},
  title   = {Observations of bispectra of ocean surface gravity waves},
  journal = {Journal of Fluid Mechanics},
  volume  = {161},
  pages   = {425--448},
  year    = {1985}
}

@article{miksad1983,
  author  = {Miksad, R. W. and Jones, F. L. and Powers, E. J.},
  title   = {Energy transfer in drift-wave instabilities},
  journal = {Physics of Fluids},
  volume  = {26},
  pages   = {1402--1410},
  year    = {1983}
}

@article{kim1979,
  author  = {Young C. Kim and Edward J. Powers},
  title   = {Digital bispectral analysis and its applications to nonlinear wave interactions},
  journal = {IEEE Trans.\ Plasma Sci.},
  volume  = {PS-7},
  number  = {2},
  pages   = {120--131},
  year    = {1979},
  doi     = {10.1109/TPS.1979.4317207}
}

@article{kim1987ocean,
  author  = {Kim, K. I. and Powers, E. J. and Ritz, C. P. and Miksad, R. W. and Fischer, F. J.},
  title   = {Bispectral analysis of nonlinear drift oscillations of moored vessels},
  journal = {IEEE Journal of Oceanic Engineering},
  volume  = {OE-12},
  number  = {4},
  pages   = {568--575},
  year    = {1987},
  doi     = {10.1109/JOE.1987.1145157}
}

@article{kim1980bispectrum,
  author  = {Kim, Y. C. and Beall, J. M. and Powers, E. J. and Miksad, R. W.},
  title   = {Bispectrum and nonlinear wave coupling},
  journal = {Physics of Fluids},
  volume  = {23},
  number  = {},
  pages   = {258--263},
  year    = {1980},
  doi     = {10.1063/1.862966}
}

@article{horton1999,
  author  = {Horton, W.},
  title   = {Drift waves and transport in magnetized plasmas},
  journal = {Reviews of Modern Physics},
  volume  = {71},
  number  = {3},
  pages   = {735--778},
  year    = {1999},
  doi     = {10.1103/RevModPhys.71.735}
}

@article{Melnikov2017,
  author    = {A. Melnikov and L.G. Eliseev and S.I. Lysenko and M.V. Ufimtsev and V.N. Zenin},
  title     = {Study of interactions between GAMs and broadband turbulence in the T-10 tokamak},
  journal   = {Nuclear Fusion},
  volume    = {57},
  number    = {11},
  pages     = {115001},
  year      = {2017},
  doi       = {10.1088/1741-4326/aa796c}
}

@article{bose2015inverse,
  title={Inverse mirror plasma experimental device (IMPED)--a magnetized linear plasma device for wave studies},
  author={Bose, Sayak and Chattopadhyay, PK and Ghosh, J and Sengupta, S and Saxena, YC and Pal, R},
  journal={Journal of Plasma Physics},
  volume={81},
  number={2},
  pages={345810203},
  year={2015},
  publisher={Cambridge University Press}
}

@article{sayak2015inverse,
  title={Inverse mirror plasma experimental device—A new magnetized linear plasma device with wide operating range},
  author={Bose, Sayak and Kaur, Manjit and Chattopadhyay, PK and Ghosh, J and Saxena, YC},
  journal={Review of Scientific Instruments},
  volume={86},
  number={6},
  year={2015},
  publisher={AIP Publishing}
}

@article{Rosh2025experimental,
  title={An experimental study of the Rayleigh--Taylor and Kelvin--Helmholtz instabilities, and their nonlinear interactions in a linear magnetized plasma column},
  author={Roy, Rosh and Karmakar, Tanmay and Lachhvani, Lavkesh and Khodiyar, Bhoomi and Chattopadhyay, Prabal and Sen, Abhijit},
  journal={Physics of Plasmas},
  volume={32},
  number={12},
  year={2025},
  publisher={AIP Publishing}
}

@article{burin2005transition,
  title={On the transition to drift turbulence in a magnetized plasma column},
  author={Burin, MJ and Tynan, GR and Antar, GY and Crocker, NA and Holland, C},
  journal={Physics of Plasmas},
  volume={12},
  number={5},
  year={2005},
  publisher={AIP Publishing}
}

@article{roy2025experimental,
  title={An experimental study of the existence regions and non-linear interactions of drift wave and Kelvin--Helmholtz instabilities in a linear magnetized plasma},
  author={Roy, Rosh and Karmakar, Tanmay and Lachhvani, Lavkesh and Chattopadhyay, Prabal and Sen, Abhijit and Bose, Sayak},
  journal={Physics of Plasmas},
  volume={32},
  number={2},
  year={2025},
  publisher={AIP Publishing}
}

@article{ritz1989,
  author    = {C. M. Ritz and E. J. Powers and R. D. Bengtson},
  title     = {Experimental measurement of nonlinear energy transfer in a turbulent plasma},
  journal   = {Phys. Fluids B: Plasma Phys.},
  volume    = {1},
  number    = {1},
  pages     = {153--159},
  year      = {1989},
  doi       = {10.1063/1.859187}
}

@article{Millionshchikov1941,
  author       = {Millionshchikov, M. D.},
  title        = {On the Theory of Homogeneous Isotropic Turbulence},
  journal      = {Doklady Akademii Nauk SSSR},
  volume       = {32},
  year         = {1941},
  pages        = {611--614},
  language     = {Russian}
}

@article{ritz1986,
  author    = {Ch. P. Ritz and E. J. Powers},
  title     = {Estimation of nonlinear transfer functions for fully developed turbulence},
  journal   = {Physica D: Nonlinear Phenomena},
  volume    = {20},
  number    = {2-3},
  pages     = {320--334},
  year      = {1986},
  publisher = {North-Holland},
  doi       = {10.1016/0167-2789(86)90036-9}
}

@article{Ritz1988,
  author  = {Ritz, Ch. P. and Powers, E. J. and Miksad, R. W. and Solis, R. S.},
  title   = {Nonlinear spectral dynamics of a transitioning flow},
  journal = {Physics of Fluids},
  volume  = {31},
  number  = {12},
  pages   = {3577--3588},
  year    = {1988},
  doi     = {10.1063/1.866875}
}

@techreport{TerryDiamondShaing1986,
  author      = {Terry, P. W. and Diamond, P. H. and Shaing, K. C. and Garcia, L. and Carreras, B. A.},
  title       = {The Spectrum of Resistivity Gradient Driven Turbulence},
  institution = {Institute for Fusion Studies},
  number      = {IFSR-174},
  address     = {Austin, Texas},
  year        = {1986}
}

@article{anderson2020elucidating,
  title={Elucidating plasma dynamics in Hasegawa-Wakatani turbulence by information geometry},
  author={Anderson, J. and Kim, E-j and Hnat, B. and Rafiq, T.},
  journal={Physics of Plasmas},
  volume={27},
  number={2},
  pages={022307},
  year={2020},
  publisher={AIP Publishing LLC},
  doi={10.1063/1.5122865}
}

@article{hasegawa1983plasma,
  title={Plasma edge turbulence},
  author={Hasegawa, Akira and Wakatani, Masahiro},
  journal={Physical Review Letters},
  volume={50},
  number={9},
  pages={682},
  year={1983},
  publisher={APS},
  doi={10.1103/PhysRevLett.50.682}
}

@book{krall1973principles,
  title={Principles of Plasma Physics},
  author={Krall, Nicholas A. and Trivelpiece, Alvin W.},
  year={1973},
  publisher={McGraw-Hill},
  address={New York},
  isbn={978-0070353466}
}

@article{TerryDiamond1985,
  author  = {Terry, P. W. and Diamond, P. H.},
  title   = {Resistivity-Gradient-Driven Turbulence},
  journal = {Physics of Fluids},
  volume  = {28},
  year    = {1985},
  pages   = {1419--1430}
}

@book{Nikias1993,
  author    = {C. L. Nikias and A. P. Petropulu},
  title     = {Higher-Order Spectra Analysis: A Nonlinear Signal Processing Framework},
  publisher = {Prentice Hall},
  year      = {1993},
  address   = {Englewood Cliffs, New Jersey}
}

@article{kim1996,
  author    = {J. S. Kim and R. D. Durst and R. J. Fonck and E. Fernandez and A. Ware and P. W. Terry},
  title     = {Technique for the experimental estimation of nonlinear energy transfer in fully developed turbulence},
  journal   = {Physics of Plasmas},
  volume    = {3},
  number    = {11},
  pages     = {3998--4009},
  year      = {1996},
  publisher = {American Institute of Physics},
  doi       = {10.1063/1.871572}
}

@article{hasegawa1978,
  title={Pseudo-three-dimensional turbulence in magnetized nonuniform plasma},
  author={Hasegawa, Akira and Mima, Kunioki},
  journal={Physical Fluids},
  volume={21},
  number={1},
  pages={87--92},
  year={1978},
  publisher={AIP Publishing}
}

@article{SmithPowersCaldwell1974,
  author  = {Smith, D. E. and Powers, E. J. and Caldwell, G. S.},
  title   = {Fast-Fourier-Transform Spectral-Analysis Techniques as a Plasma Fluctuation Diagnostic Tool},
  journal = {IEEE Transactions on Plasma Science},
  volume  = {PS-2},
  number  = {},
  pages   = {261--270},
  year    = {1974},
  month   = {December},
  received= {June 24, 1974}
}

@article{FreilichGuza1984,
  author  = {Freilich, M. H. and Guza, R. T.},
  title   = {Nonlinear wave interactions in ocean surface gravity waves},
  journal = {Philosophical Transactions of the Royal Society of London. Series A},
  volume  = {311},
  pages   = {1--26},
  year    = {1984}
}

@article{HasegawaMaclennanKodama1979,
  author  = {Hasegawa, A. and Maclennan, C. G. and Kodama, Y.},
  title   = {Nonlinear behavior and turbulence spectra of drift waves},
  journal = {Physics of Fluids},
  volume  = {22},
  pages   = {2122--2129},
  year    = {1979}
}

@article{xia2004,
  author    = {H. Xia and M. G. Shats},
  title     = {Spectral energy transfer and generation of turbulent structures in toroidal plasma},
  journal   = {Physics of Plasmas},
  volume    = {11},
  number    = {2},
  pages     = {561--571},
  year      = {2004},
  doi       = {10.1063/1.1637607},
  publisher = {American Institute of Physics}
}

@article{Shen2020ImprovedBispectrum,
  title   = {An improved method for bispectral analysis of nonlinear wave interaction system},
  author  = {Shen, Yu-Hang and Li, Jia-Yin and Li, Tian and Li, Jia},
  journal = {Physica Scripta},
  volume  = {95},
  number  = {5},
  pages   = {055202},
  year    = {2020},
  doi     = {10.1088/1402-4896/ab725f},
  publisher = {IOP Publishing}
}

@article{Xu2009Pop,
  author       = {M. Xu and G. R. Tynan and C. Holland and Z. Yan and S. H. Muller and J. H. Yu},
  title        = {Study of nonlinear spectral energy transfer in frequency domain},
  journal      = {Physics of Plasmas},
  volume       = {16},
  number       = {4},
  pages        = {042312},
  year         = {2009},
  publisher    = {American Institute of Physics},
  doi          = {10.1063/1.3098538}
}

@article{Xu2010PoP,
  author  = {Xu, M. and Tynan, G. R. and Holland, C. and Yan, Z. and Muller, S. H. and Yu, J. H.},
  title   = {Fourier-domain study of drift turbulence driven sheared flow in a laboratory plasma},
  journal = {Physics of Plasmas},
  volume  = {17},
  number  = {3},
  pages   = {032311},
  year    = {2010},
  doi     = {10.1063/1.3325397}
}

@article{Manz2008PPCF,
  author = {P. Manz and M. Ramisch and U. Stroth and V. Naulin and B. D. Scott},
  title = {Bispectral experimental estimation of the nonlinear energy transfer in two-dimensional plasma turbulence},
  journal = {Plasma Physics and Controlled Fusion},
  volume = {50},
  number = {3},
  pages = {035008},
  year = {2008},
  doi = {10.1088/0741-3335/50/3/035008}
}

@article{Itoh2017PFR,
  author = {S.-I. Itoh and K. Itoh and Y. Nagashima and Y. Kosuga},
  title = {On the Application of Cross Bispectrum and Cross Bicoherence},
  journal = {Plasma and Fusion Research},
  volume = {12},
  pages = {1101003},
  year = {2017},
  doi = {10.1585/pfr.12.1101003}
}

@article{Taylor1938,
  author    = {Taylor, G. I.},
  title     = {The Spectrum of Turbulence},
  journal   = {Proceedings of the Royal Society A: Mathematical, Physical and Engineering Sciences},
  year      = {1938},
  volume     = {164},
  number     = {919},
  pages      = {476--490},
  doi        = {10.1098/rspa.1938.0032}
}

@article{Howes2014,
  author    = {Howes, G. G. and Klein, K. G. and TenBarge, J. M.},
  title     = {Validity of the Taylor Hypothesis for Linear Kinetic Waves in the Weakly Collisional Solar Wind},
  journal   = {The Astrophysical Journal},
  year      = {2014},
  volume     = {789},
  number     = {2},
  pages      = {106},
  doi        = {10.1088/0004-637X/789/2/106}
}

@article{Klein2014Violation,
  author  = {Klein, K. G. and Howes, G. G. and TenBarge, J. M.},
  title   = {The Violation of the Taylor Hypothesis in Measurements of Solar Wind Turbulence},
  journal = {The Astrophysical Journal Letters},
  year    = {2014},
  volume  = {790},
  number  = {2},
  pages   = {L20},
  doi     = {10.1088/2041-8205/790/2/L20}
}
\endgroup
\end{document}